\documentclass[aps,floats,amssymb,amsmath,prb,10pt,nofootinbib,twocolumn,superscriptaddress,longbibliography]{revtex4-2}

\usepackage[utf8]{inputenc}
\usepackage{graphicx}
\usepackage{color}
\usepackage{units}
\usepackage[export]{adjustbox}
\usepackage[unicode=true,pdfusetitle,
 bookmarks=true,bookmarksnumbered=false,bookmarksopen=false,
 breaklinks=false,pdfborder={0 0 0},backref=false,colorlinks=true]
 {hyperref}
\usepackage{geometry}
\pdfpageattr {/Group << /S /Transparency /I true /CS /DeviceRGB>>} %necessary for transparent plots

\newcommand{\mc}{m^*_\mathrm{c}}
\newcommand{\mtr}{m^*_\mathrm{tr}}
\newcommand{\mtot}{m^*_\mathrm{tr}}

\usepackage{pdfpages}
\makeatletter
\AtBeginDocument{\let\LS@rot\@undefined}
\makeatother

\begin{document}

\title{Hubbard vs. Emery model: spectra, transport and relevance for cuprates}
\author{J. Vu\v ci\v cevi\'c}
\affiliation{Scientific Computing Laboratory, Center for the Study of Complex Systems, Institute of Physics Belgrade,
University of Belgrade, Pregrevica 118, 11080 Belgrade, Serbia}
\author{R. \v{Z}itko}

\affiliation{Jo\v{z}ef Stefan Institute, Jamova 39, SI-1000 Ljubljana,
Slovenia}
\affiliation{Faculty  of Mathematics and Physics, University of
Ljubljana, Jadranska 19, SI-1000 Ljubljana, Slovenia}

\begin{abstract}
Understanding the transport properties of cuprate superconductors is one of the central challenges in the physics of strongly correlated electrons.
The most common approach is to define and solve a low-energy lattice model,
but it is still unclear what the minimal model is to capture all relevant mechanisms and provide quantitative predictions.
The main uncertainty concerns the choice of the orbital degrees of freedom to be included in the model, as well as the definition of the effective coupling.
In this paper, we study the two most commonly considered models, namely the single-orbital Hubbard model and the three-orbital Emery model.
We investigate and compare their spectral and transport properties,
and find that the two models provide a similar, but not identical, physical picture. 
We identify several strong quantitative differences which might allow one to discriminate between the two models by comparing theory with experiments.
We compare our results for several physical quantities with 7 different experiments on 3 different La$_2$CuO$_4$-based cuprates, and in general find good quantitative agreement.
The dc resistivity and the effective mass results suggest that the coupling constant in the effective Hubbard model is larger than expected.
We find several more properties that are sensitive to the precise value of the coupling constant, including the critical doping for the Lifshitz transition, and the local spectral weight in the vicinity of the Fermi level; the latter provides a promising way to estimate the effective coupling constant in future photoemission experiments.
\end{abstract}

\maketitle

\section{Introduction}

Unconventional superconductivity is a strong motivation to study simplified low-energy lattice models \cite{Scalapino2012,Keimer2015,Varma2020}. 
Previous works have shown that the single-band Hubbard model can qualitatively describe the normal-state transport properties in several classes of unconventional superconductors, including the cuprates\cite{VucicevicPRL2015}, $\kappa$-organics\cite{Limelette2003,Terletska2011,vucicevic2013,Furukawa2015}, and moiré systems\cite{VucicevicPRL2021, VucicevicPRB2021,Li2021Nature}. Spectral properties of the cuprates are also qualitatively described by the Hubbard model, in particular the node-antinode dichotomy in the pseudogap regime and the superconducting phase\cite{Maier2000,Kyung2006,Civelli2008NodalAntinodalTwoGaps,Ferrero2009,Gull2013,WuPRB2017,Wu2018}. However, more recent studies of the magnitude of the superconducting critical temperature in the cuprates have suggested that other models might be more appropriate\cite{KowalskiPNAS2021,Jiang2021,VucicevicPRB2024}. One possibility is to consider a three-band Emery model \cite{Kent2008,Weber2012,KowalskiPNAS2021,VucicevicPRB2024,Malcolms2026} which includes additional orbital degrees of freedom. Another possibility is to consider a more general single-band model that involves interactions beyond the local density-density coupling\cite{Castellani1995,WangPRL2021,JiangScalapinoWhite2023_Downfolding,Lange2026} included in the Hubbard model.

The standard Hubbard model has been thoroughly studied for many decades\cite{dagotto1994, Bergeron2011,Sordi2013cAxisResistivityPRB,deng,perepelitsky2016,Huang2019,georges1996,Lee2006DopedMott,LeBlanc2015Solutions2DHubbard,Arovas2022HubbardModel,Qin2022ComputationalPerspective}, but in recent years the attention has been increasingly shifting towards the Emery model \cite{ZaanenSawatzkyAllen1985,Emery1987,Zoelfl2000EPJB_CuO2_DMFT,WeberHauleKotliar2008_LDA_DMFT,Kent2008,deMediciWangCaponeMillis2009,HankeKieselAichhornBrehmArrigoni2010,Weber2012,Weber2014,Hansmann2014,MaiEtAl2021_PRB,Liu2024EmeryScatteringRate,Mao2024,StCyrSenechal2025,Zhao2025,Tseng2025,Malcolms2026}. However, the comparison between the results obtained in the two models has been so far performed only in a limited way\cite{HankeKieselAichhornBrehmArrigoni2010, Malcolms2026}. In particular, the transport properties of the Emery model have been studied only very recently\cite{Zhao2025}, at the time of writing this manuscript.

The sensitivity of cuprate compounds to doping is one of the central aspects of their physics.
In the standard lattice-model approach, the dopant concentration is modeled indirectly, by varying the chemical potential; therefore, the carrier density becomes an input parameter of the calculation rather than the result of it\cite{Maier2000,Weber2012,Gull2013,deng,VucicevicPRL2015,Vucicevic2018,Vucicevic2019,VucicevicPRB2021,VucicevicPRL2021,Zhao2025}. 
Yet, the relation between carrier density and dopant concentration is not necessarily simple. 
For this reason, more \emph{ab initio} approaches that avoid formulating a lattice model are of great interest. 
For example, charge-self-consistent schemes that combine the density functional theory (DFT) and the dynamical mean field theory (DFT+DMFT\cite{BhandaryPRB2016,She2026} or eDMFT \cite{HaulePRB2010,BacqLabreuil2025}) allow one to model the dopant concentration directly using virtual crystal approximation \cite{Bellaiche2000,BacqLabreuil2025}. 
However, charge-self-consistent schemes run into the same problem of having to choose the correlated orbitals and the effective coupling constants\cite{BacqLabreuil2025}. In fact, computing the coupling constants has been one of the most difficult problems in the field, as it formally represents a quantum many-body problem in its own right \cite{Imada2010JPSJ_RMS_Review}.

Namely, downfolding an \emph{ab initio} Hamiltonian to a few-band lattice model involves integrating out interacting degrees of freedom; the resulting action should in principle involve all possible time-dependent many-body interactions\cite{Profe2026}. Nevertheless, the common wisdom is that the dominant interactions can be taken as two-body, local and instantaneous (e.g., the Hubbard interaction), and that one can choose an effective coupling constant $U$ that captures the major effects of the screening processes omitted in the downfolded model. There are many theoretical approaches to downfolding and estimating the effective Hubbard-$U$ \cite{Dederichs1984PRL_cDFT,McMahan1988_PRB38_6650,White2002,Aryasetiawan2004PRB_cRPA_freqU,kotliar2006,Yanai2006,Miyake2008,Aryasetiawan2009PRL_DownfoldedSelfEnergy,Imada2010JPSJ_RMS_Review,Honerkamp2012,Vaugier2012,Changlani2015,Shinaoka2015,Werner2016DynamicalScreening,Bauman2019,JiangScalapinoWhite2023_Downfolding,Sharma2023,Chang2024,Scott2024,Chang2024b,Canestraight2025}, but true control over error bars is not feasible at present. The most widely adopted way of computing $U$ is the constrained random phase approximation (cRPA)\cite{Springer1998,Miyake2008,Vaugier2012,Shinaoka2015,Scott2024,Chang2024b}; it yields a frequency-dependent coupling amplitude, going from a highly renormalized small value at low frequency, to the very high bare-Coulomb value at infinite frequency. The procedure to convert this result to a single Hubbard-$U$ value for an instantaneous interaction is not obvious\cite{PauliPRB2025}. The overall uncertainty is reflected in the range of $U$-values considered in the literature: in different studies of the cuprates $U$ has been taken to be anywhere from about \unit[3]{eV} to about \unit[14]{eV} \cite{Hybertsen1989,Kancharla2008,Sakakibara2010,Ahmed2012,Sakakibara2012a,Sakakibara2012b,Sakakibara2014,Werner2015,Choi2016,Ivashko2019,Tadano2019,Hirayama2018,Nilsson2019,Sheshadri2023,BacqLabreuil2025}. Certainly, the correct $U$ must be material-dependent, and different models (single-band vs. three-band) require different effective values of $U$.
The most appropriate value of $U$ to describe the cuprates is one of the field's central, long-standing open questions \cite{Emery1987}. The problem is exacerbated by the lack of clear $U$-sensitive features that would allow one to fix the value of $U$ by comparing theoretical and experimental results. The size of the energy gap above the conduction band is one obvious $U$-dependent feature\cite{Ahmed2012,OMahony2022}, but its precise relation with the value of $U$ has not been studied systematically enough\cite{Carrasquilla2010,HafezTorbati2021,Zhang2007}, i.e. across different models and for different compounds.

In this paper, we study the standard Hubbard and Emery models, aiming to resolve which one of these models provides the best description of the La$_2$CuO$_4$-based cuprates and to constrain the correct coupling in each case.

We compute the model parameters following the standard procedure of downfolding a DFT band structure \cite{kotliar2006,Held2007}, and do so for the La$_2$CuO$_4$ parent compound. We use the method of maximally localized Wannier functions (MLWF) \cite{marzari1997,Marzari2012MLWF,Nakamura2016}, starting from the Cu $3d_{x^2-y^2}$ and the appropriate O $2p$ orbitals of the copper and oxygen atoms that form the copper-oxide planes. We then use DMFT with the numerical renormalization group (NRG) impurity solver \cite{wilson1975,krishna1980a,bulla2008} to compute the local spectral function and the direct-current (dc) resistivity. At the level of DMFT, the vertex corrections vanish\cite{khurana1990,VucicevicPRB2021} and the dc resistivity can be computed using the Kubo bubble, without invoking additional approximations. The NRG is formulated in the real-frequency domain and thus avoids the issues of analytic continuation; the downside is that NRG is approximate, but it has been successfully benchmarked against numerically exact methods many times in the past \cite{deng,Kugler2020_PRL_016401,Stadler2015_PRL_136401,LaBollita2026,Kugler2026,Merker2012}.

Using DMFT+NRG, we scan the doping-temperature phase diagrams and generate an extensive dataset. 
We analyze these results and start by identifying systematic differences in the predictions of the Emery and Hubbard models.
We find that there is a robust quantitative difference between the local spectra in those models.
Most importantly, the size of the energy gap varies with doping much more strongly in the Hubbard than in the Emery model; in the Emery model it is roughly constant.
On the other hand, the density of states near the Fermi level has a strong dependence on the doping level, similarly in both models. We identify a qualitative feature, clearly evident in the case of the Hubbard model: the local density of occupied states, integrated over a small window of energy just below the Fermi level, reaches a maximum at the value of doping that is strongly dependent on $U$. The position of this maximum, if observed in a photoemission experiment, could be used to estimate the value of the effective coupling constant.

We also compute the momentum-resolved spectral function and compare it directly with available angle-resolved photoemission spectroscopy (ARPES)\cite{Damascelli2003,Sobota2021,Zhang2022,Vishik2018} measurements, in particular Ref.~\onlinecite{Yoshida2009}.
With both the Hubbard and Emery models, we find good agreement, but only after broadening our results with an additional $\sim$\unit[0.5]{eV} imaginary self-energy. The need for additional broadening could reflect the significant disorder in the cuprates\cite{Zhou2004,Alloul2024}, which is not considered at the level of our (clean) lattice models.

We compute the dc resistivity $\rho_\mathrm{dc}$ and find that it is significantly higher in the Emery model than in the Hubbard model when 
the size of the energy gap in the spectral function is well matched between the two models, which is when $U^\mathrm{Hubbard} \approx \frac{2}{3}U^\mathrm{Emery}$.
Similar dc resistivity in both models is only obtained when $U^\mathrm{Hubbard} \approx \frac{5}{4}U^\mathrm{Emery}$, but then the spectral functions are no longer similar.
Interestingly, the best matching between the local spectra of Emery and Hubbard models at energies around the Fermi level is observed when $U^\mathrm{Hubbard} \approx U^\mathrm{Emery}$.
This provides a clear example where the effective scattering rate that determines the transport properties is not simply related to the width of the quasiparticle peak in the single-particle spectrum.

We compare our results to the experimental measurements on single crystals of Sr-doped (LSCO)\cite{Cooper2009}, Ba-doped (LBCO)\cite{Tee2017} and Ce-doped (LCCO)\cite{Naito} La$_2$CuO$_4$ (the latter being grown as a thin film on a substrate using molecular beam epitaxy).
The Emery model $\rho_\mathrm{dc}$ results (assuming the conventional $U^\mathrm{Emery}=\unit[8]{eV}$) have the correct magnitude and are in excellent agreement with LSCO and LBCO, but less so with LCCO. The Hubbard model resistivity is too small when $U^\mathrm{Hubbard}$ is taken to be of order \unit[4--5]{eV}, as suggested by cRPA estimates\cite{Tadano2019,Sheshadri2023,Schmid2023}. However, with a larger value of $U^\mathrm{Hubbard}\approx \unit[8-10]{eV}$ it is possible to obtain an excellent and very systematic agreement with experiment, for all three compounds. 
The agreement with LCCO, however, is good only at low temperature below about \unit[150]{K} where $\rho_\mathrm{dc}(T)$ dependence (at a fixed doping) is linear.
At higher temperatures, one observes $\rho_\mathrm{dc}(T)$ with significantly larger curvature in the experiment, which likely comes from some effects not accounted for in our theory (e.g. the effective retardation of interactions or electron-phonon coupling), or even temperature-dependent effective doping, for example due to oxygen loss\cite{takagi1992}.

Finally, we look at two more important quantities that have been previously studied in experiment. First, LSCO is known to undergo a Lifshitz transition (LT) at about 21\% doping \cite{Ino2002,yoshida2006,Zhong2022}; at that point, the Fermi surface changes topology, from hole-like to electron-like. We observe such a LT in our results; the precise value of doping where it occurs, $\delta_\mathrm{LT}$, depends on the coupling $U$, but ostensibly not so much on the model. The dependence of $\delta_\mathrm{LT}$ on $U$ is relatively weak, but it can be used to narrow down the range of experimentally relevant $U$ values. More importantly, both the Hubbard and the Emery model, when their parameters are fixed so that $\rho_\mathrm{dc}$ results agree with experiment, they also (roughly) reproduce the correct value of $\delta_\mathrm{LT}$. This is consistent with the number of holes ($\delta$) in the model being equal to the dopant concentration $x$ in the experiment, at least when it comes to LSCO. This is an important finding that justifies the lattice-model approach, as $\delta$ is not expected to be the same as $x$ in general, especially in more complicated multi-layer cuprates\cite{BacqLabreuil2025}.

The second quantity of great interest is the effective (cyclotron, or thermodynamic) mass, $\mc$. In our theory, $\mc$ can be extracted from the momentum-resolved spectral function. Again, we find that when our results for $\rho_\mathrm{dc}$ are in good agreement with experiment, the results for $\mc$ are in good agreement with experiment, too, at least for LSCO. Our findings are relevant for the Planckian dissipation hypothesis \cite{Zaanen2004,Zaanen2019,Hartnoll2022}. This hypothesis aims to explain linear-in-temperature strange-metallic resistivity in terms of a saturated $T$-linear bound for the scattering rate. Multiple experiments on various materials  \cite{martin1990,Batlogg1991,takagi1992,Marel2003,Bruin2013,Legros2018,Grissonnanche2021} have shown that in strange-metal regimes, the (inelastic) scattering rate $\hbar/\tau$ indeed scales as $\alpha k_\mathrm{B} T$ with $\alpha$ around 1. From the knowledge of $\rho_\mathrm{dc}$ and $\mc$ in our theory, we reproduce the value $\alpha \approx 1$, but find that it is also strongly $U$-dependent. This raises questions about the universality of this experimentally observed phenomenon. 
More importantly, we find inconsistencies in the way the inelastic scattering rate has been extracted from experiment. More specifically, it appears that the effective cyclotron mass appearing in the Sommerfeld expression for the specific heat coefficient has been used interchangeably with the effective transport mass that appears in the Drude formula. We show with a simple example that in the vicinity of a Lifshitz transition, it is not justified to identify the two effective masses. Our findings might have implications for the validity of the conclusions made in previous experimental studies of Planckian dissipation:
under the same assumptions, the effective scattering rate might be as much as 5 times higher than was previously thought.
However, we also recognize that the definition of scattering rate is ambiguous and we discuss this in detail.
We find that scattering rate estimates based on dc transport alone will always depend on the assumptions made for the effective mass and the number of carriers, or at least their ratio.
On the other hand, the estimates based on frequency-dependent optical conductivity may be better defined, but we find such estimates may be inconsistent with the assumptions of Planckian dissipation.

Overall, our results provide a systematic assessment of how well the standard lattice models capture the spectral and transport properties of the La$_2$CuO$_4$-based cuprates. We show that quantitative agreement with experimental results is possible across multiple compounds and physical quantities. Our work suggests future directions for both theoretical and experimental work. 
Further investigations are necessary to understand our unexpected finding that a large coupling constant in the Hubbard model is needed to reproduce the experimental results.
On the other hand, momentum-integrated photoemission experiments could be devised to narrow down the possible values of the effective coupling in the standard Hubbard and Emery models.

The rest of the paper is organized as follows. In Sec.~\ref{sec:models} we describe our models and the procedure to parametrize them. In Sec.~\ref{sec:methods} we explain our methodology, i.e. the details of the DFT calculations we used as a starting point for our lattice models (Sec.~\ref{sec:dft_and_wannierization}), the DMFT solution of those models (Sec.~\ref{sec:dmft}, App.~\ref{app:nrg_numerical}), the computation of the resistivity (Sec.~\ref{sec:Kubo_bubble}, App.~\ref{app:velocity_matrix}), and our approach to phase diagram scans (Sec.~\ref{sec:phase_diagram_scan}). In Sec.~\ref{sec:results} we present our numerical data. In Sec.~\ref{sec:hubbard_vs_emery_spectra} we examine the spectra of the Hubbard and Emery models and compare them with ARPES experiments, with additional details given in App.~\ref{app:orbital_resolved}. In Sec.~\ref{sec:hubbard_vs_emery_transport} we compare the resistivity between Hubbard and Emery models, and then compare our transport data with experiment in Sec.~\ref{sec:transport_vs_experiment}. 
Then we study the Lifshitz transition in our data and compare to experiments in Sec.~\ref{sec:lifshitz}.
Finally, we compare our effective mass results to experiment in Sec.~\ref{sec:effective_mass} with further details given in App.~\ref{app:effective_masses} and App.~\ref{app:Drude_and_Planck}. In Sec.~\ref{sec:discussion} we discuss our findings in detail, connect with existing literature, and envision prospects for future work.

\section{Models}
\label{sec:models}

\begin{figure}[t]
\centering
\includegraphics[width=0.4\textwidth, trim=0 0 0 0.4cm, clip]{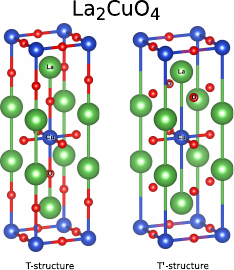}
\caption{Two possible crystal structures for compounds obtained by doping La$_2$CuO$_4$, denoted T and T'. T-structure is relevant for LSCO and LBCO; T'-structure is relevant for LCCO above about 5\% doping\cite{Naito}.
}
\label{fig:crystal_structures}
\end{figure}

\begin{figure*}
    \includegraphics[width=0.4\textwidth,valign=t,trim=0 0 0 1.5cm]{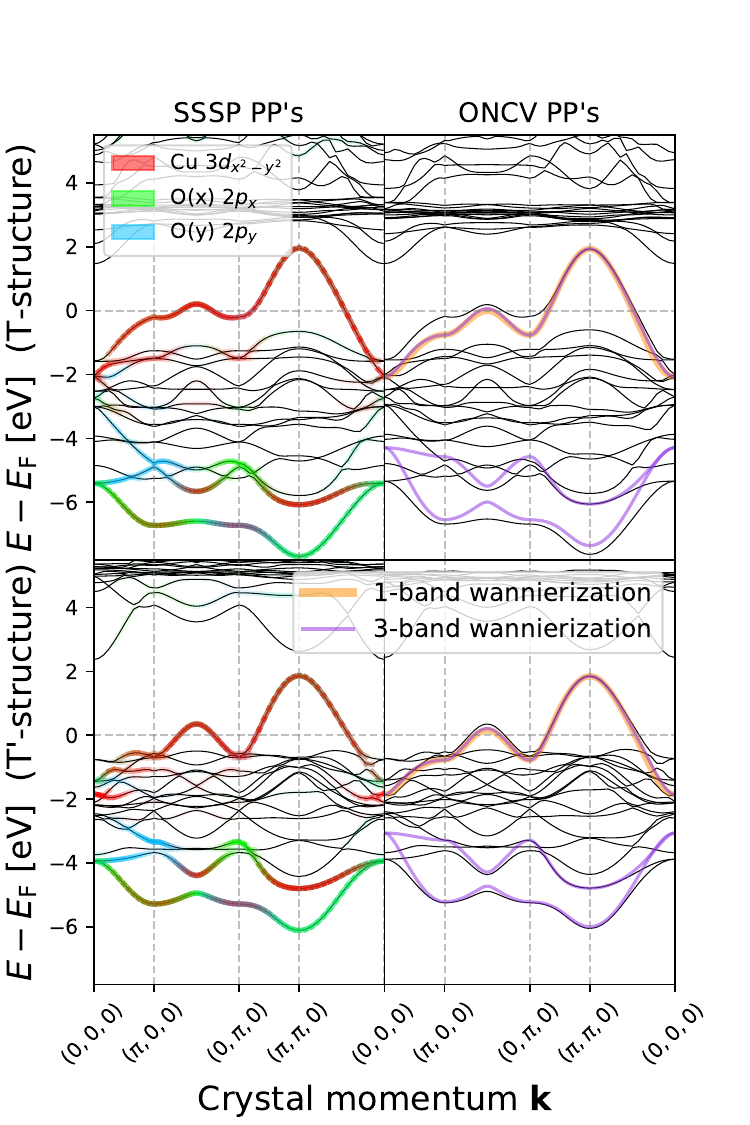}
    \includegraphics[width=0.4\textwidth,valign=t]{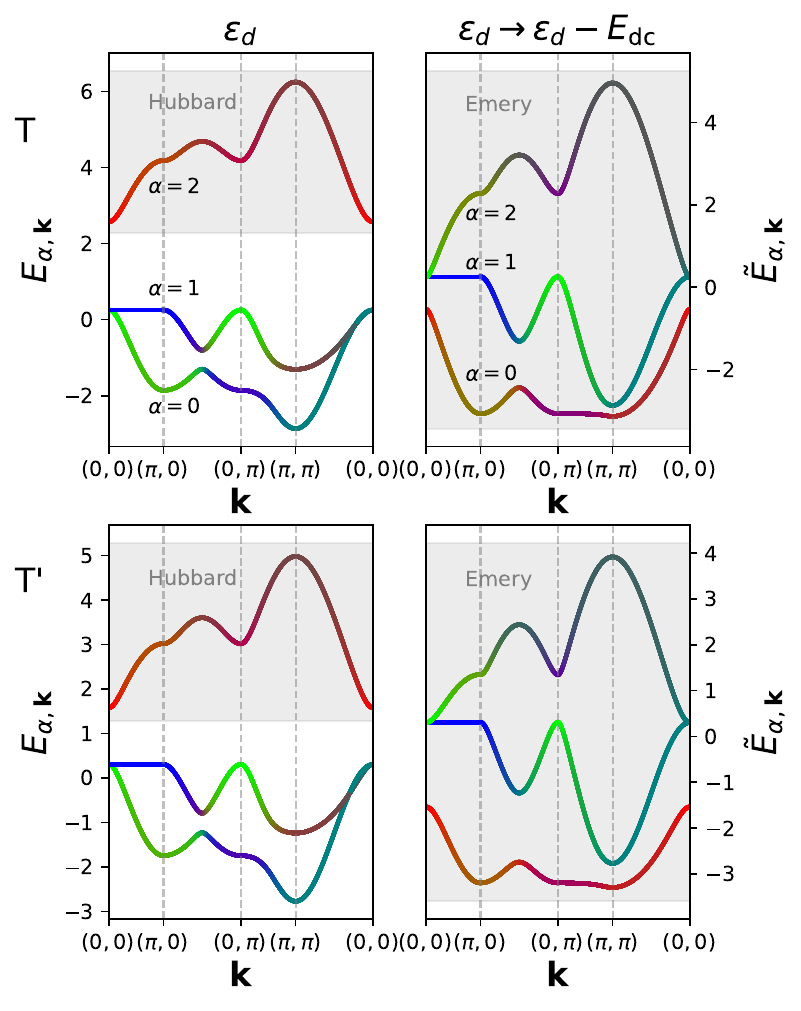}
\caption{Left: Orbital-projected DFT band structure and wannierized band structures for the two possible crystal structures of La$_2$CuO$_4$-based compounds (top: T, bottom: T'). Right: non-interacting band structures for our models to be solved using DMFT; truncating long-range hoppings in the three-band picture slightly modifies the band structure. Subtracting the double-counting shift further modifies the band structure. The single-band Hubbard model dispersion is identified with the corresponding band in the three-band model, before subtracting the double-counting shift.}
\label{fig:dft}
\end{figure*}

We start by considering the crystal structure of the parent compound La$_2$CuO$_4$. There are two relevant crystal structures, usually referred to as the T and T' structures\cite{Das, Naito}, shown in Fig.~\ref{fig:crystal_structures}. The main difference is that in the T-structure one finds ``apical oxygens'' placed just below and above the copper atoms (along the c-axis), while in the T'-structure the oxygens are instead found on the faces of the conventional unit cell. The T-structure is relevant for the hole-doped compounds LSCO and LBCO, while the T'-structure is relevant for the electron-doped LCCO, which goes through a structural phase transition from T to T' at a small doping of around 5\%\cite{Naito}.
Another subtlety is that LSCO and LBCO undergo a structural phase transition at intermediate temperatures $\sim \unit[300]{K}$ (depending on doping) to a slightly different crystal structure that features buckling and enlargement of the unit cell (LTO phase)\cite{Fabbris2013,Tidey2022}, but is overall similar to the high-temperature HTT phase that we are considering for the sake of simplicity. LBCO goes through another structural phase transition at even lower temperature $\sim\unit[50]{K}$ and about 1/8 doping\cite{Karapetyan2012}, but we will not be focusing on that part of the phase diagram.

We take the crystal structures from The Materials Project database \cite{Horton2025,Jain2013}, with IDs \texttt{mp-19735} for the T-structure and \texttt{mp-1077929} for the T'-structure. We perform the DFT calculation using Quantum Espresso \cite{giannozzi2009,Giannozzi2017}, to obtain the electronic band structure for both crystal structures. The band structures are obtained with two sets of pseudopotentials (SSSP\cite{prandini2018precision} and ONCV\cite{Hamann2013}) and we have found no significant difference between the respective results. The DFT band structures are shown in Fig.~\ref{fig:dft} on the left. Using SSSP pseudopotentials, we were also able to compute the orbital character of the Bloch states - the contributions of the relevant Cu $3d$ and O $2p$ orbitals are color-coded on top of the band structure in Fig.~\ref{fig:dft} (first column). We then use standard wannierization tools to obtain the parameters for the tight-binding matrix (TB-matrix) of the three-band Emery model (more details given in Section~\ref{sec:dft_and_wannierization}). The TB-matrix for the Hubbard model can be obtained in the same way by wannierization of a single band at the Fermi level; as expected, we find that the resulting downfolded dispersion coincides almost perfectly with the corresponding band in the three-band Emery model. The downfolded band structures are overlaid on top of the DFT band structures in Fig.~\ref{fig:dft} (second column).

Formally, the Emery model Hamiltonian in the orbital basis $l=0,1,2\equiv d,p_x,p_y$ reads
\begin{eqnarray} \nonumber
 \hat{H}^{\mathrm{Emery}} &=& \hat{H}^{\mathrm{Emery}}_0 + \hat{H}^{\mathrm{Emery}}_\mathrm{int} \\
          \hat{H}^{\mathrm{Emery}}_0                &=& \sum_{\sigma,\mathbf{r}\mathbf{r}'} \mathbf{\Psi}^\dagger_{\sigma,\mathbf{r}} \mathbf{h}_{\mathbf{r}\mathbf{r}'} \mathbf{\Psi}_{\sigma,\mathbf{r}'}\\ \nonumber
          \hat{H}^{\mathrm{Emery}}_\mathrm{int} &=& U^{\mathrm{Emery}}\sum_\mathbf{r} c^\dagger_{l=0,\uparrow,\mathbf{r}}c^\dagger_{l=0,\downarrow,\mathbf{r}} c_{l=0,\downarrow,\mathbf{r}} c_{l=0,\uparrow,\mathbf{r}}
\end{eqnarray}
where $\mathbf{r}$ denotes the position of the unit cell (set to the position of the $d$-orbital within the unit cell) and $\sigma=\uparrow,\downarrow$ is the spin projection.
\begin{widetext}
For convenience, we have defined the row vector of creation operators
\begin{equation}
\mathbf{\Psi}^\dagger_{\sigma,\mathbf{r}} = ( c^\dagger_{l=0,\sigma,\mathbf{r}}, c^\dagger_{l=1,\sigma,\mathbf{r}}, c^\dagger_{l=2,\sigma,\mathbf{r}} )
\end{equation}
and a $3\times 3$ TB-matrix
\begin{equation}\label{eq:Horig_real_space}
\footnotesize
 \mathbf{h}_{\mathbf{r}\mathbf{r}'} = \left(
 \begin{array}{ccc}
    \varepsilon_d \delta_{\mathbf{r}\mathbf{r}'}
       & t_{pd}(\delta_{\mathbf{r}\mathbf{r}'}-\delta_{\mathbf{r}+\mathbf{e}_x,\mathbf{r}'})
       & t_{pd}(\delta_{\mathbf{r}\mathbf{r}'}-\delta_{\mathbf{r}+\mathbf{e}_y,\mathbf{r}'}) \\
    t_{pd}(\delta_{\mathbf{r}\mathbf{r}'}-\delta_{\mathbf{r}-\mathbf{e}_x,\mathbf{r}'})
        & \varepsilon_p \delta_{\mathbf{r}\mathbf{r}'} + t'_{pp}(\delta_{\mathbf{r}+\mathbf{e}_x,\mathbf{r}'} + \delta_{\mathbf{r}-\mathbf{e}_x,\mathbf{r}'})
        & t_{pp}(\delta_{\mathbf{r}\mathbf{r}'}-\delta_{\mathbf{r}+\mathbf{e}_y,\mathbf{r}'}
                 -\delta_{\mathbf{r}-\mathbf{e}_x,\mathbf{r}'}+\delta_{\mathbf{r}-\mathbf{e}_x+\mathbf{e}_y,\mathbf{r}'})\\
    t_{pd}(\delta_{\mathbf{r}\mathbf{r}'}-\delta_{\mathbf{r}-\mathbf{e}_y,\mathbf{r}'})
        & t_{pp}(\delta_{\mathbf{r}\mathbf{r}'}-\delta_{\mathbf{r}-\mathbf{e}_y,\mathbf{r}'}
                 -\delta_{\mathbf{r}+\mathbf{e}_x,\mathbf{r}'}+\delta_{\mathbf{r}+\mathbf{e}_x-\mathbf{e}_y,\mathbf{r}'})
        & \varepsilon_p \delta_{\mathbf{r}\mathbf{r}'} + t'_{pp}(\delta_{\mathbf{r}+\mathbf{e}_y,\mathbf{r}'} + \delta_{\mathbf{r}-\mathbf{e}_y,\mathbf{r}'})
 \end{array}
 \right)
\end{equation}
\end{widetext}

\begin{figure}
    \includegraphics[width=0.8\columnwidth,valign=t,trim=0 0 0 0cm]{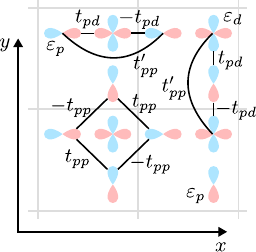}    
\caption{Illustration of the physical meaning of the tight-binding matrix elements in the Emery model.}
\label{fig:real_space_illustration}
\end{figure}

We parametrize the TB part of the Emery model Hamiltonian by the onsite energies $\varepsilon_d$ and $\varepsilon_p$, and the hopping amplitudes $t_{pd}$, $t_{pp}$ and $t'_{pp}$. The physical meaning of these hopping amplitudes is illustrated in Fig.~\ref{fig:real_space_illustration}. The non-interacting Hamiltonian resulting from the wannierization of the DFT band structure also contains longer-range $\mathbf{h}_{\mathbf{r}\mathbf{r}'}$ components. However, for the sake of reproducibility, it is important that our models can be defined using a small set of numbers. For this reason, we truncate all longer range hoppings. This introduces only a minor modification of the downfolded band structure, comparable in size to the discrepancy between the DFT band structure and the original downfolded one. The computed values of the TB-parameters for both T and T' crystal structures are summarized in Table~\ref{tab:my_table}.

\begin{table}[h!]
    \centering
    \caption{Tight-binding parameters for the Emery model ($\varepsilon_d-\varepsilon_p$ is given before double-counting subtraction)}
    \begin{tabular}{c|cccc} % Five centered columns with vertical lines
        structure & $\varepsilon_d-\varepsilon_p$ & $t_{pd}$ & $t_{pp}$ & $t'_{pp}$ [eV]\\
        \hline
        T & 2.579 & 1.335 & 0.654 & 0.126 \\ % Second row (data)
        T' & 1.582 & 1.094 & 0.616 & 0.152 \\ % Second row (data)
    \end{tabular}
    \label{tab:my_table}
\end{table}

We can diagonalize the non-interacting part of the Hamiltonian to obtain:
\begin{equation}
 \hat{H}^{\mathrm{Emery}}_0 = \sum_{\sigma,\alpha,\mathbf{k}} E_{\alpha,\mathbf{k}} d^\dagger_{\alpha,\sigma,\mathbf{k}} d_{\alpha,\sigma,\mathbf{k}}
\end{equation}
where $\alpha=0,1,2$ enumerates the eigenbands so that $\alpha'>\alpha \Rightarrow E_{\alpha',\mathbf{k}}>E_{\alpha,\mathbf{k}}$, and $\mathbf{k}$ denotes the crystal momenta in the first Brillouin zone (BZ).

Finally, instead of parameterizing the Hubbard model based on an independent wannierization, we adopt the dispersion relation of the highest-energy band ($\alpha=2$) in the Emery model as the dispersion in the Hubbard model. This is a negligible modification of the single-band wannierization results, yet it allows for a more straightforward comparison between the two models. The Hubbard model Hamiltonian then reads
\begin{eqnarray} \nonumber
 \hat{H}^{\mathrm{Hubbard}} &=& \hat{H}^{\mathrm{Hubbard}}_0 + \hat{H}^{\mathrm{Hubbard}}_\mathrm{int} \\
 \hat{H}^{\mathrm{Hubbard}}_0 &=&  \sum_{\sigma,\mathbf{k}} E_{\alpha=2,\mathbf{k}} c^\dagger_{\sigma,\mathbf{k}}  c_{\sigma,\mathbf{k}}  \\ \nonumber
 \hat{H}^{\mathrm{Hubbard}}_\mathrm{int} &=& U^{\mathrm{Hubbard}}\sum_\mathbf{r} c^\dagger_{\uparrow,\mathbf{r}}c^\dagger_{\downarrow,\mathbf{r}} c_{\downarrow,\mathbf{r}} c_{\uparrow,\mathbf{r}}
\end{eqnarray}
Note that $\hat{H}^{\mathrm{Hubbard}}_0$ can be Fourier transformed back to the real-space basis to yield
$$\hat{H}^{\mathrm{Hubbard}}_0=\sum_{\sigma,\mathbf{r}\mathbf{r}'} t_{\mathbf{r}-\mathbf{r}'} c^\dagger_{\sigma,\mathbf{r}}  c_{\sigma,\mathbf{r}'}$$
and $t_{\mathbf{r}-\mathbf{r}'}$ will in general have infinite range, even though we have truncated the range of hopping in the corresponding Emery model.

In both models we only keep the on-site density-density coupling, and in the Emery model, the interaction is retained only on the $d$-orbitals. As already mentioned, the effective coupling constants $U$ for both models are not well known, but the common approach is to compute them approximately using cRPA. 
One can find a wide range of estimates in the available
literature\cite{Hybertsen1989,Kancharla2008,Sakakibara2010,Weber2010,Weber2012,Ahmed2012,Sakakibara2012a,Sakakibara2012b,Sakakibara2014,Werner2015,Jang2016,Choi2016,Ivashko2019,Tadano2019,Nilsson2019,Sheshadri2023,BacqLabreuil2025}, but the values appear to cluster around \cite{Weber2010,Weber2012,Hirayama2018,Nilsson2019,Tadano2019,Schmid2023,Sheshadri2023}
\begin{equation}
 U^{\mathrm{Hubbard}}=4\mathrm{eV}, \;\;\;\; U^{\mathrm{Emery}}=8\mathrm{eV}
\end{equation}
The large difference between the coupling constants in the two effective models is easily understood---downfolding to a larger number of bands (orbitals per unit cell) yields Wannier orbitals that are more localized and thus have effectively stronger on-site couplings.
However, as we will see, the half-filled T-structure Hubbard model turns out not to be a Mott insulator at $U^{\mathrm{Hubbard}}=\unit[4]{eV}$, at least at the level of our DMFT solution.
We will study how the results in the Hubbard model depend on the coupling constant over a wide range of values, and we will consider the Emery model at $U^{\mathrm{Emery}}=8\mathrm{eV}$.

An important subtlety in the downfolding procedure is that some effects of interactions are already included at the level of DFT calculation. Solving a downfolded model using many-body techniques can thus lead to double counting of interaction effects. This is not an issue in the case of the Hubbard model, but in the Emery model one needs to appropriately shift the onsite energy on the interacting $d$-orbital to subtract the Hartree shift that was already introduced at the level of DFT. This shift is not well known. We take the estimate from the literature $E_\mathrm{dc}=\unit[3.12]{eV}$\cite{Weber2012}, consistent with the estimate of $U^{\mathrm{Emery}}=\unit[8]{eV}$ that we use.

\section{Methods}
\label{sec:methods}

\subsection{Extraction of effective lattice-model parameters from DFT calculations}
\label{sec:dft_and_wannierization}
The DFT band structures that we have downfolded (to get our effective models, as discussed in the previous section) have been obtained using Quantum Espresso. We have used SG15 ONCV PBE pseudopotentials\cite{Hamann2013}. In the calculation we have used the primitive unit cell, a $6\times6\times6$ $\mathbf{k}$-point mesh, and the kinetic energy cutoffs were set to \unit[80]{Ry} and \unit[400]{Ry} for the wavefunctions and charge density, respectively. We used Gaussian smearing, and \unit[0.02]{Ry} spreading. For wannierization we have used \texttt{RESPACK}\cite{Nakamura2016,Nakamura2021} and have checked that \texttt{Wannier90}\cite{Mostofi2014} gives the same result. In the wannierization we have not used the frozen window.

\subsection{DMFT solution for the self-energy}
\label{sec:dmft}

We solve our models using DMFT. The underlying approximation is that the self-energy is fully local, $\Sigma_\mathbf{k}(\omega)\approx \Sigma(\omega)$. It is computed within an effective Anderson impurity problem using NRG. NRG introduces additional approximations, but these have been shown in multiple previous works to be small by benchmarking with numerically exact quantum Monte Carlo impurity solvers and other methods \cite{deng,Kugler2020_PRL_016401,Stadler2015_PRL_136401,Merker2012}. The advantage of NRG is that it produces results on the real-frequency axis; in this way we avoid the ill-defined analytic continuation, which is particularly important when studying dynamical response functions. We also emphasize that the DMFT treatment of the Emery model does not involve any approximations beyond those already present in the Hubbard-model calculations.
DMFT is not expected to work well at very low temperature because it omits nonlocal correlations. Our calculations are restricted to translationally invariant paramagnetic normal-state solutions.
For a full explanation of the DMFT and NRG methods we refer the reader to the ample literature on these subjects\cite{georges1992,georges1996,krishna1980b,bulla2008,resolution}.

The solution of DMFT for the Emery model follows the following steps:
\begin{enumerate}
 \item Choose an initial guess for the self-energy (if no better approximation is available, we start from $\Sigma_{dd}(\omega)$ with an approximate Hartree shift and a moderate frequency-independent imaginary part; starting from $\Sigma_{dd}=0$ does not converge well or at all)
 \item Compute the $dd$-component of the lattice local Green's function as
\begin{equation} \label{eq:Gddloc}
  G_{\mathrm{loc}, dd}(\omega) = \int_\mathrm{BZ} \frac{\mathrm{d}\mathbf{k}}{(2\pi)^2} G_{\mathbf{k},dd}(\omega)
\end{equation}
 where
 \begin{equation} \label{eq:Gddk}
   \mathbf{G}_{\mathbf{k}}(\omega) = [(\omega+\mu)\mathbf{I}_{3\times3} - \mathbf{h}_\mathbf{k} - \boldsymbol{\Sigma}(\omega)]^{-1}
 \end{equation}
 is the momentum-dependent retarded Green's function, expressed as a $3\times 3$-matrix in the orbital space.
 The chemical potential $\mu$ can be either preset, or tuned in this step to enforce the desired level of doping.
 In the same basis, the self-energy reads
 $$ \boldsymbol{\Sigma}(\omega) = \left(\begin{array}{ccc}
                                   \Sigma_{dd}(\omega) & 0 & 0 \\
                                   0 & 0 & 0 \\
                                   0 & 0 & 0
                                  \end{array}\right) $$
  and the non-interacting Hamiltonian is given by
  \begin{equation}
  \mathbf{h}_\mathbf{k} = \left(
  \begin{array}{ccc}
    \varepsilon_d & 2t_{pd}\sin\frac{ k_x}{2} & -2t_{pd}\sin\frac{k_y}{2} \\
    \mathrm{H.c.} & \varepsilon_p+2t'_{pp} \cos k_x & -4t_{pp} \sin\frac{k_x}{2}\sin\frac{k_y}{2} \\
    \mathrm{H.c.} & \mathrm{H.c.} &\varepsilon_p+2t'_{pp}\cos k_y
  \end{array}
  \right).
  \end{equation}
  We emphasize again that $\varepsilon_d$ here is shifted by the double-counting energy, $E_\mathrm{dc}$

 \item Compute the new hybridization function using
 $$\Delta(\omega) = \omega+\mu-\varepsilon_d-\Sigma_{dd}(\omega) - 1/G_{\mathrm{loc},dd}(\omega)$$

 \item Solve the impurity problem with the hybridization function $\Delta$ and compute the impurity self-energy $\Sigma^\mathrm{imp}$

 \item Approximate the lattice self-energy as $\Sigma_{dd}(\omega)\approx \Sigma^\mathrm{imp}$ and go to Step 2
\end{enumerate}
The DMFT loop is iterated until convergence.

The implementation of Step 2 is numerically non-trivial. The spectral function $\mathrm{Im}G_{\mathbf{k},dd}(\omega)$ is in many cases very sharply peaked, and thus converging the integral over momenta in Eq.~\eqref{eq:Gddloc} is difficult. We make use of the \texttt{cubepy}\cite{cubepy} adaptive integration library which implements the Genz-Malik adaptive quadrature algorithm\cite{Genz1980} using Berntsen's error-estimation scheme\cite{Berntsen1991}. Moreover, we avoid doing the full inverse of the $3\times 3$ matrix in Eq.~\eqref{eq:Gddk}, as only the $dd$-component is actually needed. Any other components of the Green's function are computed only after convergence, if needed (e.g. to inspect the orbital-resolved spectral function).

\begin{figure}
\centering
    \includegraphics[width=0.45\textwidth]{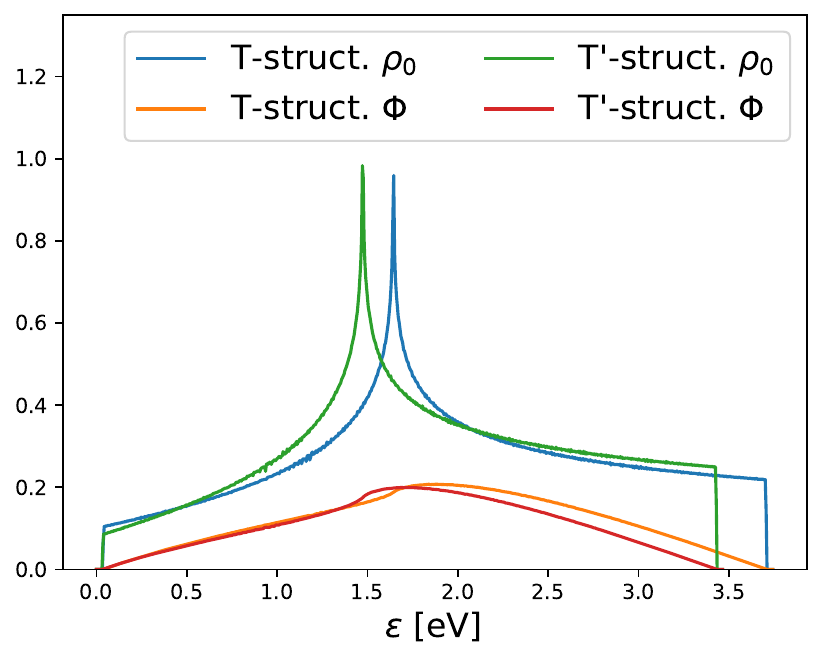}
\caption{Bare density of states $\rho_0$ and the velocity-weighted density of states $\Phi$ in the Hubbard model for the two
different crystal structures.}
\label{figphi}
\end{figure}

For the Hubbard model, the computation of the local Green's function in Step 2 is significantly simpler because it is a scalar (as there is only one orbital per unit cell) and one can convert the momentum integral into a frequency integral
\begin{equation}
 G_\mathrm{loc} = \int \mathrm{d}\varepsilon \frac{\rho_0(\varepsilon)}{\omega+\mu-\varepsilon-\Sigma(\omega)}
\end{equation}
where $\rho_0(\varepsilon)$ is the bare density of states, computed numerically as
\begin{equation}
 \rho_0(\varepsilon) = \int_\mathrm{BZ} \frac{\mathrm{d}\mathbf{k}}{(2\pi)^2} \delta(\varepsilon-E_{\alpha=2,\mathbf{k}})
\end{equation}
The densities of states for the two Hubbard models used in this work are shown in Fig.~\ref{figphi}.

\subsection{Kubo bubble for the dc conductivity}
\label{sec:Kubo_bubble}
We compute the conductivity using the Kubo bubble expression. This step does not introduce any additional approximations beyond those we already make in DMFT and NRG. Within single-site DMFT, the self-energy is approximated by a local quantity, which in turn leads to cancellation of vertex corrections to uniform conductivity.
The vanishing of vertex corrections in DMFT was established long ago for the Hubbard model\cite{khurana1990}, and it also holds in the presence of magnetic fields\cite{VucicevicPRB2021}. It is not immediately obvious that the same line of argument applies to the Emery model, but we show that it does in the following.

In the Emery model, the Kubo bubble for the DC conductivity can be expressed as
\begin{eqnarray} \label{eq:Kubo_bubble}
 &&\sigma_\mathrm{dc} \\ \nonumber
 &&= -\frac{2e^2}{\pi\hbar} \frac{N_\mathrm{layers}}{c}
                 \int_0^{2\pi} \frac{\mathrm{d}\mathbf{k}}{(2\pi)^2} \sum_{l_1,l_2,l_3,l_4} \\ \nonumber
 &&\times    \tilde{v}_{x,\mathbf{k},l_1l_2}\tilde{v}_{x,\mathbf{k},l_3l_4}\int\mathrm{d}\omega
              \mathrm{Im}G_{\mathbf{k},l_2l_3}(\omega)\mathrm{Im}G_{\mathbf{k},l_4l_1}(\omega)n'_\mathrm{F}(\omega)
\end{eqnarray}
where $G_\mathbf{k}$ is the momentum-dependent Green's function in the orbital basis defined in Eq.~\eqref{eq:Gddk}, $e$ is the electron charge, $c$ is the thickness of the conventional unit cell (T-structure: \unit[1.322]{nm}, T'-structure: \unit[1.259]{nm}), and $N_\mathrm{layers}$ is the number of CuO$_2$ planes per conventional unit cell (in our case $N_\mathrm{layers}=2$).
The first derivative of the Fermi-Dirac distribution is denoted $n'_\mathrm{F}$. 
Note that the conductivity does not explicitly involve the in-plane lattice constant $a$.
For the sake of simplicity, throughout the paper we take the BZ to extend from (0,0) to ($2\pi$,$2\pi$); we keep all the crystallographic quantities and fundamental constants in the prefactor and define the reduced velocity in the $x$-direction by the (matrix-valued) derivative
\begin{equation}\label{eq:v_matrix_definition}
 \tilde{\mathbf{v}}_{x,\mathbf{k}} = 
 -\frac{\hbar}{e}\left. (\partial_{A_x} \mathbf{h}_{\mathbf{k}}) \right|_{\mathbf{A}\rightarrow 0}
\end{equation}
To perform this differentiation, one needs to revisit the preceding step and introduce in the TB matrix a uniform vector potential in the $x$-direction $A_x(\mathbf{q}=0)$. The derivation is outlined in App.~\ref{app:velocity_matrix}, and here we only present the final result:
\begin{equation}\label{eq:vxk_from_momentum}
 \tilde{\mathbf{v}}_{x,\mathbf{k}} = \left(
 \begin{array}{ccc}
  0 & t_{pd}\cos\frac{k_x}{2} & 0 \\
  \mathrm{H.c.} & - 2 t'_{pp} \sin (k_x) & -2t_{pp} \cos\frac{k_x}{2}\sin\frac{k_y}{2} \\
  0 & \mathrm{H.c.} & 0
 \end{array}
 \right).
\end{equation}

\begin{figure}
    \includegraphics[width=\columnwidth]{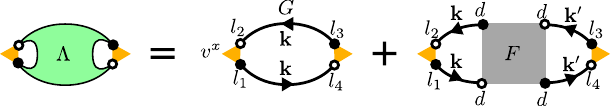}
\caption{Illustration of the proof of the cancellation of vertex corrections in the DMFT solution of the Emery model. Spin indices are omitted for the sake of simplicity, without loss of generality.}
\label{fig:vertex}
\end{figure}

The vertex corrections to conductivity are obtained from the connected part of the uniform current-current correlation function\cite{Kovacevic2025}. Up to a prefactor, it is expressed as\cite{Rohringer2012}
\begin{eqnarray} 
 &&\Lambda^{\mathrm{conn}}_{\mathbf{q}=0}(i\nu) \\ \nonumber
 &&\sim \iint_\mathrm{BZ} \mathrm{d}\mathbf{k}\mathrm{d}\mathbf{k}' 
         \sum_{l_1,l_2,l_3,l_4} \sum_{l'_1,l'_2,l'_3,l'_4} 
         \sum_{i\omega,i\omega'}
         \tilde{v}_{x,\mathbf{k},l_1l_2}\tilde{v}_{x,\mathbf{k}',l_3l_4}
 \\ \nonumber
 &&\;\;\;\;\times    
              G_{\mathbf{k},l'_1l_1}(i\omega+i\nu)
              G_{\mathbf{k},l_2l'_2}(i\omega) 
 \\ \nonumber
 &&\;\;\;\;\times          
              F_{\mathbf{k}\mathbf{k}',\mathbf{q}=0,l'_1l'_2l'_3l'_4}(i\omega,i\omega',i\nu)
\\ \nonumber
 &&\;\;\;\;\times    
              G_{\mathbf{k}',l'_3l_3}(i\omega'+i\nu)
              G_{\mathbf{k}',l_4l'_4}(i\omega')              
\end{eqnarray}
where $i\omega$ and $i\nu$ denote the fermionic and bosonic Matsubara frequencies, respectively, and $F$ is the full vertex function. 
We have omitted spin indices for the sake of brevity, without loss of generality.
At the level of DMFT approximation, the full vertex depends only on the transfer momentum $\mathbf{q}$ (which is here set to zero, as we are interested in uniform conductivity)\cite{VucicevicPRB2021}.
Furthermore, because we only keep interactions on the $d$-orbitals, it holds that $F_{l_1l_2l_3l_4}=\delta_{l_1=l_2=l_3=l_4=d}F_{dddd}$, thus the sums over the orbital indices $l'_1,l'_2,l'_3,l'_4$ reduce to a single term
\begin{eqnarray}
 &&\Lambda^{\mathrm{conn}}_{\mathbf{q}=0}(i\nu) \\ \nonumber
 &&\sim \iint_\mathrm{BZ} \mathrm{d}\mathbf{k}\mathrm{d}\mathbf{k}' 
         \sum_{l_1,l_2,l_3,l_4} 
         \sum_{i\omega,i\omega'}
         \tilde{v}_{x,\mathbf{k},l_1l_2}\tilde{v}_{x,\mathbf{k}',l_3l_4}
 \\ \nonumber
 &&\;\;\;\;\times    
              G_{\mathbf{k},dl_1}(i\omega+i\nu)
              G_{\mathbf{k},l_2d}(i\omega) 
 \\ \nonumber
 &&\;\;\;\;\times          
              F_{\mathbf{q}=0,dddd}(i\omega,i\omega',i\nu)
\\ \nonumber
 &&\;\;\;\;\times    
              G_{\mathbf{k}',dl_3}(i\omega'+i\nu)
              G_{\mathbf{k}',l_4d}(i\omega')              
\end{eqnarray}
This expression is illustrated in Fig.~\ref{fig:vertex}.

In the Hubbard model, the proof of the cancellation of the vertex corrections relies on the velocity being antisymmetric with respect to the wave vector, $v^x_\mathbf{k} = -v^x_{-\mathbf{k}}$. In the Emery model, the velocity is a matrix $\tilde{v}_{x,\mathbf{k},ll'}$ and we get multiple terms summing over $l_1,l_2,l_3,l_4$. We see in Eq.~\eqref{eq:vxk_from_momentum} that some of $\tilde{\mathbf{v}}_{x,\mathbf{k}}$ elements are not antisymmetric, but rather symmetric. Nevertheless, the corresponding components of the Green's function are antisymmetric and the cancellation still proceeds in much the same way as in the Hubbard model.
The main point is that, for fixed $\mathbf{k}',l_3,l_4$ and all the Matsubara frequencies, we always have two terms with $\mathbf{k}$ from the opposite halves of the Brillouin zone,  say $\mathbf{k}=\mathbf{k}_o$ and $\mathbf{k}=-\mathbf{k}_o$; whether these two terms cancel out will depend on the symmetry properties of the product $G_{\mathbf{k},l_1d}(i\omega+i\nu)
              G_{\mathbf{k},d l_2}(i\omega) \tilde{v}_{x,\mathbf{k},l_1l_2}$ with respect to $\mathbf{k}\rightarrow-{\mathbf{k}}$.
In the following table, we summarize those symmetry properties for each factor and the full product, using shorthand notation AS=antisymmetric and  S=symmetric. 
%Note that the full vertex $F$ is diagonal in the orbital indices (because we only keep interactions on the $d$-orbital) and only depends on the transfer momentum because of the DMFT approximation ( $F_{l_1l_2l_3l_4,\mathbf{k}\mathbf{k}'\mathbf{q}}\approx\delta_{l_1=l_2=l_3=l_4=d}F_{dddd,\mathbf{q}}$ ).
\begin{equation}
 \begin{array}{c||c|c|c||c}
  l_1,l_2 & v_{\mathbf{k},l_1l_2} & G_{\mathbf{k},dl_1} & G_{\mathbf{k},l_2d} & \mathrm{overall}\\ \hline
d, d & 0 & S & S & 0\\
d, p_x & S & S & AS & AS \\
d, p_y & 0 & S & AS & 0\\
p_x, p_x & AS & AS & AS & AS \\
p_x, p_y & AS & AS & AS & AS\\
p_y, p_y & 0 & AS & AS & 0
 \end{array}
\end{equation}
We see that for all choices of $l_1,l_2$, the two terms with $\mathbf{k}=\mathbf{k}_o$ and $\mathbf{k}=-\mathbf{k}_o$ will be either exactly zero, or exactly the opposite, thus leading to $\Lambda^{\mathrm{conn}}_{\mathbf{q}=0}(i\nu)=0$, which we set out to prove. The vertex corrections generated within single-site DMFT indeed vanish. 

Going one step back, however, there is an important conceptual subtlety in the way conductivity is formulated in lattice models downfolded from DFT band structures. 
The velocity of the momentum eigenstates in lattice models might not match well the velocity of the corresponding eigenstates of the Kohn-Sham Hamiltonian in DFT, which might skew the physical meaning of conductivity in the lattice model. Ultimately, the problem arises when the dispersion relation in a downfolded model does not match perfectly any of the bands in the DFT band structure (as is the case here, see Fig.~\ref{fig:dft}) and the Wannier space has non-zero projection on multiple bands in DFT. The mismatch between the band structure of the downfolded model and the DFT band structure is known to cause issues in cRPA calculations, and solutions for this problem have been proposed recently\cite{Vucicevic2026arXiv,Kaltak2025_cRPA_spectral}. Following analogous reasoning, one could evaluate the velocity for each momentum eigenstate in our lattice model, not by using the Peierls substitution\cite{Peierls1933} (as we do), but by projecting the current operator to the Wannier orbitals in a continuous space calculation (i.e. at the level of DFT). Such calculations are, however, beyond the scope of the present paper.

To numerically evaluate the dc resistivity via Eq.~\eqref{eq:Kubo_bubble}, we perform adaptive integration over the BZ and compute Green's functions on the fly from the self-energy. The integral over frequency $\omega$ is performed using a uniform grid within cutoffs proportional to temperature, set so that we capture 99.9\% of the weight of $n'_\mathrm{F}(\omega)$.

We see in Eq.~\eqref{eq:Kubo_bubble} that the sum over orbital indices $\sum_{l_1,..,l_4}$ has \emph{a priori} $3^4=81$ terms, but many of them will not contribute, and some are connected by symmetry. Based on the form of $\mathbf{v}_{x,\mathbf{k}}$ ($v_{x,\mathbf{k},ll'}=v_{x,\mathbf{k},l'l}$) and the fact that some components are zero, and making use of the symmetry $G_{\mathbf{k},ll'} =G_{\mathbf{k},l'l}$ (this can be proven by writing the matrix inverse of a symmetric matrix explicitly, term by term), we find that the only independent non-zero terms are
\begin{equation}
 \begin{array}{c|c||c}
  l_1,l_2 & l_3,l_4 & \mathrm{symmetry\;prefactor}\\ \hline
d , p_x & d , p_x & 2 \\
d , p_x & p_x , d & 2 \\
d , p_x & p_x , p_x & 4 \\
d , p_x & p_x , p_y & 4 \\
d , p_x & p_y , p_x & 4 \\
p_x , p_x & p_x , p_x & 1 \\
p_x , p_x & p_x , p_y & 4 \\
p_x , p_y & p_x , p_y & 2 \\
p_x , p_y & p_y , p_x & 2 \\
 \end{array}
\end{equation}
However, it turns out that the contributions $p_x , p_y | p_x , p_y$ and $p_x , p_y | p_y , p_x$ are also the same. Similarly, $d , p_x | p_y , p_x$ and $d , p_x | p_x , p_y$, as well as $d , p_x | d , p_x$ and $d , p_x | p_x, d$ are the same. We use these symmetries to further speed up the calculation. Note also that some components give a negative contribution to conductivity. 

For the Hubbard model, Eq.~\eqref{eq:Kubo_bubble} simplifies because there are no sums over orbital indices, the velocity and the Green's function are scalars (fully analogously, $\tilde{v}_{x,\mathbf{k}}=(1/\hbar) \partial_{k_x} E_{\alpha=2,\mathbf{k}}$), and the momentum integral can be further simplified by transforming it into a single integral over frequency
\begin{equation}\label{eq:ksum_to_epsilon_integral}
  \int_\mathrm{BZ} \frac{\mathrm{d}\mathbf{k}}{(2\pi)^2} \tilde{v}_{x,\mathbf{k}}^2 \rightarrow \int \mathrm{d}\varepsilon \Phi(\varepsilon)
\end{equation}
with
\begin{equation}\label{eq:transport_function}
  \Phi(\varepsilon) = \int_\mathrm{BZ} \frac{\mathrm{d}\mathbf{k}}{(2\pi)^2} \tilde{v}_{x,\mathbf{k}}^2 \delta(\varepsilon-E_{\alpha=2,\mathbf{k}})
\end{equation}
The velocity-weighted densities of states $\Phi(\varepsilon)$ (also known as the transport functions) for the two Hubbard model parametrizations used in this work are shown in Fig.~\ref{figphi}.

\subsection{Definition of doping and the approach to phase diagram scans}
\label{sec:phase_diagram_scan}
We solve the Hubbard and the Emery model on fixed chemical-potential and temperature grids. The doping level is determined \emph{a posteriori} by computing the total occupancy per site per spin.

In the Emery model we define
\begin{eqnarray}
\langle n_\sigma \rangle
  &\equiv& \sum_{l} \langle c^\dagger_{l, \sigma, \mathbf{r}=0} c_{l, \sigma, \mathbf{r}=0}\rangle  \\ \nonumber
  &=& -\frac{1}{\pi}\sum_l \int_\mathrm{BZ}\frac{\mathrm{d}\mathbf{k}}{(2\pi)^2} \int\mathrm{d}\omega\mathrm{Im}G_{\mathbf{k},ll}(\omega) n_\mathrm{F}(\omega)
\end{eqnarray}
and the doping level as the number of holes per site
\begin{equation}
 \delta \equiv 2\left(\frac{5}{2}-\langle n_\sigma \rangle\right)
\end{equation}
Therefore, positive $\delta$ corresponds to hole-doping, negative $\delta$ corresponds to electron-doping. 
%As $\delta\in[-1,1]$, 
We usually express it in terms of (signed) percentages of unit doping.

Similarly, in the Hubbard model we define
\begin{eqnarray}
\langle n_\sigma \rangle
  &\equiv& \langle c^\dagger_{\sigma, \mathbf{r}=0} c_{\sigma, \mathbf{r}=0}\rangle  \\ \nonumber
  &=& -\frac{1}{\pi} \int_\mathrm{BZ}\frac{\mathrm{d}\mathbf{k}}{(2\pi)^2} \int\mathrm{d}\omega\mathrm{Im}G_{\mathbf{k}}(\omega) n_\mathrm{F}(\omega)
\end{eqnarray}
and
\begin{equation}
 \delta \equiv 2\left(\frac{1}{2}-\langle n_\sigma \rangle\right).
\end{equation}

\begin{figure*}[t!]
    \includegraphics[width=\textwidth,valign=t,trim=0 0 0 0cm]{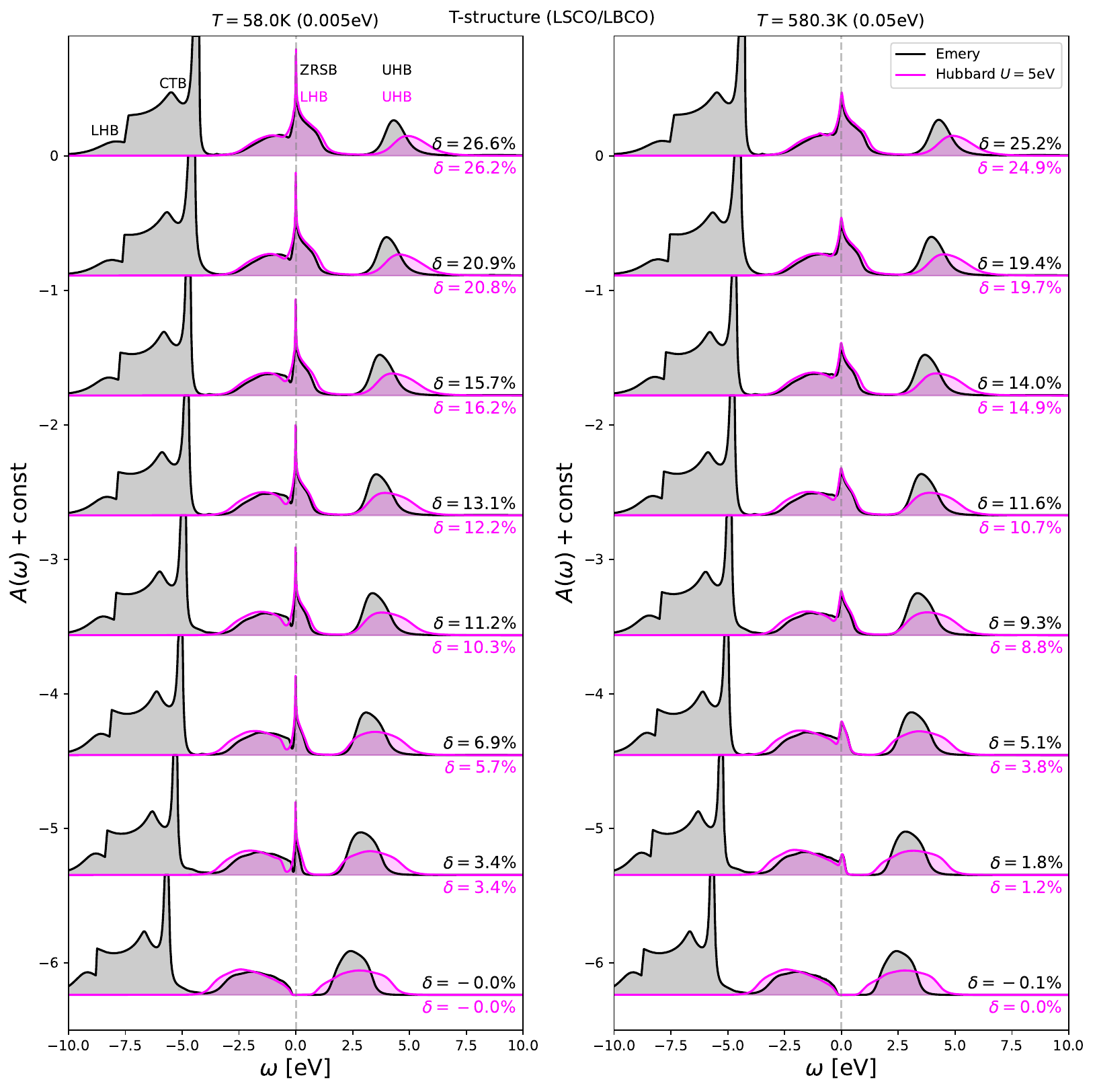}
\caption{Comparison of the local spectral function between the Hubbard and Emery models for the T-structure, relevant for LSCO and LBCO, at various dopings and two different temperatures. The coupling constants in the two models are: $U^\mathrm{Hubbard}=\unit[5]{eV}$, $U^\mathrm{Emery}=\unit[8]{eV}$. Different bands are referred to as:
lower Hubbard band (LHB), charge transfer band (CTB), Zhang-Rice singlet band (ZRSB), upper Hubbard band (UHB).
}
\label{fig:Emery_vs_Hubbard_Aloc_Tstruct}
\end{figure*}

\begin{figure}[t!]
    \includegraphics[width=0.9\columnwidth,valign=t,trim=0 0 0 0cm]{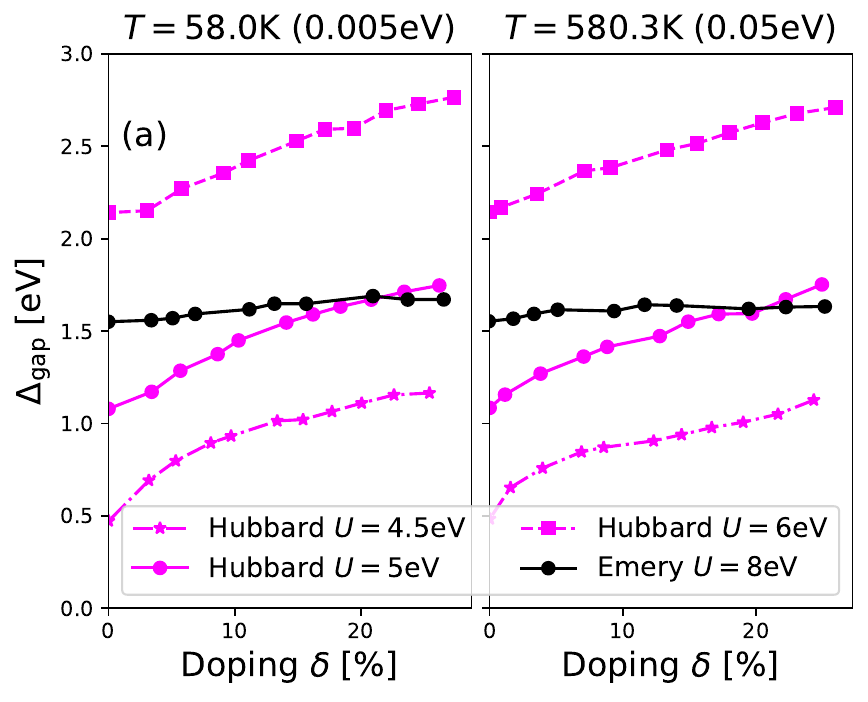}
    \includegraphics[width=0.9\columnwidth,valign=t,trim=0 0 0 0cm]{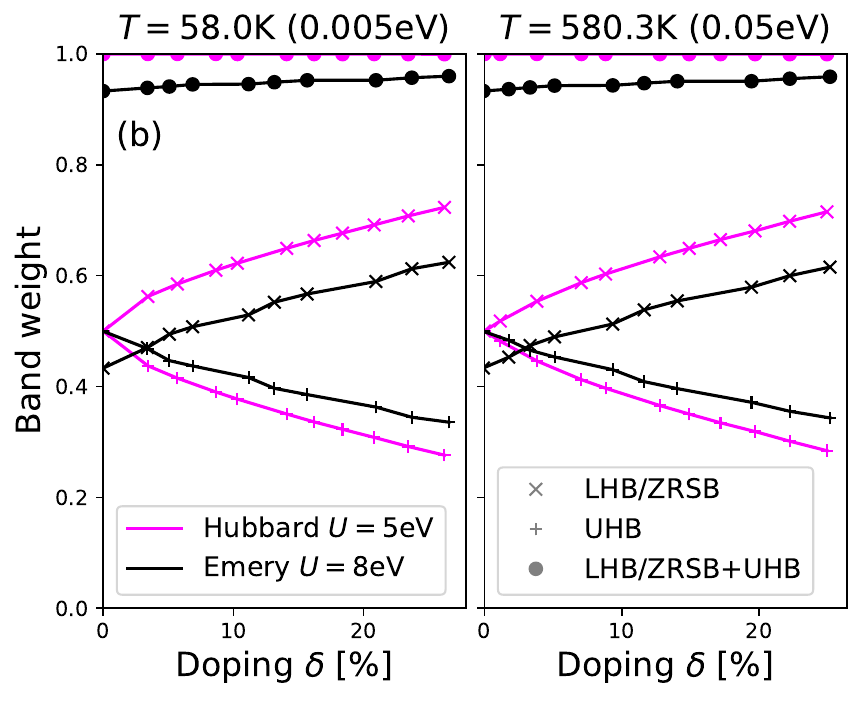}
    \includegraphics[width=0.9\columnwidth,valign=t,trim=0 0 0 0cm]{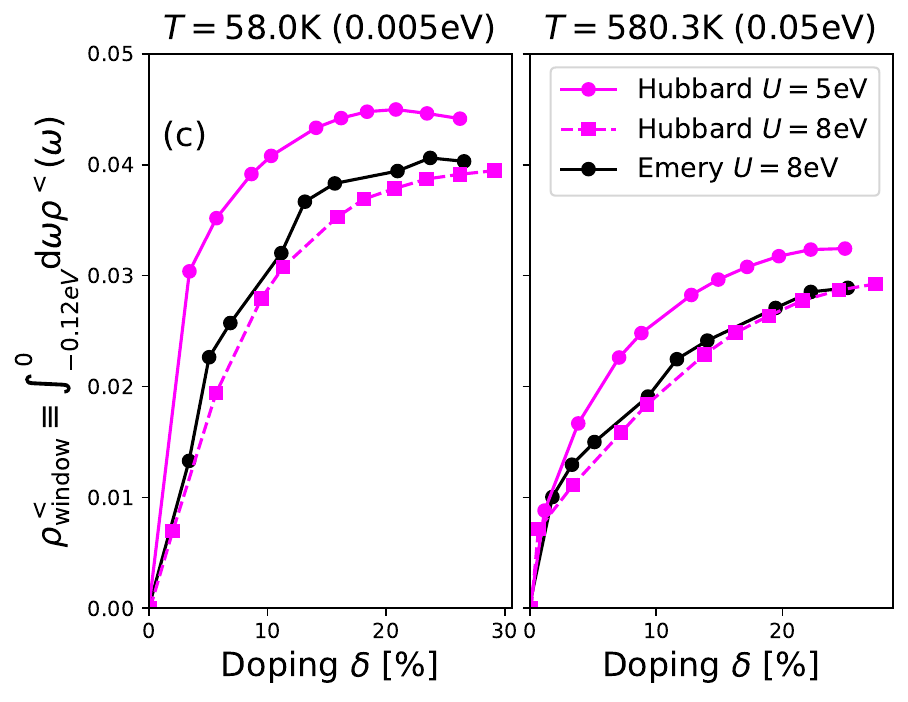}
\caption{Quantifying spectral properties in the T-structure models relevant for LSCO/LBCO: comparison between the Emery and the Hubbard model as a function of doping, for two different temperatures, and for different choices of $U^\mathrm{Hubbard}$. $U^\mathrm{Emery}=\unit[8]{eV}$ everywhere. (a) Comparison between the gap-size $\Delta_\mathrm{gap}$ of the Mott gap in the Hubbard model and the charge-transfer gap in the Emery model. (b) Comparison of the spectral weight between the corresponding bands in the two models. (c) Comparison of the occupied spectral weight just below the Fermi level, as defined in the text, $\rho^<_\mathrm{window}$.}
\label{fig:Emery_vs_Hubbard_bandweights}
\end{figure}

Since we are ultimately interested in the dc resistivity as a function of temperature at a fixed \emph{doping}, we interpolate our results. For a fixed temperature, we first interpolate $\mu(\langle n_\sigma \rangle)$ to get $\mu$ for the desired doping, and then do a 1D interpolation of $\rho_\mathrm{dc}(\mu,T)$ at that value of the chemical potential $\mu$. 
In some cases we also perform a calculation for the preset doping level, and check that the errors due to interpolation of $\rho_\mathrm{dc}$ are negligible.

We do not interpolate the spectral functions; the curves shown are  obtained directly from the DMFT calculations. For this reason, the matching between doping values in the Hubbard and Emery models on some figures will not be perfect, but the resulting mismatch does not affect the conclusions of our analyses.

\begin{figure}
    \includegraphics[width=\columnwidth,valign=t,trim=0 0 0 0cm,page=2]{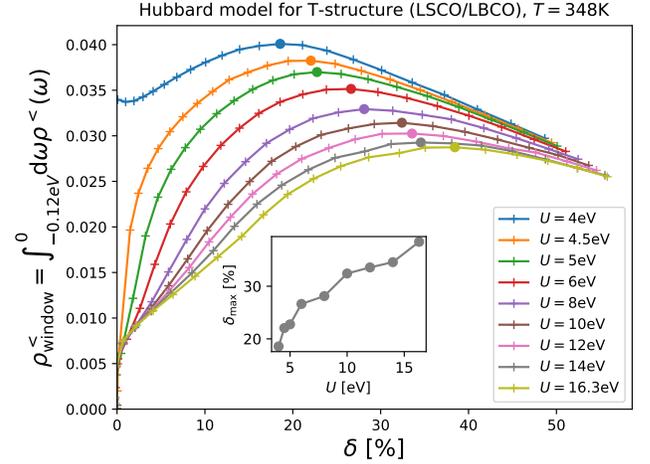}
\caption{Dependence of the occupied spectral weight just below the Fermi level, $\rho^<_\mathrm{window}$, in the Hubbard model for the T-structure (LSCO/LBCO), as a function of doping $\delta$ and the coupling constant $U$. $\rho^<_\mathrm{window}(\delta)$ reaches a maximum at $\delta_\mathrm{max}(U)$, denoted by colored dots, and summarized in the inset.}
\label{fig:Hubbard_PES_prediction_signal}
\end{figure}

\begin{figure*}
    \includegraphics[width=\textwidth,valign=t,trim=0 0 0 0cm,page=2]{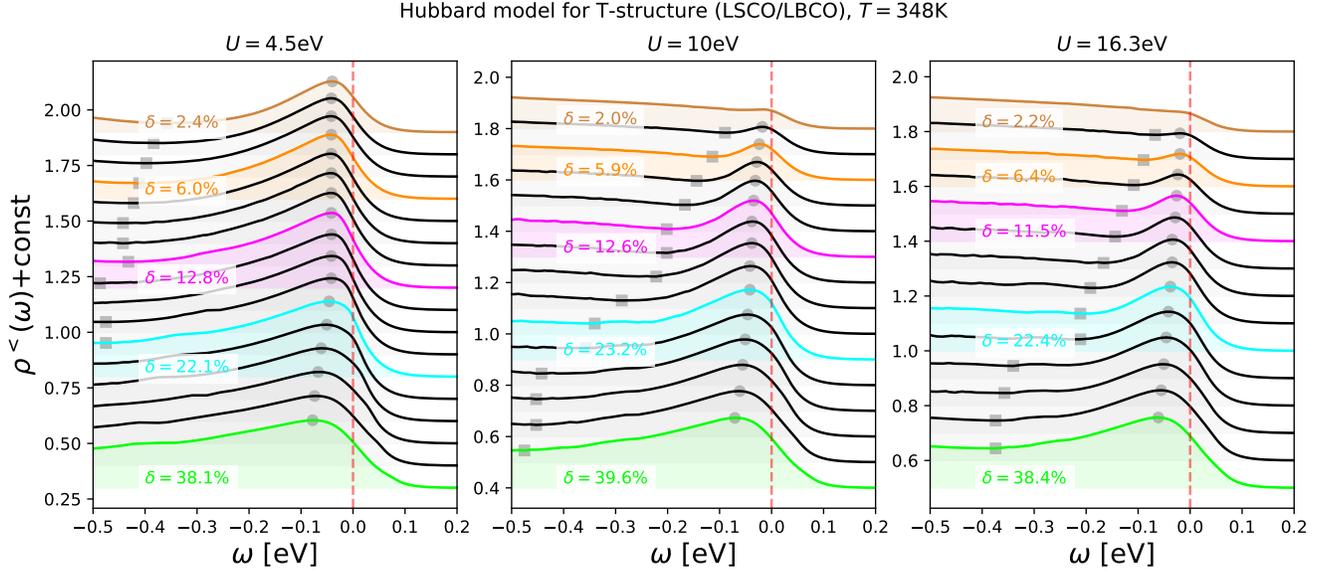}
\caption{Density of occupied states $\rho^<(\omega)$ in the Hubbard model for the T-structure (LSCO/LBCO) in the vicinity of the Fermi level, as a function of doping, and for three different values of the coupling constant. Gray circles/gray squares denote local maxima/minima in the presented frequency range and highlight that the quasiparticle peak becomes narrower with the increasing coupling constant. }
\label{fig:Hubbard_PES_prediction_rholess}
\end{figure*}

\begin{figure*}
\centering
    \includegraphics[width=\columnwidth,trim=0 0 0 0cm]{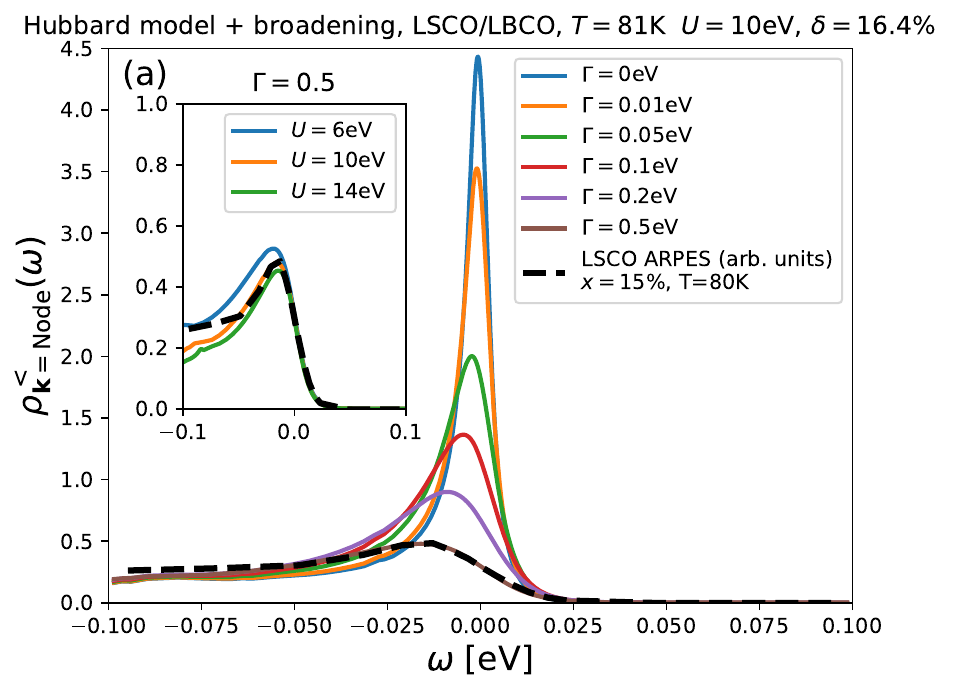}
    \includegraphics[width=0.96\columnwidth,trim=0 0 0 0cm]{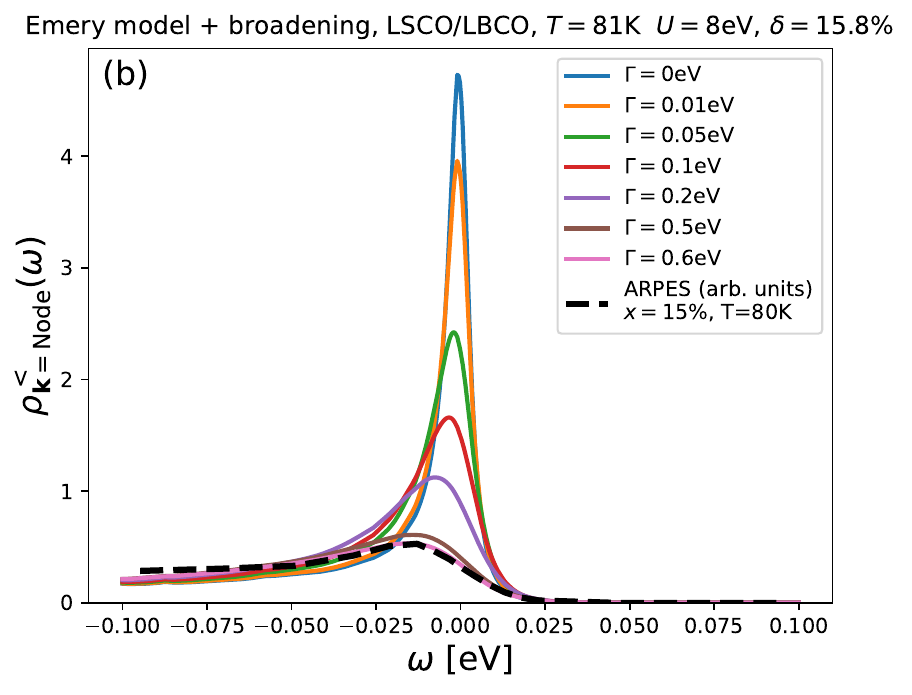}
\caption{Comparison of the DMFT momentum-resolved spectral function with ARPES data for LSCO from Ref.~\onlinecite{Yoshida2009}. (a) Hubbard model; (b) Emery model. The spectral function is computed at the node, i.e. precisely at the Fermi surface in the $\theta=\pi/4$ radial direction in the $k$-space.
Different curves correspond to different levels of phenomenological broadening $\Gamma$ added to the theoretical results.
The inset in (a) shows the dependence on $U^\mathrm{Hubbard}$ for a fixed $\Gamma$.
}
\label{fig:arpes_momentum_resolved}
\end{figure*}

\section{Results}
\label{sec:results}
\subsection{Hubbard vs. Emery: spectral properties}
\label{sec:hubbard_vs_emery_spectra}
\subsubsection{$T$-structure (LSCO, LBCO)}

In Fig.~\ref{fig:Emery_vs_Hubbard_Aloc_Tstruct} we compare the local spectral function in the Hubbard and Emery models for the hole-doped T-structure.
The local spectral function is defined as:
\begin{eqnarray}\nonumber
A(\omega) &=& -\frac{1}{\pi}\sum_l\int_\mathrm{BZ} \frac{\mathrm{d}\mathbf{k}}{(2\pi)^2} \mathrm{Im}G_{\mathbf{k},ll}(\omega)\;\;\;\mathrm{(Emery)}\\
A(\omega) &=& -\frac{1}{\pi}\int_\mathrm{BZ} \frac{\mathrm{d}\mathbf{k}}{(2\pi)^2} \mathrm{Im}G_{\mathbf{k}}(\omega)\;\;\;\mathrm{(Hubbard)}
\end{eqnarray}
Here we take $U^\mathrm{Hubbard}=\unit[5]{eV}$ and $U^\mathrm{Emery}=\unit[8]{eV}$. For the given $U^\mathrm{Emery}$ value, this choice of $U^\mathrm{Hubbard}$ yields the best match between the spectra of the two models in terms of the size of the gap, as we will see.
We find that for $U^\mathrm{Hubbard}=\unit[4]{eV}$ the Hubbard model is not yet a Mott insulator and does not match well the behavior of the $U^\mathrm{Emery}=\unit[8]{eV}$ Emery model at low doping values.

In both models one recognizes two bands of similar shape, one around the Fermi level and one above it.
The upper one is referred to as the upper Hubbard band (UHB).
The one at the Fermi level is referred to as the lower Hubbard band (LHB) in the context of the Hubbard model, but as the Zhang-Rice singlet band (ZRSB) in the context of the Emery model.
 The gap between the LHB and UHB in the Hubbard model is referred to as the Mott gap; the gap between the ZRSB and the UHB in the Emery model is referred to as the charge transfer gap (CTG).

\begin{figure*}
\centering
    \includegraphics[width=\textwidth,trim=0 0 0 0cm]{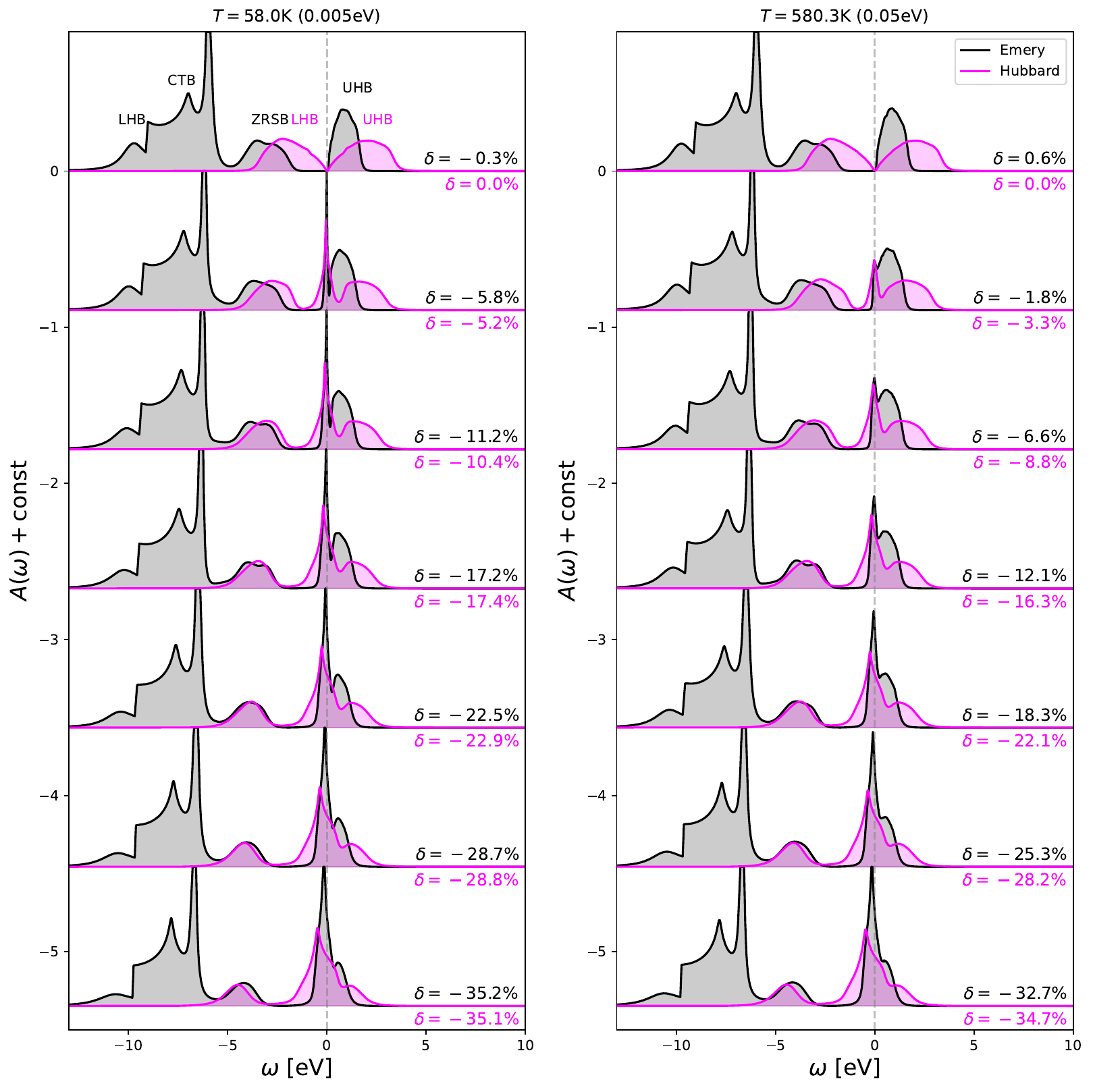}
\caption{
Comparison of the local spectral function between the Hubbard and Emery models for the T'-structure, relevant for LCCO, at various dopings and two different temperatures. The coupling constants are taken to be $U^\mathrm{Hubbard}=\unit[4]{eV}$, $U^\mathrm{Emery}=\unit[8]{eV}$. Band labels are the same as in Fig.~\ref{fig:Emery_vs_Hubbard_Aloc_Tstruct}.
}
\label{fig:Emery_vs_Hubbard_Aloc_Tpstruct}
\end{figure*}

\begin{figure}
    \includegraphics[width=\columnwidth,valign=t,trim=0 0 0 0cm,page=2]{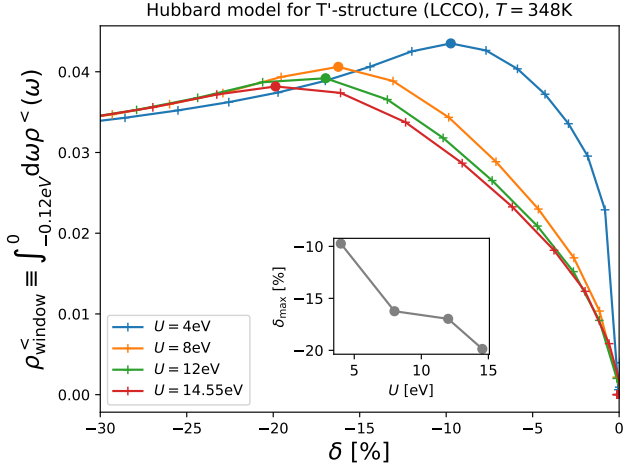}
\caption{Same as Fig.~\ref{fig:Hubbard_PES_prediction_signal}, but for the electron-doped T'-structure (LCCO). The maxima occur at lower (absolute) values of doping.
}
\label{fig:Hubbard_PES_prediction_signal_LCCO}
\end{figure}

At finite doping and low temperature, a sharp quasiparticle (QP) peak around the Fermi level is present in both models and has a similar shape.
The QP peak evolves with doping and temperature in a similar fashion in both models.

We observe two obvious differences between the models. First, in the Hubbard model, the UHB is wider and has lower spectral weight.
Second, the size of the gap between the top two bands is not the same.
We define the edges of the bands as frequencies where $A(\omega)$ drops below 0.02, and the size of a gap as the energy difference between the two edges, $\Delta_\mathrm{gap}$. We find that the size of the Mott gap in the Hubbard model varies with doping much more than the CTG in the Emery model, see Fig.~\ref{fig:Emery_vs_Hubbard_bandweights}(a).
At low doping the Mott gap is smaller than CTG, while at high doping the Mott gap is larger than the CTG. Clearly, $\Delta_\mathrm{gap}$ is strongly dependent on $U$ in both models; in Fig.~\ref{fig:Emery_vs_Hubbard_bandweights} we see that the best matching in terms of $\Delta_\mathrm{gap}$ is precisely when $U^\mathrm{Hubbard}=\unit[5]{eV}$ (given that $U^\mathrm{Emery}=\unit[8]{eV}$).

In the Hubbard model, the spectral weights of the LHB and UHB add up to 1, by construction. In the Emery model, ZRSB and UHB weights do not add up to 1. The comparison of relevant band weights in the two models is shown in Fig.~\ref{fig:Emery_vs_Hubbard_bandweights}(b).
UHB in the Hubbard model has less spectral weight compared to the Emery model, while LHB in the Hubbard model has more weight than Emery ZRSB. At zero doping (Mott insulator) in the Hubbard model, LHB and UHB have equal weights. In the Emery model, the UHB and ZRSB have equal weights at doping of around 3\%; in the undoped Mott insulator, the UHB is stronger than ZRSB, and it has a weight of precisely 1/2 (i.e., exactly one hole per cell, as expected).

In Fig. ~\ref{fig:Emery_vs_Hubbard_bandweights}(c) we compare the density of occupied states just below the Fermi level for the two models. The density of occupied states is defined as
\begin{equation}
 \rho^{<}(\omega) = \int_\mathrm{BZ} \frac{\mathrm{d}\mathbf{k}}{(2\pi)^2}  \rho^{<}_\mathbf{k}(\omega)
\end{equation}
where
\begin{eqnarray}
 \rho^{<}_\mathbf{k}(\omega) &=& A_\mathbf{k}(\omega)n_\mathrm{F}(\omega) \\
 A_\mathbf{k}(\omega) &=& -\frac{1}{\pi}\sum_l \mathrm{Im}G_{\mathbf{k},ll}(\omega)\;\;\;\mathrm{(Emery)} \\
 A_\mathbf{k}(\omega) &=& -\frac{1}{\pi}\mathrm{Im}G_{\mathbf{k}}(\omega)\;\;\;\;\;\;\mathrm{(Hubbard)}
\end{eqnarray}

and we are interested in its integral in the energy window $\omega\in[\unit[-0.12]{eV},\unit[0]{eV}]$.
In both models, $\rho^{<}_{\mathrm{window}}\equiv \int_{\unit[-0.12]{eV}}^{0}\mathrm{d}\omega\rho^{<}(\omega)$ grows with doping, but in a slightly different fashion. This difference is particularly pronounced at low temperature. In the Hubbard model the growth between 3.5\% and 21\% doping is slower: $\rho^{<}_{\mathrm{window}}$ grows by a factor of about 1.5; in the Emery model it grows by a factor of about 3.
However, this difference appears to be related more to the value of $U$ in the model than to the choice of model. As can be seen in Fig.~~\ref{fig:Emery_vs_Hubbard_bandweights}(c) the doping dependence of $\rho^{<}_{\mathrm{window}}$ is much more similar between the Hubbard and the Emery model when both are evaluated for $U=\unit[8]{eV}$.
We inspect in detail the $U$-dependence of $\rho^{<}_{\mathrm{window}}(\delta)$ in the Hubbard model and show the results in Fig.~\ref{fig:Hubbard_PES_prediction_signal}. We see that there is a pronounced maximum $\rho^{<}_{\mathrm{window}}(\delta)$ at a doping value $\delta_\mathrm{max}$ that is strongly dependent on $U$: this characteristic doping $\delta_\mathrm{max}$ grows roughly linearly from about 20\% to about 40\% as $U$ is increased from \unit[4]{eV} to about \unit[16]{eV}. This effect is only weakly dependent on temperature (data not shown).

We now look more closely at the behavior of $\rho^{<}(\omega)$ in the vicinity of the Fermi level. This is shown in Fig.~\ref{fig:Hubbard_PES_prediction_rholess}. We see that the value of $U$ strongly affects the width of the quasiparticle peak. At about 22\% doping, the distance between the maximum and the minimum  of $\rho^{<}(\omega)$ (gray circle and gray square, respectively) goes from about \unit[0.45]{eV} to about \unit[0.15]{eV} as $U$ is increased from about \unit[4]{eV} to about \unit[16]{eV}.

We believe that both Fig.~\ref{fig:Hubbard_PES_prediction_signal} and Fig.~\ref{fig:Hubbard_PES_prediction_rholess} document strong quantitative trends that could be sought experimentally to help deduce the correct effective value of $U$. However, we are unaware of any existing experimental data that could be compared to our results directly. We find ARPES results in Ref.~\onlinecite{Yoshida2009} that present $\rho^<_\mathbf{k}(\omega)$ at momenta on the Fermi surface corresponding to the node and the antinode. The measurement at temperature \unit[80]{K} and 15\% doping shows that the node-antinode dichotomy is weak, and our local self-energy approximation is expected to be reasonable in that parameter regime. In Fig.~\ref{fig:arpes_momentum_resolved}, we compare our results to measurements reported in Ref.~\onlinecite{Yoshida2009} taken at the node. We find that DMFT for both models predicts much more coherent quasiparticles than are observed in the experiment. This discrepancy could be related to the strong disorder that is present in real samples, but not in our clean lattice-model theories. Good line-shape agreement is obtained when we broaden our spectral function with a phenomenological frequency-independent imaginary self-energy, $\Sigma(\omega)\rightarrow\Sigma(\omega)-i\Gamma$, with $\Gamma=\unit[0.5]{eV}$ in the Hubbard model, and $\Gamma=\unit[0.6]{eV}$ in the Emery model.
We also note that in the case of the Hubbard model, the best agreement with experiment is obtained when we take $U^\mathrm{Hubbard}\approx\unit[10]{eV}$ (see inset of Fig.~\ref{fig:arpes_momentum_resolved}). 

However, it is not immediately clear whether disorder can be responsible for such a large broadening of the quasiparticle peak. In fact, if that interpretation is correct, disorder is the leading cause of broadening: the imaginary part of the self-energy at zero frequency in both our models is about \unit[0.05]{eV}, i.e. 10 times smaller.
One might therefore estimate the elastic scattering rate due to disorder and compare it with the $\Gamma$ required to reproduce the ARPES line shapes.
One could assume the validity of the Drude formula $\rho_\mathrm{dc}=\mtr/(ne^2\tau)$ \cite{BerthodApplicationsManyBodyFormalism}, which connects the scattering rate to $\rho_\mathrm{dc}$ and then estimate the scattering rate from the residual resistivity $\rho_\mathrm{dc}(T=0)$, under the assumption that inelastic e-e scattering does not contribute to the residual resistivity.
However, to derive this kind of estimate, one would have to assume a value of $m^*/n$ that enters the effective Drude theory for the disordered system at $T=0$. Even under the assumption that $n$ is the total electron concentration coming from a single valence band, we still cannot estimate the relevant $m^*$ using our clean lattice model theory, because the renormalization of mass in this context comes entirely from disorder. Estimates of the elastic scattering rate can be found in experimental literature and are of order \unit[20]{meV}\cite{Lemberger2010}, which is apparently much smaller than the phenomenological broadening required to reproduce the ARPES line shape.
However, the elastic scattering rate is not necessarily in a simple relation with spectral broadening. In Bi2212, for example, the residual ARPES linewidth has been reported to significantly exceed the transport-relevant elastic rate, by up to an order of magnitude \cite{Evtushinsky2006}.
The understanding is that out-of-plane disorder leads to predominantly forward scattering, which contributes strongly to the single-particle self-energy (i.e. spectral broadening), yet does not contribute much to resistivity and the transport scattering rate.
Other features of the spectral function, such as the superconducting gap, are also in complex relations with disorder. The broadening of the superconducting gap (Dynes parameter) and the broadening of quasiparticle features in ARPES are not necessarily the same scale \cite{kondo2015}: it has been proposed that two different lifetimes need to be introduced, a pair-conserving and a pair-breaking one. The Dynes parameter includes only the latter, while the broadening in ARPES includes both \cite{herman2017}.

To summarize, an experiment aimed at estimating $U$ based on $\rho^{<}(\omega)$ or $\rho^{<}_\mathrm{window}(\delta)$ results would have to devise a way to account for the effects of disorder.
In particular, one must take into account that the amount of disorder is ultimately sample-dependent, even if there are rough general trends in terms of the doping dependence (perhaps counterintuitively, LSCO is known to become less disordered with increasing doping\cite{Yoshida2009}). 
Estimating the scattering rate from residual resistivity is one possible route, but there is an ambiguity in the choice of $m^*/n$ that enters the Drude formula, and, furthermore, the relation between the transport scattering rate and the spectral broadening is likely not simple.
Our results provide clear motivation for further theoretical studies of the effects of disorder on the broadening of spectral lines.

\subsubsection{$T'$-structure (LCCO)}

Finally, we study the Emery and the Hubbard models for the electron-doped LCCO, downfolded from the T'-structure La$_2$CuO$_4$ DFT result. In this case, the difference in spectra between the two models is much greater, see Fig.~\ref{fig:Emery_vs_Hubbard_Aloc_Tpstruct}. The best matching in terms of the position of the LHB in the Hubbard model and the corresponding band in the Emery model is obtained if we take $U^{\mathrm{Hubbard}}=\unit[4]{eV}$ (given that $U^{\mathrm{Emery}}=\unit[8]{eV}$). The size of the gap $\Delta_\mathrm{gap}$ variation with $\delta$ in the Hubbard model is much more pronounced. The shape of the QP peak and the UHB is not matched well. The behavior of $\delta_\mathrm{max}$ in the Hubbard model is similar to that in the T-structure (see Fig.~\ref{fig:Hubbard_PES_prediction_signal_LCCO}), but it is generally lower in value (the maximum occurs at a lower doping).

We conclude this section by noting that the Emery and the Hubbard model spectra share many qualitative features, but they do not describe exactly the same physics.
This means that the single-band Hubbard model, as it is usually defined and parameterized (based on the DFT band structure), is not a particularly good effective single-band description of the corresponding Emery model at low energies, in all situations. A different single-band model, possibly more complicated, is likely needed.
In App.~\ref{app:orbital_resolved} we discuss the orbital-resolved and non-interacting-eigenband-resolved spectral functions in the Emery model and show that the latter provides indications of what an effective single-band model might be to describe the Emery model at low energies. These results might be relevant for attempts to further downfold the Emery model, e.g. as in Ref.~\onlinecite{JiangScalapinoWhite2023_Downfolding}.

\begin{figure}
\centering
    \includegraphics[width=\columnwidth,trim=0 0 0 0cm]{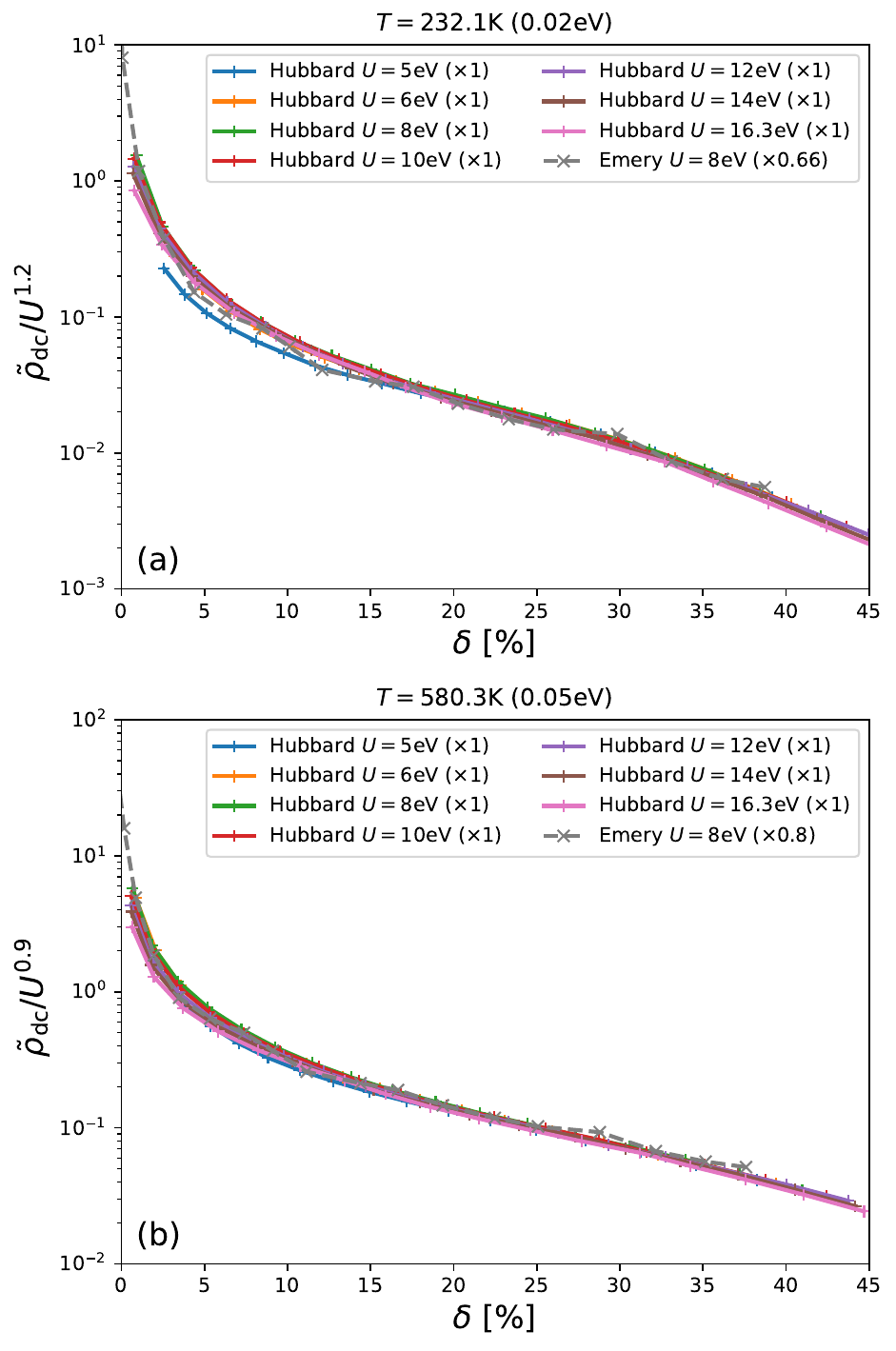}
\caption{
Scaling of dc resistivity in the hole-doped Hubbard and Emery models for the T-structure (LSCO/LBCO), at two different temperatures. The numerical values presented are obtained without the constant prefactor in the definition of conductivity, Eq.~\eqref{eq:Kubo_bubble}. Different curves correspond to different models and different values of $U$. Each theory curve is multiplied by the factor stated in the legend. The collapse of data points is excellent, except close to the Mott transition. 
}
\label{fig:Emery_vs_Hubbard_rhodc_vs_delta}
\end{figure}

\subsection{Hubbard vs. Emery: dc resistivity}
\label{sec:hubbard_vs_emery_transport}

\begin{figure}
\centering
    \includegraphics[width=\columnwidth,trim=0 0 0 0cm]{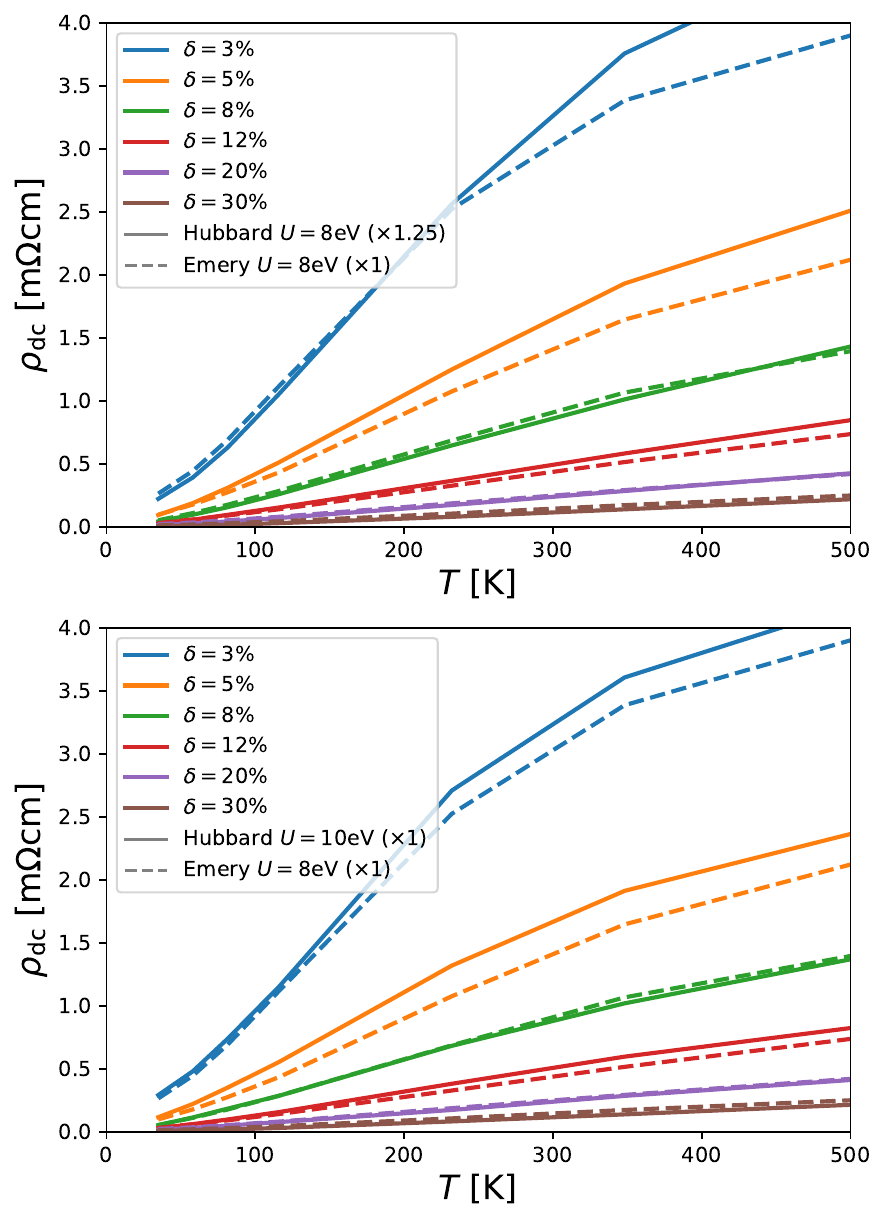}
\caption{
Comparison of the resistivity between the hole-doped Hubbard and Emery models for the T-structure (LSCO/LBCO). Different colors represent different dopings. Each theory curve is multiplied by a model-dependent constant indicated in the legend. The best agreement between the two models is obtained when $U^\mathrm{Hubbard}=\frac{5}{4}U^\mathrm{Emery}$ (bottom panel). When $U^\mathrm{Hubbard}=U^\mathrm{Emery}$, the result is qualitatively similar, but Hubbard model yields a smaller resistivity (top panel).
}
\label{fig:Emery_vs_Hubbard_rhodc_vs_T}
\end{figure}

We start by plotting the dc resistivity as a function of the doping in Fig.~\ref{fig:Emery_vs_Hubbard_rhodc_vs_delta}. We observe that the dc resistivity scales roughly as a power law of $U$, with a temperature-dependent exponent close to unity. This simple scaling does not hold at low values of $U$ where the undoped system is metallic. Furthermore, we see that the qualitative behavior is practically indistinguishable between the two models.
However, there is some quantitative difference between the two models, the resistivity in the Emery model being higher. 
The best matching of the resistivity is obtained when $U^\mathrm{Hubbard}=1.2 U^\mathrm{Emery}$ (see Fig.~\ref{fig:Emery_vs_Hubbard_rhodc_vs_T}). However, then the spectral functions are not well matched in terms of $\Delta_\mathrm{gap}$.

We note that the recent determinantal quantum Monte Carlo (DQMC) study\cite{Zhao2025} also finds that the Emery model has a larger resistivity than the Hubbard model, but the difference observed in that paper is somewhat less pronounced than what we see. This is likely due to a different parametrization of the models in that study, but could also be related to difference in systematic errors (finite-size effects and analytic continuation in DQMC vs. the neglect of vertex corrections in our DMFT approach\cite{Vucicevic2019}).

\subsection{Theory vs. experiments: dc resistivity}
\label{sec:transport_vs_experiment}
\begin{figure*}
\centering
    \includegraphics[width=\textwidth,trim=0 0 0 0cm]{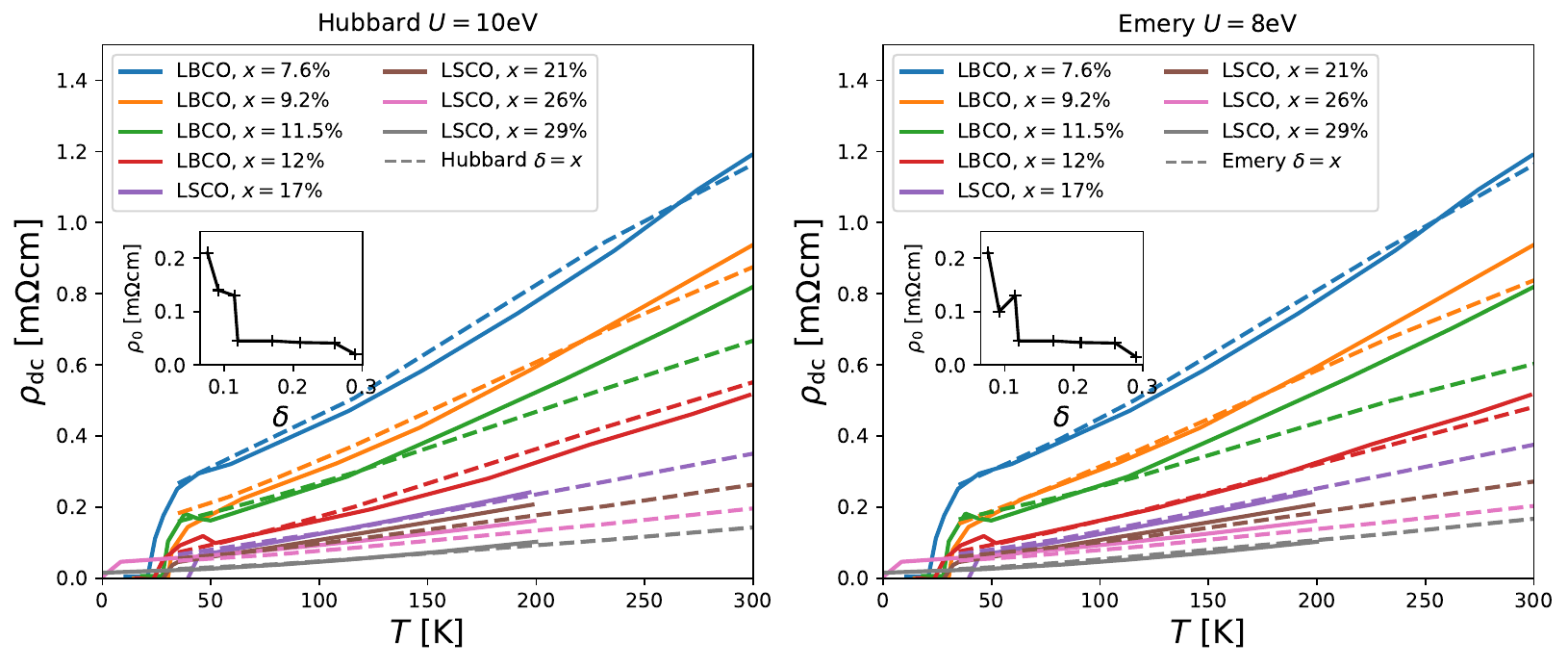}
    \includegraphics[width=\textwidth,trim=0 0 0 0cm]{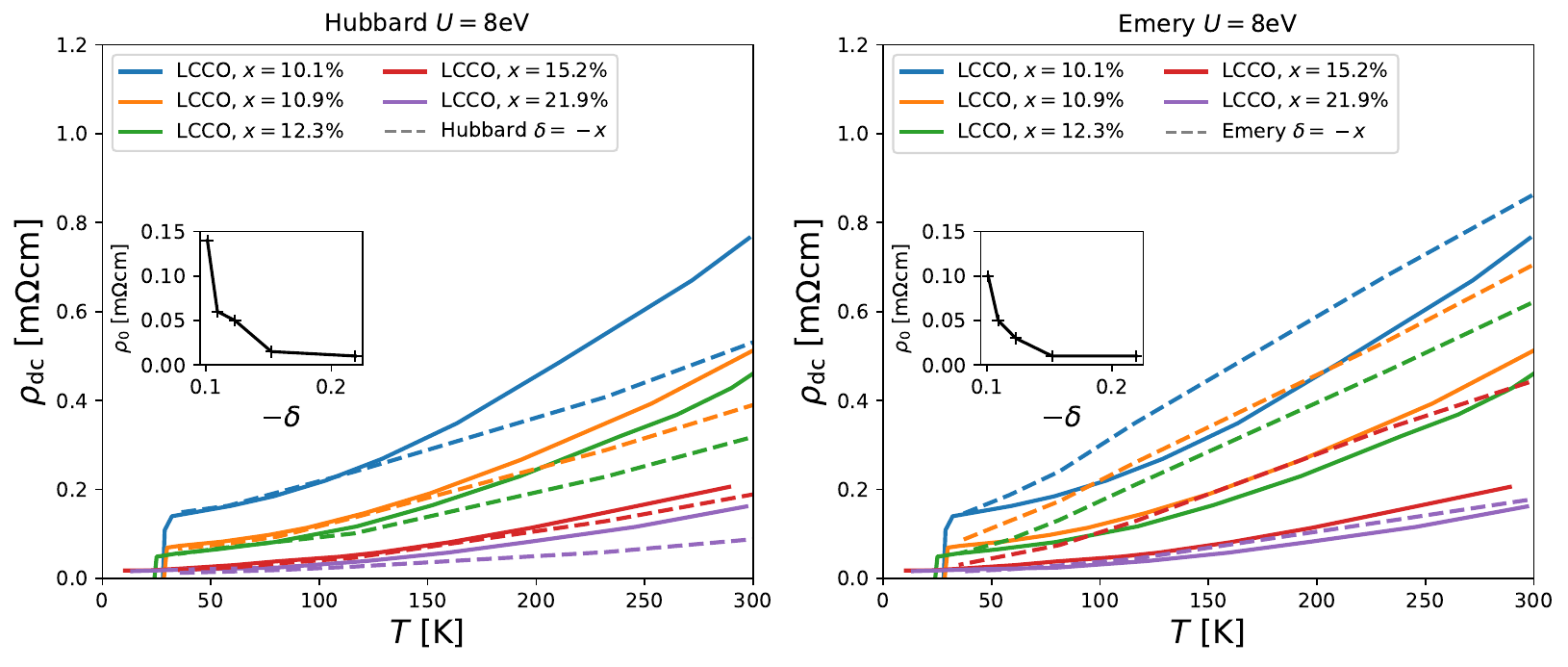}
\caption{
Comparison between our DMFT theory results and the experimental measurements for the dc resistivity in three La$_2$CuO$_4$-based cuprates. Left column: Hubbard model with $U^\mathrm{Hubbard}$ chosen to obtain the best agreement with experiment; 
%with $U^\mathrm{Hubbard}=\unit[10]{eV}$
Right column: Emery model with $U^\mathrm{Emery}=\unit[8]{eV}$. Top row: LSCO and LBCO, experimental data taken from Ref.~\onlinecite{Cooper2009,Tee2017}; Bottom row: LCCO, experimental data taken from Ref.~\onlinecite{Naito}. 
Solid curves are experiment, dashed curves are theory. Different colors of curves correspond to different dopings. Each theoretical curve is vertically shifted by a doping-dependent amount shown in the corresponding inset to account for the residual resistivity due to disorder in real samples.
}
\label{fig:rhodc_vs_experiment}
\end{figure*}

In Fig.~\ref{fig:rhodc_vs_experiment} we compare our theoretical $\rho_\mathrm{dc}(T)$ to the data measured on single-crystal LSCO\cite{Cooper2009}, LBCO\cite{Tee2017} and LCCO\cite{Naito}; the latter, in particular, was grown on a substrate using molecular beam epitaxy.
We again emphasize that to compare with LSCO and LBCO we use the hole-doped models for the T-structure, and to compare with LCCO we use the electron-doped models for the T'-structure.

As already mentioned, the real cuprates are disordered materials with a large residual resistivity $\rho_\mathrm{dc}(T=0)$. In the DMFT solutions for the clean, doped Hubbard and Emery models, the resistivity vanishes at $T=0$. To be able to compare with the experiment directly we shift the theoretical curves by adding a constant term $\rho_0$; the shift for each doping is shown in the insets. Shifting the curves is well motivated by Matthiessen's rule\cite{Ziman1960ElectronsPhonons,AshcroftMermin1976SolidStatePhysics,Ataei2022,Alloul2024}, according to which disorder makes an approximately 
temperature-independent contribution to the resistivity. However, this is unlikely to be applicable at very low doping, close to the Mott insulator. We consider only dopings above $|\delta|=3\%$ where the curves show a clear metallic trend, $\mathrm{d}\rho_\mathrm{dc}/\mathrm{d}T >0$, at all temperatures.

The experiments with LSCO and LBCO are performed at different ranges of doping, but combined results appear to describe a smooth evolution. This suggests that in those materials it is the doping level that determines the resistivity, not so much the mass or the atomic number of the dopant, at least below \unit[200]{K} where the data are available for both LSCO and LBCO. However, LBCO does exhibit a more complicated behavior at even higher dopings: there are two superconducting domes and the resistivity is not a monotonically decreasing function of the dopant (Ba) concentration $x$ \cite{Tee2017}. For that reason, we do not take into consideration the highest doping curve for LBCO shown in Ref.~\onlinecite{Tee2017}.

Upon inspecting Fig.~\ref{fig:rhodc_vs_experiment}, an immediate observation is that the hole-doped Emery model with $U^\mathrm{Emery}=\unit[8]{eV}$ describes the experimental LSCO/LBCO results very well.
However, it gives neither qualitatively nor quantitatively good results for LCCO, except at the highest doping. The Hubbard model with $U^\mathrm{Hubbard}=\unit[5]{eV}$ predicts far too low values for $\rho_\mathrm{dc}$ for all compounds.
However, with $U^\mathrm{Hubbard}=\unit[10]{eV}$, we get excellent agreement with LSCO.
Using $U^\mathrm{Hubbard}=\unit[8]{eV}$ we get excellent agreement with LCCO as well, but only up to about \unit[150]{K}. The experimental data show upward curvature of $\rho_\mathrm{dc}(T)$ above that temperature, which is not present in either of our models.
It is perhaps expected that the effective $U$ in LCCO is smaller, because we find that the bare Coulomb interaction in our models (obtained from wannierization) is about \unit[16.3]{eV} for LSCO/LBCO and about \unit[14.5]{eV} for LCCO (computed using \texttt{COQUI}\cite{Yeh2023,Yeh2024}). 

\begin{figure*}[t]
\centering
    \includegraphics[width=\textwidth,trim=0 0 0 0cm]{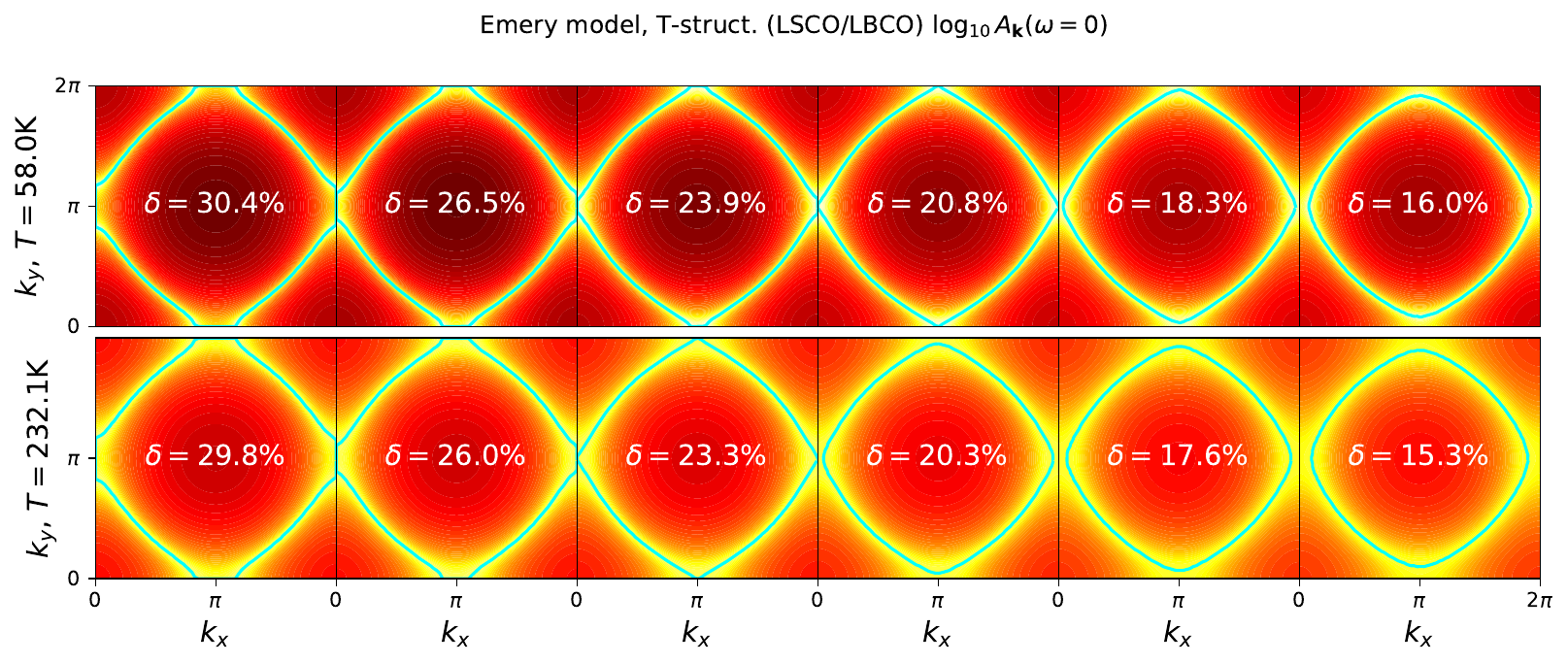}
\caption{
Illustration of the Lifshitz transition using the Emery-model results for the T-structure with $U^\mathrm{Emery}=\unit[8]{eV}$. The plots present the momentum-resolved spectral function at zero frequency; color mapping is logarithmic. Cyan line denotes the Fermi surface inferred from the maxima of the spectral function. Different columns correspond to different dopings, different rows to different temperatures. The change in topology of the Fermi surface is obvious, but $\delta_\mathrm{LT}$ is only known with the systematic error given by our $\delta$-resolution.
}
\label{fig:Lifshitz1}
\end{figure*}

\begin{figure}[ht!]
\centering
    \includegraphics[width=\columnwidth,trim=0 0 0 0cm]{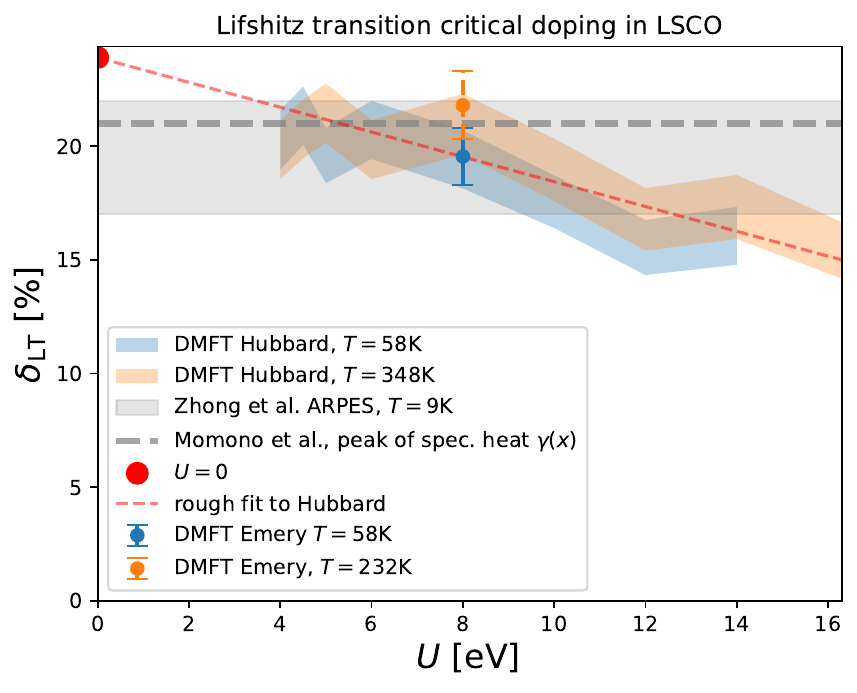}
    \caption{
Dependence of the critical doping for the Lifshitz transition $\delta_\mathrm{LT}$ on the model choice and $U$. Theoretical $\delta_\mathrm{LT}$ is compared to the experimental results: gray shaded horizontal stripe is the $x_\mathrm{LT}$ estimate from LSCO ARPES data in \emph{Zhong et al.}, Ref.~\onlinecite{Zhong2022}; dashed horizontal line is the $x_\mathrm{LT}$ estimate based on the maximum of the specific-heat coefficient $\gamma(x)$, as reproduced by multiple experiments\cite{Momono2002,Girod2021} and analyzed in \emph{Zhong et al}. At $U=0$, the LT occurs when the chemical potential is precisely at the van Hove singularity, in which case we can determine $\delta_\mathrm{LT}$ with high precision. In other cases, we present data with error bars set by our $\delta$-resolution. The dashed red line is a rough fit to theoretical data, indicating that the comparison with experiment is compatible with $U$ between 4 and \unit[12]{eV}.
    }
\label{fig:Lifshitz2}
\end{figure}

Most importantly, we see that the DMFT theory reproduces the roughly linear temperature dependence of resistivity at low temperatures and in a broad range of dopings. As $T\rightarrow 0$, the DMFT solution reaches a Fermi-liquid behavior and $\rho_\mathrm{dc}\sim T^2$, but this happens at much lower temperatures, below those that are relevant for experiments addressing transport properties of the normal state. In doped Mott insulators at strong coupling, the characteristic Fermi-liquid temperature $T_\mathrm{FL}$ is expected to be low (as was shown in the $t'=0$ Bethe lattice Hubbard model\cite{deng}), and longer range hoppings that we have in our model might suppress $T_\mathrm{FL}$ even further. Our results show that strange metallic ($T$-linear) resistivity can exist at experimentally relevant low temperatures, even if the ground state is a Fermi liquid. This is relevant for the hypothesis that linear resistivity in the cuprates would extend to $T=0$ were it not for the superconducting phase below $T_c$\cite{Legros2018}.

Finally, it is important to note that our DMFT results for $\rho_\mathrm{dc}$ are subject to systematic errors. In our previous work\cite{Vucicevic2019}, we have shown that the main error comes from neglecting vertex corrections to $\rho_\mathrm{dc}$. They were found to mostly amount to an overall downward shift of the $\rho_\mathrm{dc}(T)$, at least at high enough temperatures. Therefore, it might be possible to absorb the vertex corrections in our constant shift of the curves, and they might not affect our conclusions based on the comparison with experiment. In addition, we are more likely to overestimate than underestimate $\rho_\mathrm{dc}$. That means that the necessity for taking a higher $U$ in the Hubbard model is likely a robust result. Furthermore, in the case of the Emery model, it is possible that vertex corrections would push our $\rho_\mathrm{dc}(T)$ results downward, and towards a better agreement with LCCO.

\subsection{Theory vs. experiment - Lifshitz transition in LSCO}
\label{sec:lifshitz}

LSCO is known to undergo a Lifshitz transition (LT) upon doping\cite{Ino2002,yoshida2006,Zhong2022} where the topology of the Fermi surface changes from hole-like to electron-like at around $x_\mathrm{LT}=21\%$. The ARPES data presented in Ref.~\onlinecite{Zhong2022} narrow down $x_\mathrm{LT}$ to between 17\% and 22\%, but the sharp maximum in the $T$-linear specific-heat coefficient $\gamma$ that occurs at precisely $x=21\%$ suggests that this is the exact value of $x_\mathrm{LT}$.

We inspect the shape of the Fermi surface in our DMFT solutions. At finite temperatures, we may define the Fermi surface as the line connecting the maxima of the (total) spectral function at zero frequency $A_\mathbf{k}(\omega=0)$ along each radial direction $\theta$, i.e. $k_\mathrm{F}(\theta)= \mathrm{argmax}_k A_{\mathbf{k} = (k\cos\theta,k\sin\theta)}(\omega=0)$ [where such a maximum exists away from $\mathbf{k}=(\pi,k_y)$ and $\mathbf{k}=(k_x,\pi)$ lines]. In Fig.~\ref{fig:Lifshitz1} we illustrate the LT in our Emery model solution. We color code $A_\mathbf{k}(\omega=0)$ in the first Brillouin zone, and the Fermi surface is indicated by a cyan line. We can determine $\delta_\mathrm{LT}$ only up to a systematic uncertainty set by the $\delta$-resolution we have in our data. 
For $U=0$, the LT occurs when the chemical potential coincides precisely with the van Hove singularity in the non-interacting density of states, and this we can calculate with high precision.

In Fig.~\ref{fig:Lifshitz2} we summarize our finite-temperature estimates of $\delta_\mathrm{LT}$.
It appears that $\delta_\mathrm{LT}$ is pushed downwards as $U$ is increased, roughly linearly. 
Importantly, when we take the value of $U$ that yields good agreement with the experimental $\rho_\mathrm{dc}$, we are also consistent with the experimental value of $x_\mathrm{LT}$.
However, we see that $\delta_\mathrm{LT}$ depends slightly on the temperature at which we compute the estimate. The experiment in Ref.~\cite{Zhong2022} was performed at a very low temperature where the DMFT solution is unlikely to give reasonable results for the spectral function; our estimate at the higher temperature is perhaps the most meaningful.

\subsection{Theory vs. experiment - effective mass and Planckian dissipation}
\label{sec:effective_mass}

\begin{figure*}
\centering
    \includegraphics[width=\columnwidth,trim=0 0 0 0cm]{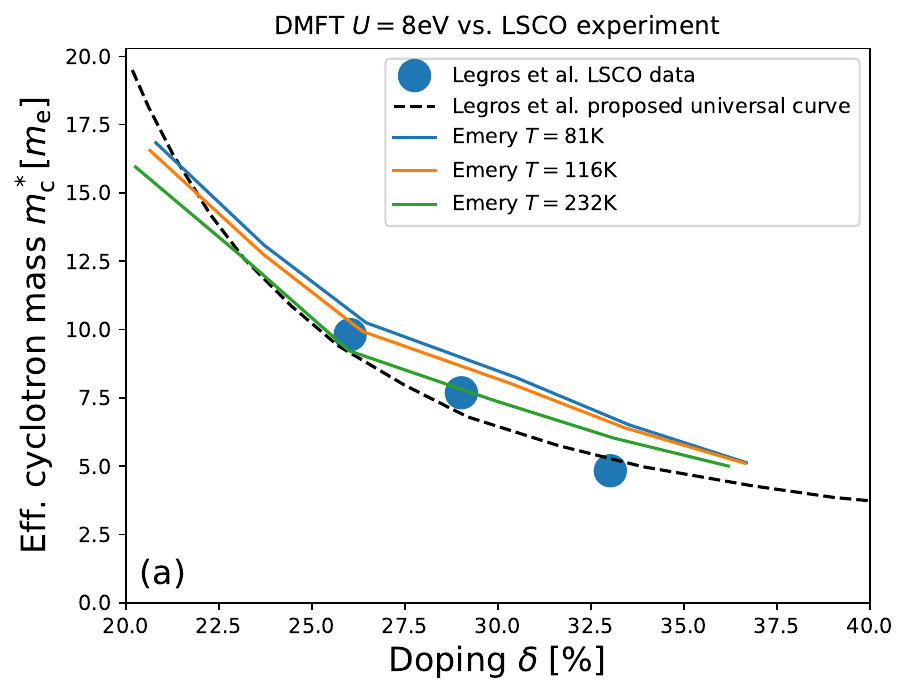}
    \includegraphics[width=\columnwidth,trim=0 0 0 0cm]{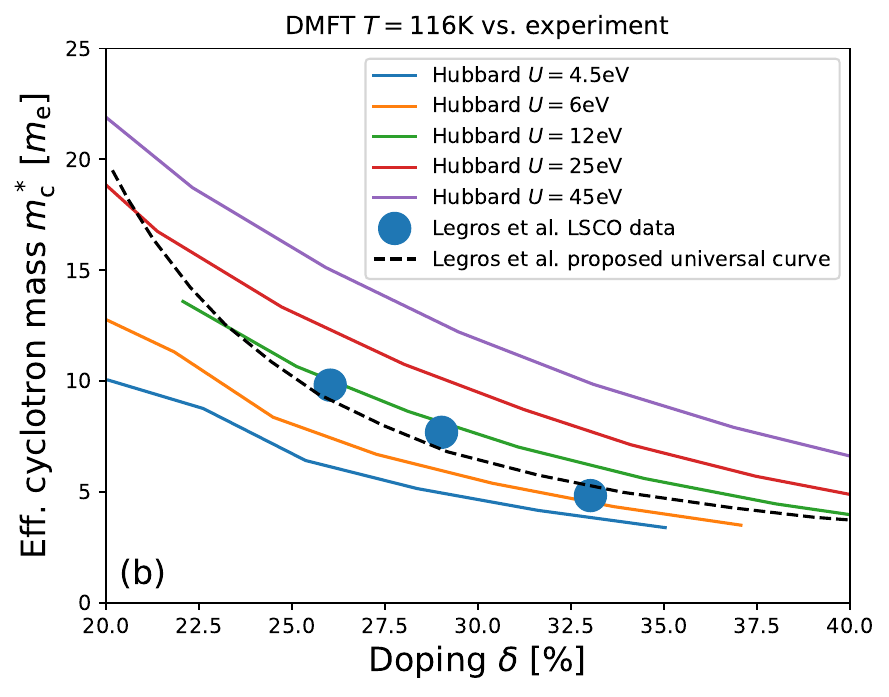}
\caption{Comparison between theoretical and experimental results for the effective cyclotron mass $\mc$ as a function of doping. LSCO data are taken from \emph{Legros et al.}, Ref.~\onlinecite{Legros2018} and presented with large blue symbols, roughly the size of the experimental error bars. 
Dashed black line is a hypothetical universal curve, proposed in \emph{Legros et al}. On the left panel we show that the Emery-model result for $\mc$ is in good agreement with experiment, and that it depends weakly on $T$ at which it is extracted from the momentum-resolved spectral functions. On the right panel we illustrate how $\mc$ depends on $U^\mathrm{Hubbard}$, and that the best agreement is obtained with $U^\mathrm{Hubbard}$ in the range \unit[6-12]{eV}, which is consistent with the estimates from the analysis of both $\rho_\mathrm{dc}$ and $\delta_\mathrm{LT}$.
}
\label{fig:effective_mass}
\end{figure*}

\begin{figure}
\centering
    \includegraphics[width=\columnwidth,trim=0 0 0 0cm]{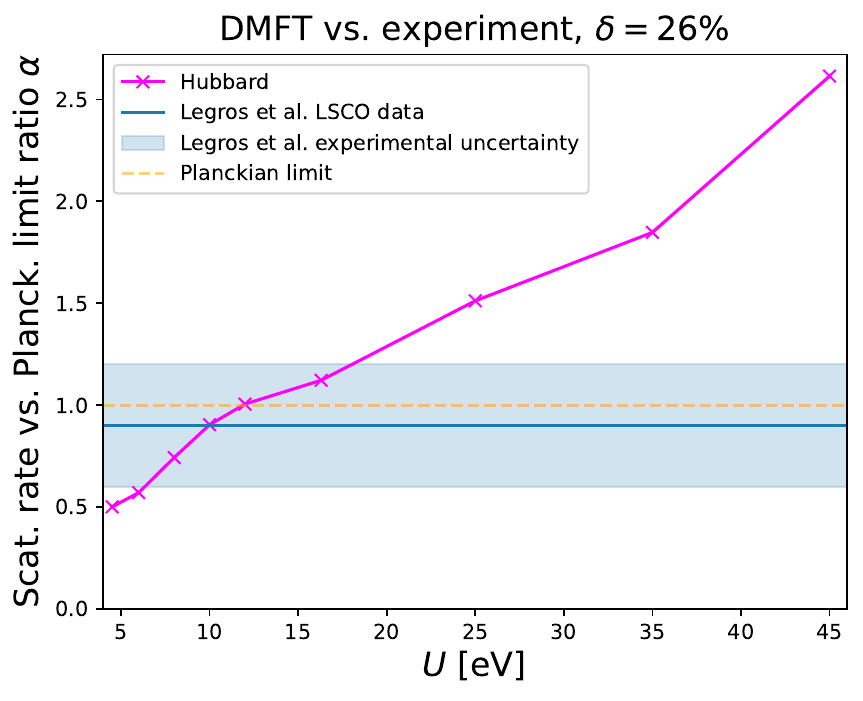}
\caption{
Comparison between our Hubbard model theory and the experiment\cite{Legros2018} in terms of the effective scattering rate. The factor $\alpha$ is the ratio between the effective scattering rate and the Planckian limit. The theoretical results are computed from the slope of the sheet resistivity $\partial_T \rho_\mathrm{dc}^\square(T)$ and the cyclotron mass $\mc$, \emph{following closely} the procedure used in \emph{Legros et al.}, Ref.~\onlinecite{Legros2018}. Using this approach, the result for the Hubbard model with $U^\mathrm{Hubbard}=\unit[10]{eV}$ 
reproduces the experimental extraction under the same cyclotron-mass normalization. A critique of the approach is, however, presented in the text and in App.~\ref{app:effective_masses}.
Namely, we observe that the procedure in \emph{Legros et al.} contains an inconsistency in the definition of the effective mass - it appears that the effective mass in the Sommerfeld formula (i.e., thermodynamic/cyclotron mass) has been identified with the effective (transport) mass in the Drude formula, even though these quantities are in general quite different, even in the absence of interactions. Using the transport mass estimated from $\langle K \rangle$ to extract the effective scattering rate, we find a scattering rate that is about 5 times bigger than the Planckian limit, i.e. $\alpha\approx 5$. See text and App.~\ref{app:effective_masses} for details.
}
\label{fig:alpha}
\end{figure}

Finally, we consider the effective \emph{cyclotron} mass, $\mc$. It can be extracted from the spectral function as\cite{Tamai2019}
\begin{equation}
 \mc = \frac{\hbar}{2\pi}\int \mathrm{d}\theta k_F(\theta)/v^*_F(\theta)
\end{equation}
where $\hbar k_F$ is the Fermi momentum, and $v_F^*$ is the renormalized Fermi velocity in the direction $\theta$ of reciprocal space which can be deduced from the line connecting the maxima of $A_{\mathbf{k}=(k\cos\theta,k\sin\theta)}(\omega)$ for each $\theta$. Our results can be readily compared with the LSCO measurements reported in Ref.~\cite{Legros2018}, see Fig.~\ref{fig:effective_mass}. In this particular experiment, $m^*$ was estimated from specific heat data, assuming the validity of the Sommerfeld expression (see App.~\ref{app:effective_masses} for details). 
Most importantly, we find that $\mc$ is strongly $U$ dependent (see Fig.~\ref{fig:effective_mass}(b)). The Hubbard model with $U^\mathrm{Hubbard}=\unit[6-12]{eV}$ yields the best agreement with experimental data. 
The Emery model with $U^\mathrm{Emery}=\unit[8]{eV}$ is also in very good agreement with experiment.
Once again, the models that reproduce well the experimental $\rho_\mathrm{dc}$, also reproduce well $\mc$. This finding is not trivial: at the level of our theory, $\mc$ is not simply related to $\rho_\mathrm{dc}$ and it is not subject to vertex corrections. In principle, $\mc$ ought to characterize a zero-temperature Fermi liquid; our calculation is performed at a finite temperature, but we find that the estimate of $\mc$ is only slightly dependent on temperature (see Fig.~\ref{fig:effective_mass}(a)). Again, we choose here intermediate temperatures, as DMFT is not expected to work well at very low temperature.

We have performed the same calculation for the LCCO. The experimental data are available only for a single doping. The Emery model strongly overestimates this $m^*$ value. Good agreement is obtained with the Hubbard model only when we take $U^{\mathrm{Hubbard}}=\unit[4]{eV}$. This is inconsistent with our observations related to dc resistivity, where we had to take a much larger $U^\mathrm{Hubbard}$ to reproduce experimental results. It is important to note that in the case of LCCO the cyclotron mass $m^*$ in Ref.~\onlinecite{Legros2018} was extracted from quantum oscillations, rather than specific heat data. It is not clear that the effective mass extracted from specific heat and quantum oscillations must be exactly the same; they could differ e.g. due to $k_z$-dependent Fermi-surface dispersion or field dependence \cite{Merino2000cyclotron,Joss1987}.

The effective mass $m^*$ plays an important role in the Planckian dissipation theory of linear resistivity. The idea is that, under certain conditions (e.g. in the vicinity of a quantum critical point), the scattering rate saturates a $T$-linear Planckian upper bound, given by $\alpha k_\mathrm{B} T/\hbar$, with $\alpha=1$, universally. (Here $k_\mathrm{B}$ is the Boltzmann constant.) This is supported by data from numerous experiments, although $\alpha$ is typically extracted with a large error bar. In our data at around 20--30\% doping, we indeed observe a strange metallic behavior where $\rho_\mathrm{dc}\sim T$ in the range of temperature \unit[50--200]{K}. We follow precisely the procedure laid out in \emph{Legros et al.} (Ref.~\onlinecite{Legros2018}) and compute the scattering factor $\alpha$ (the ratio between the scattering rate and the Planckian limit) as
\begin{equation} \label{eq:A1_vs_alpha}
\alpha = \frac{e^2 \hbar}{k_B} \frac{1-\delta}{a^2} \frac{A_1^\square }{m^*}
\end{equation}
where $m^*$ is taken to be $\mc$, the slope of the sheet resistivity is defined as $A_1^\square = A_1/d = \partial_T \rho_\mathrm{dc}(T)/d$, $a$ is the in-plane lattice spacing (T-structure: 3.81\AA; T'-structure: 4.01\AA), and $d$ is the thickness of a single CuO$_2$ layer ($d=c/2$, where $c$ is the thickness of the conventional unit cell stated earlier). The results are shown in Fig.~\ref{fig:alpha}. Unsurprisingly, in our Hubbard model with $U^\mathrm{Hubbard}=\unit[10]{eV}$, given that we have excellent agreement of both $\rho_\mathrm{dc}$ and $m^*$ with the experiment, we can reproduce $\alpha\approx 1$ at $\delta=26\%$ which was reported in \emph{Legros et al.} However, we see that this value is not universal and that it strongly depends on the coupling constant $U$. In fact, increasing the coupling $U$ seems to increase $\alpha$ beyond 1. This contrasts with the findings of Ref.~\onlinecite{fratini2025strangemetaltransportcoupling} in the $tJ$ model where $\alpha$ approaches 1 from above as $J$ goes to zero (which roughly corresponds to taking $U\rightarrow\infty$ in the Hubbard model). Super-Planckian materials with $\alpha>1$ are known to exist\cite{Hwang2019,Sadovskii2021,Poniatowski2021,Hartnoll2022}, but the value of $\alpha$ rarely exceeds 1 by much. Indeed, we do not see $\alpha$ become large, even for values of the effective coupling that are well beyond what one might expect to have in a real material.

However, we observe that Eq.~\eqref{eq:A1_vs_alpha} used in \emph{Legros et al.} follows directly from the Drude expression
\begin{equation} \label{eq:drude_for_show}
\rho_\mathrm{dc} = \frac{m^*}{ne^2} \frac{1}{\tau}
\end{equation}
under the assumptions that $1/\tau=\alpha k_B T/\hbar$, $\rho_\mathrm{dc}^\square(T) = A_1^\square T$, and that the concentration of electrons per unit cell $n$ corresponds to all of the electrons in the (single) valence band (one per unit cell, minus the doping).
Under those assumptions (see Appendixes~\ref{app:effective_masses} and ~\ref{app:Drude_and_Planck} for details), the effective mass $m^*$ in Eq.~\eqref{eq:drude_for_show} must have the meaning of \emph{transport} effective mass $\mtr$, which is in general quite different from the \emph{cyclotron} mass $\mc$, even in the absence of interactions, and especially so in the vicinity of Lifshitz transitions (see App.~\ref{app:effective_masses} for details). 
Our result $\alpha \approx 1$ was thus obtained using cyclotron mass in place of transport mass in Eq.~\eqref{eq:A1_vs_alpha}. 
Only in this way are we able to reproduce the $\alpha \approx 1$ result reported in \emph{Legros et al.} 
However, at the level of our theory, $\mtr$ can be easily obtained from the kinetic-energy (stress-tensor) operator expectation value.
We find that at $\delta=26\%$, $\mtr$ is about $1.5-2m_e$ (depending on $U$), i.e. about five times smaller than $\mc$ (which is about 7.5-10$m_e$ in the same $U$-range). We believe that using $\mc$ and $\mtr$ interchangeably may have led to a poor evaluation of the scattering rate in some of the experimental literature, including in \emph{Legros et al.} By using the transport mass $\mtr$ in Eq.~\eqref{eq:A1_vs_alpha} we come to the conclusion that the inelastic scattering rate is much higher than the Planckian limit, i.e. $\alpha\approx 5$.

However, we emphasize that there are other possible definitions of the effective scattering rate \cite{Hartnoll2022}. 
The estimate of scattering rate based on the imaginary part of the self-energy is consistent with $\alpha\approx 5$ (see App.~\ref{app:effective_masses}).
If the estimate is made for a small subset of the electrons that contribute only to the low-energy Drude peak in the optical conductivity spectrum, then a much smaller value, closer to the $\alpha=1$ Planckian limit, is obtained. However, such an estimate cannot be obtained solely based on $\rho_\mathrm{dc}$ - rather, the frequency-dependent optical conductivity results are needed. 
In addition, we find that this scattering rate is not quite $T$-linear or simply proportional to the dc resistivity.
We discuss in detail the different estimates of the scattering rate, the optical conductivity and the underlying assumptions of the Planckian dissipation hypothesis in Appendixes \ref{app:effective_masses} and \ref{app:Drude_and_Planck}.

\section{Discussion, conclusions and prospects for future work}
\label{sec:discussion}

The central question our work poses is: Which model and which choice of the effective coupling are most appropriate for the cuprates?
Despite good agreement between our data and experimental results across multiple physical quantities and cuprate materials, we cannot give a definite answer to this question.
In terms of the dc resistivity, both Emery and Hubbard models yield excellent agreement with the LSCO/LBCO experimental results. 
In the Emery model, the conventional estimate of the coupling constant $U^\mathrm{Emery}=\unit[8]{eV}$\cite{Weber2012,Sheshadri2023} indeed appears to be the correct one.
However, the Hubbard model only works if the coupling constant is taken large, $U^\mathrm{Hubbard}\approx\unit[10]{eV}$. 
In the case of LCCO, the Hubbard model with $U^\mathrm{Hubbard}\approx\unit[8]{eV}$ yields dc resistivity in excellent agreement with experiment, but only at temperatures below \unit[150]{K}; at higher temperatures, the experimental curves bend upwards sharply, unlike what is seen in LSCO/LBCO. The Emery model (with $U^\mathrm{Emery}=\unit[8]{eV}$) overestimates resistivity in LCCO, and similarly fails to reproduce the curving of $\rho_\mathrm{dc}(T)$ observed in experiment.
These observations can lead to any of the following conclusions:
\begin{itemize}
\item The conventional estimate of the effective $U$ in the MLWF downfolded Hubbard model (around 4eV\cite{Das,Werner2015,Sheshadri2023}) is wrong. When parametrized correctly, both Hubbard and Emery models give good predictions for the dc resistivity. In addition, the effective coupling can vary significantly between compounds, and in LCCO it is smaller than in LSCO/LBCO because of a slightly different crystal structure (a trend which is already observed in the bare Coulomb value in the basis of the corresponding Wannier orbitals).

\item The estimate of electron velocity obtained via Peierls substitution in single-band models is poor, leading to an underestimation of the dc resistivity. The problem is alleviated already with the three-orbital Emery model which correctly captures the relevant conduction processes, which certainly involve hopping over the oxygen sites in the copper-oxide planes. However, in the Hubbard model one must take an unjustifiably high value of coupling to reproduce the experimental results.

\item Upon doping, in reality, some of the holes go to the oxygen sites. This means that the effective doping $\delta(x)$ in the Hubbard model should be less than the dopant concentration $x$, which would lead to higher resistivity estimates, and one would then need a more moderate $U^\mathrm{Hubbard}$ to reproduce experimental results.

\end{itemize}
Based on the dc resistivity data alone, it would be difficult to deduce which one of the above conclusions is correct.
However, the data we have obtained for other physical quantities in LSCO corroborate that the correct $U$ in the Hubbard model is relatively high: both the critical doping for the Lifshitz transition $\delta_\mathrm{LT}$ and the effective cyclotron mass $m^*_\mathrm{c}$ narrow down the range of possible $U^\mathrm{Hubbard}$ to between about 6 and \unit[12]{eV}. 
We emphasize that these quantities are computed directly from the spectral function and are not subject to vertex corrections that are missing in our computation of resistivity.
Even our comparison with LSCO ARPES data suggests that $U^\mathrm{Hubbard}$ is about \unit[10]{eV}. 
Still, one has to consider that the optimal $U^\mathrm{Hubbard}$ might be different for different physical quantities, and that no single model can reproduce the entire physics of a given cuprate compound.
In addition, extracting from the comparison with experiment both the coupling constant $U$ and the effective doping $\delta(x)$, at the same time, is difficult. If we assume that the doping in the Hubbard model is some percentage $c$ of the dopant concentration, say $\delta=cx$, the estimate of $U$ will depend on the precise value of $c$. However, we have checked that $c$ would have to be unreasonably small for the $U^\mathrm{Hubbard}$ estimate to be below \unit[6]{eV}. Ultimately, the above three conclusions are not necessarily mutually exclusive - the effective $U^\mathrm{Hubbard}$ might be a bit higher than expected, the Peierls substitution might introduce some systematic error, and the effective doping in the Hubbard model might be a bit less than the dopant concentration, altogether leading to the overestimation of $U^\mathrm{Hubbard}$ based on our data. By contrast, the Emery model with the conventional estimate $U^\mathrm{Emery}=\unit[8]{eV}$ works well, indicating perhaps that inclusion of oxygen orbitals alleviates some of the issues present in the case of the Hubbard model.

We end the discussion of the (correct) effective coupling in our lattice models by summarizing our results in Table~\ref{tab:UHubbard_vs_experiment}. 
By comparing results of the Hubbard model with experiments, we constrain the correct value of $U^\mathrm{Hubbard}$; depending on the quantity, we a priori get different ranges: in the case of LSCO/LBCO the estimated ranges are consistent, but it is not so in the case of LCCO.

\begin{table*}
    \centering
    \begin{tabular}{c|c|c}
        Quantity & \multicolumn{2}{|c}{$U^\mathrm{Hubbard}$ [eV] to match experiment} \\        
        & LSCO/LBCO & LCCO \\
        \hline
        dc resistivity $\rho_\mathrm{dc}(T; x=\delta)-\rho_0$ & $\approx 10$ & $\approx 8$\\
        shape of ARPES spectrum at the node $\rho^{<}_{\mathbf{k}=\mathrm{node}}(-0.1\mathrm{eV}<\omega<0.1\mathrm{eV})$ & $8-12$ &  \\
        doping for Lifshitz transition $\delta_\mathrm{LT}=x_\mathrm{LT}$ & $5-12$ & \\
        cyclotron mass $m^*_\mathrm{c}$ & $6-12$ & $\approx 4$    
    \end{tabular}
    \caption{Summary of findings regarding the optimal $U^\mathrm{Hubbard}$ to reproduce experimental results for various quantities.}
    \label{tab:UHubbard_vs_experiment}
\end{table*}

\begin{table*}
    \centering
    \begin{tabular}{c|c}
        Quantity & $U^\mathrm{Hubbard}/U^\mathrm{Emery}$ to match results between two models \\
        \hline
        dc resistivity $\rho_\mathrm{dc}(T; \delta)$& $\approx 5/4$ \\
        spectral weight at zero frequency $A(\omega=0)$ & $\approx 1$ \\
        size of the spectral gap $\Delta_\mathrm{gap}$ & $\approx 2/3$
        
    \end{tabular}
    \caption{Summary of findings regarding the matching between Hubbard and Emery models in terms of various quantities.}
    \label{tab:UHubbard_vs_UEmery}
\end{table*}

Another important aspect of the cuprates is the presence of significant disorder, and at this point we are not accounting for it in our models - rather, we are assuming the validity of Matthiessen's rule and making comparisons with experimental resistivity data only up to an overall temperature-independent shift.
Disorder is also expected to play a role in spectral line shapes, and our DMFT indeed predicts much more coherent quasiparticles than is observed in ARPES.
It is unclear how much of this discrepancy can be attributed to disorder in real samples and how much to systematic errors in DMFT and our modelling scheme.

However, we are unaware of other readily available methods that are both less approximate and feasible for all the computations that we perform in this work. For example, cluster DMFT can be used to solve both the Hubbard and Emery models and introduce some nonlocal components of the self-energy\cite{KowalskiPNAS2021}.
If an exact-diagonalization solver is used for the impurity problem, one obtains a discrete self-energy which might be unsuitable for calculation of transport or detailed spectral properties considered here, unless an uncontrolled broadening is performed.
If quantum Monte Carlo solvers are used instead, then analytic continuation to real-frequency domain
might introduce a more significant systematic error than the one we make with our single-site DMFT solution\cite{Vucicevic2019,jarrell1996,Gunnarsson2010}.
Determinantal quantum Monte Carlo methods\cite{Zhao2025, Huang2019} also require analytic continuation, cannot be used for detailed inspection of the Fermi surface due to finite-size effects (as needed for the estimates of effective mass and the study of Lifshitz transition), and most likely cannot be pushed to the range of relatively low temperatures we consider here (say $T<0.1t$ in the Hubbard model). Methods such as the finite-temperature Lanczos method\cite{kokalj2017,Vucicevic2019} and numerical linked cluster expansion\cite{Pham2026arXiv} avoid analytic continuation, but similarly suffer from strong finite-size effects and cannot be used at sufficiently low temperature. Another possible approach is the diagrammatic Monte Carlo method\cite{vucicevic2020,vucicevic2021,Kovacevic2025}. It would be, however, very difficult to converge the interaction-expansion series at the values of the coupling relevant for the cuprates, and the computation of conductivity would either require analytic continuation\cite{Eom2025arXiv}, or a prohibitively strong scaling of complexity with the perturbation order\cite{Kovacevic2025}, at least with the currently available methods. It is unclear to us whether tensor-based methods like finite-temperature DMRG might be feasible, but finite-size effects and breaking of rotational symmetry\cite{Yang2016,Wietek2021,Zhang2026prb} could still be a significant source of systematic error.

We thus conclude that the DMFT+NRG approach we employ here is well suited to the broad real-frequency survey undertaken here. 
Indeed, this approach has been applied very recently to other perovskite oxides, with good success\cite{LaBollita2026,Kugler2026}.
Moreover, the single-site DMFT solution of the Hubbard model has been well understood for decades, which allows us to better interpret our results. It is well known that this solution exhibits no quantum critical points at finite doping. Still, as we see in the case of our models, the DMFT does reproduce the strange-metal regime, even regardless of the coupling constant; with the appropriately set coupling, we reproduce the strange-metallic regime in LSCO, LBCO and LCCO even \emph{quantitatively}. 
We reproduce the effective mass and the $\alpha=1$ value hailed as the hallmark of the Planckian dissipation mechanism for linear-in-temperature resistivity, but show that both are sensitive to model parametrization. These results raise the question of whether a more mundane explanation of linear resistivity is possible, perhaps one that does not involve any quantum criticality or universal phenomena. 
Planckian dissipation and other critical phenomena could still be associated even with quantum critical points in a wider parameter space of a more general system, as suggested earlier\cite{VucicevicPRL2015}.
However, we show that a more rigorous interpretation of the LSCO experimental data leads to a drastically different conclusion regarding the value of $\alpha$. Namely, we show that in the estimate of the inelastic scattering rate one should use the effective transport mass instead of the cyclotron mass (which was used in  \emph{Legros et al.}). One then obtains an effective scattering rate as high as 5 times the Planckian rate in LSCO at 26\% doping. 
We also show that, ultimately, the estimate of the scattering rate depends crucially on its definition, and that alternative definitions can yield lower values, closer to the Planckian limit. We find, however, that such definitions are not consistent with the underlying assumption that $\rho_\mathrm{dc}\sim\Gamma$ or $\Gamma\sim T$. 
This highlights the limitations of the Planckian dissipation paradigm and brings into question whether it applies to LSCO at all.

Finally, our study highlights some robust differences in the spectral functions predicted by the Hubbard and the Emery model which might have been previously overlooked. The difference in the doping-dependence of the energy gap might be large enough to be resolvable in photoabsorption experiments.
We also find robust similarities between the predictions of the two models, but, interestingly, different quantities are matched best using different ratios of the effective couplings, $U^\mathrm{Hubbard}/U^\mathrm{Emery}$.
The matching between the two models in terms of resistivity is achieved by taking $U^\mathrm{Hubbard}\approx \frac{5}{4}U^\mathrm{Emery}$;
the matching between sizes of energy gaps in the two models is obtained when $U^\mathrm{Hubbard}\approx \frac{2}{3}U^\mathrm{Emery}$, whereas matching in the spectrum around the Fermi level is obtained when $U^\mathrm{Hubbard}\approx U^\mathrm{Emery}$. These results are summarized in Table~\ref{tab:UHubbard_vs_UEmery}. 
In addition, we find that the doping dependence of the spectral weight in the vicinity of the Fermi level is strongly sensitive to the coupling strength, apparently regardless of the model. Our findings might serve as a basis for constraining the value of the effective coupling in photoemission experiments. 
However, the picture is additionally complicated by disorder in cuprate materials which we do not account for in our clean-lattice-model theory. 

Overall, our results suggest ways to distinguish between different theoretical scenarios in experimental data. This sets the stage for a combination of theory and experiments that could resolve the question of the correct or minimal model to describe the normal-phase transport in the cuprates. This remains, however, an immensely difficult question, as the answer might ultimately depend on the specific compound, physical quantity we are interested in, as well as on the regime of physical parameters. Nevertheless, the goal should be to obtain a consistent description, with agreement between theoretical and experimental results for a wide range of quantities.

\appendix

\section{Effective masses}
\label{app:effective_masses}

\begin{figure*}
\centering
    \includegraphics[width=\columnwidth]{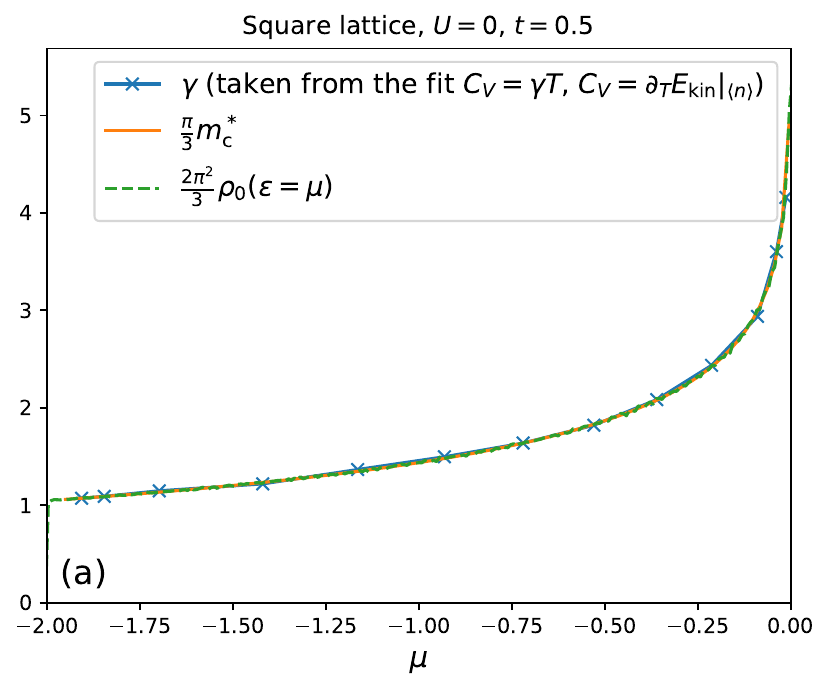}
    \includegraphics[width=\columnwidth]{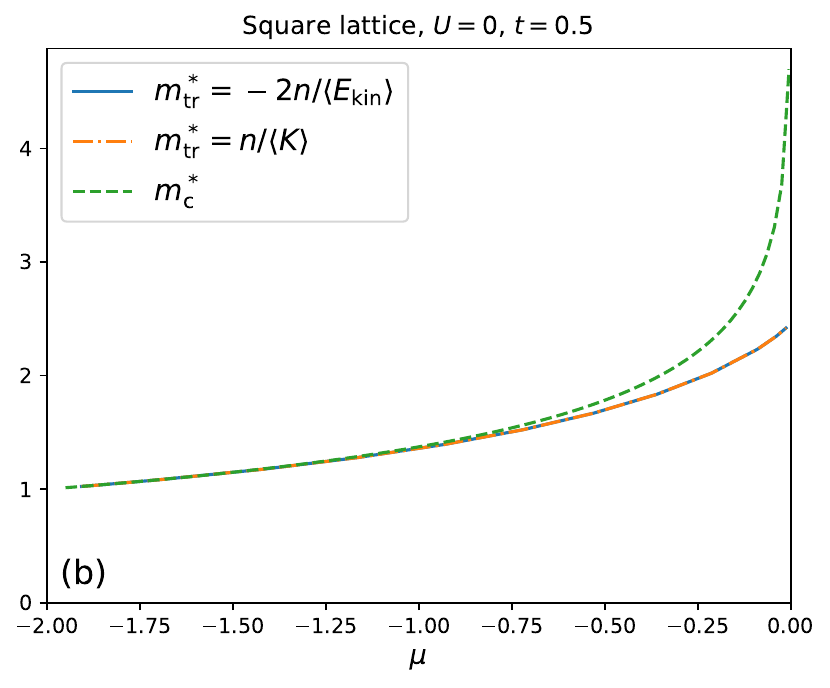}
\caption{Simple non-interacting square-lattice theory as an illustration of the difference between the cyclotron mass $\mc$ and the transport mass $\mtr$ appearing in the Sommerfeld and the Drude formulas, respectively. Here $\hbar=m_e=k_\mathrm{B}=2t=1$. (a) The results for the $T$-linear specific-heat coefficient $\gamma$, the corresponding Sommerfeld formula depending on the (computed) cyclotron mass, Eq.~\eqref{A5}, and the Fermi-level DOS [$2\rho_0(\varepsilon=\mu)$, where $\rho_0(\varepsilon)$ is normalized to 1] multiplied by $\pi^2/3$, Eq.~\eqref{A4}, showing perfect agreement for all values of the chemical potential $\mu$, as expected. 
(b) Comparison between $\mc$, Eq.~\eqref{eq:mstar_vanilla}, and $\mtr$ computed via Eq.~\eqref{eq:transport_mass}, as well as making use of the simplification arising in the present case, Eq.~\eqref{eq:K_vs_Ekin} and $\langle n_\mathbf{k} \rangle = n_\mathrm{F}(\varepsilon_\mathbf{k}-\mu)$; at the Lifshitz transition, the cyclotron mass $\mc$ diverges, whereas the transport mass $\mtr$ does not, which leads to a very large difference between the two effective masses.
}
\label{fig:U0_effective_mass}
\end{figure*}

\begin{figure}
\centering
    \includegraphics[width=\columnwidth]{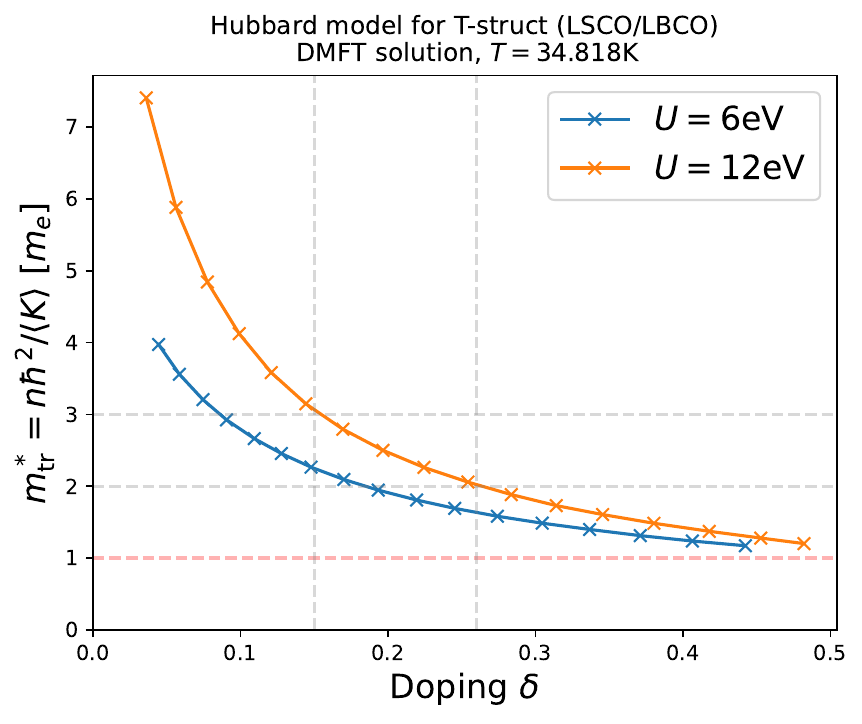}
\caption{
Effective transport mass as a function of doping, at a fixed low temperature and two values of $U$ in the DMFT solution of the Hubbard model for the T-structure. The effective transport mass is significantly less renormalized than the cyclotron mass (see Fig.~\ref{fig:effective_mass}). At $\delta=26\%$, we observe $\mc/\mtot\approx 5$, regardless of $U$ (in the range \unit[6-12]{eV}).
}
\label{fig:Hubbard_transport_mass}
\end{figure}

Effective mass is an important parameter characterizing Fermi-liquid states. However, there is no unique definition of the effective mass: different properties of the system depend on different "kinds" of effective mass. Here we recall two main definitions, relevant for our analyses:

\begin{itemize}
\item the \emph{cyclotron} (also referred to as the \emph{thermodynamic}) mass; it enters the Lifshitz-Kosevich theory\cite{LK} of quantum oscillations (de Haas-van Alphen effect), as well as the Sommerfeld expression for the specific heat\cite{imada1998}. In two dimensions, the cyclotron mass relates directly to the change of the area of the Fermi sea, $A_\mathrm{FS}$, with respect to the Fermi level
\begin{equation}
    \mc = \frac{\hbar^2}{2\pi} \frac{\partial A_\mathrm{FS}}{\partial E_\mathrm{F}}
\end{equation}
If we express $A_\mathrm{FS}$ as an integral in polar coordinates $(k,\theta)$, $A_\mathrm{FS} = \int \mathrm{d}\theta \int_0^{k_F(\theta)} k \mathrm{d}k$, and take a derivative of this expression with respect to $E_\mathrm{F}$ using the chain rule, we can write
\begin{equation}
    \mc = \frac{\hbar^2}{2\pi} \int_0^{2\pi} \mathrm{d}\theta \frac{k_\mathrm{F}(\theta)}{\partial\varepsilon_{k}(\theta)/\partial k|_{k=k_\mathrm{F}(\theta)}}.
\end{equation}
The denominator is nothing but the velocity in the radial direction, at the Fermi surface, i.e.
\begin{equation}\label{eq:mstar_vanilla}
    \mc = \frac{\hbar}{2\pi} \int_0^{2\pi} \mathrm{d}\theta \frac{k_\mathrm{F}(\theta)}{v_\mathrm{F}(\theta)}
\end{equation}
In strongly correlated Fermi liquids, the dispersion relation needs to be replaced by the effective dispersion $\varepsilon^*_\mathbf{k}$ formed by the zeros of $\omega+\mu-\varepsilon_\mathbf{k}-\mathrm{Re}\Sigma_\mathbf{k}(\omega)$ (or rather the $\omega$'s of the maxima of $-\mathrm{Im}G_\mathbf{k}(\omega)$ for each given $\mathbf{k}$); the effective velocity $v_\mathrm{F}^*(\theta)=(1/\hbar) \partial_k \varepsilon^*_{k}(\theta)$ is then to be used in Eq.~\eqref{eq:mstar_vanilla} instead of $v_\mathrm{F}(\theta)$.

In Fermi liquids at low temperature, the total energy is a quadratic function of temperature $\langle E(T)\rangle = \langle E(T=0)\rangle+\gamma T^2/2$, if the number of particles is held fixed. The specific heat is the derivative of the total energy with respect to temperature, 
$C=\partial \langle E(T)\rangle/\partial T = \gamma T$, i.e. the specific heat is a linear function of temperature, with the coefficient $\gamma$. The Sommerfeld expression for $\gamma$ reads \cite{Legros2018}
\begin{equation}
\gamma = \frac{\pi^2}{3} N^*(E_F) k_B^2
\label{A4}
\end{equation}
and taking the expression for the renormalized Fermi-level DOS $N^*(E_F)$ (including the spin degeneracy factor) in 2D systems we find
\begin{equation}
\gamma = \frac{\pi}{3} \frac{k_B^2}{\hbar^2} \mc.
\label{A5}
\end{equation}
Precisely this expression was inverted in \emph{Legros et al.} to extract the effective mass from specific-heat data.

\item the \emph{transport} mass; it enters the Drude expression for conductivity.
The Drude form of the real part of the optical conductivity reads:
\begin{equation}\label{eq:Drude_optical}
    \mathrm{Re}\,\sigma(\omega) = \frac{ne^2\tau}{\mtr} \frac{1}{1+\omega^2\tau^2}
\end{equation}
where $n$ is the density of conduction electrons, $e$ is the electron charge, $\tau$ is the scattering time (inversely proportional to the scattering rate $\hbar/\tau=\Gamma$).
In systems with Galilean invariance \cite{imada1998}, $\mtr$ is the actual electron mass $m_e$; in lattice systems 
this is an effective transport mass.

From Eq.~\eqref{eq:Drude_optical} it is straightforward to obtain
\begin{equation}\label{eq:Drude_integral}
    \int_0^\infty \mathrm{d}\omega\, \mathrm{Re}\,\sigma(\omega) = \frac{\pi n e^2}{2\mtr}.
\end{equation}
For the given concentration of electrons $n$, the transport mass therefore relates directly to the integral over optical conductivity (the f-sum), independently of the scattering rate.

On the other hand, in a single-band lattice model, the f-sum relates to the $xx$-component of the kinetic-energy (stress-tensor) operator $\langle K \rangle$ \cite{mahan2000many,BerthodApplicationsManyBodyFormalism,vanderMarel2014} 
\begin{equation}\label{eq:fsum_rule}
 \int_0^\infty \mathrm{d}\omega \mathrm{Re}\sigma(\omega) = \frac{\pi e^2}{2\hbar^2} \langle K \rangle,
\end{equation}
with
\begin{equation}
    \langle K \rangle = \int \frac{\mathrm{d}\mathbf{k}}{(2\pi)^2} \langle n_\mathbf{k}\rangle \frac{\partial^2 \varepsilon_\mathbf{k}}{\partial k_x^2},
\end{equation}
which measures the curvature of dispersion relation (band mass) in the $x$-direction, averaged over the occupied states. In simple cases $\langle K \rangle$ is related to kinetic energy; e.g. in the case of hypercubic lattices with only the nearest-neighbor hopping (in any dimension $d$), one finds 
\begin{equation}\label{eq:K_vs_Ekin}
\langle K \rangle =  -\langle E_\mathrm{kin} \rangle/d, 
\end{equation}
where $\langle E_\mathrm{kin} \rangle = \int \frac{\mathrm{d}\mathbf{k}}{(2\pi)^d} \langle n_\mathbf{k}\rangle \varepsilon_\mathbf{k}$.

Under the assumption that the Drude form Eq.~\eqref{eq:Drude_optical} matches the entire optical conductivity in the lattice model and that the number of carriers in the Drude theory is equal to the number of electrons in the lattice model,
one can combine Eqs.~\eqref{eq:Drude_integral} and \eqref{eq:fsum_rule} to obtain
\begin{equation}\label{eq:transport_mass}
\mtot = \frac{n\hbar^2}{\langle K \rangle}.
\end{equation}
Here $n$ is expressed in dimensional units, so that $n=(1-\delta)/a^2$.
Thus, the transport mass $\mtot$, defined by Eq.~\eqref{eq:transport_mass}, can be extracted from the knowledge of the stress tensor and the electron density. For the 2D Hubbard model, it is trivial to compute the stress tensor from the momentum-resolved spectral function
\begin{equation}
    \langle K \rangle = 2\int \frac{\mathrm{d}\mathbf{k}}{4\pi^2} \frac{\partial^2 \varepsilon_\mathbf{k}}{\partial k_x^2}  \int \mathrm{d}\omega n_\mathrm{F}(\omega) A_\mathbf{k}(\omega),
\end{equation}
where 2 in front of the integral comes from the sum over spin.

\end{itemize}

The transport and cyclotron masses are in general not the same. They become equal under stringent conditions of noninteracting systems with a single isotropic Fermi sheet and momentum-independent scattering time $\tau(\mathbf{k})=\tau$; in other cases they can become equal only incidentally. Even in the absence of interactions, the lattice effects can make those two masses very different.
We illustrate this in the $U=0$ Hubbard model case. For the sake of simplicity, we keep here only the nearest-neighbor hopping, $t=0.5$, and we keep the unit of energy $D=2t$. This choice of the hopping amplitude simplifies the connection to Fermi-liquid expressions, as in the vicinity of the $\mathbf{k}=0$ point, we have that $\varepsilon_\mathbf{k}\approx k^2/2$, which is precisely what we need if we assume that the bare electron mass is 1. Our results are presented in Fig.~\ref{fig:U0_effective_mass}. In panel (a) we show that $\gamma=\frac{\pi}{3}\mc$ where $\mc$ has been computed via Eq.~\eqref{eq:mstar_vanilla}. It is the cyclotron mass that appears in the Sommerfeld expression, and it controls the $T$-linear specific heat coefficient. In this very simple case, both $\gamma$ and $\mc$ are just proportional to the density of states at the Fermi level, as can be seen on the plot. This means that the cyclotron mass and $\gamma$ both diverge when the Fermi level lies at the van Hove singularity in the density of states. This is easily understood, as the van Hove singularity stems from the saddle points in the dispersion relation $\varepsilon_\mathbf{k}$ (occurring at the $\mathbf{k}=(0,\pi)$ and symmetry-related points); when the Fermi surface passes through these points, one gets $v_\mathrm{F}(\theta=0)=0$, which leads to divergence of $\mc$, and through a similar mechanism, to divergence of $\gamma$. In Ref.~\onlinecite{Zhong2022} the critical doping for the Lifshitz transition in LSCO was precisely evaluated from the maximum of $\gamma(x)$. 
Finally, in panel (b) we compare $\mtot$ (computed from the Drude expression Eq.~\eqref{eq:transport_mass}) with $\mc$ (computed via Eq.~\eqref{eq:mstar_vanilla}). We see that the transport mass gets renormalized near $\mu=0$, but only slightly, and it never becomes much larger than 1. Therefore, when the Fermi level is in the vicinity of the van Hove singularity, $\mtot$ becomes much smaller than $\mc$, which diverges. We emphasize that the divergence of $\mc$ when the Fermi level is at the van Hove singularity is not a fully general result, and $\mc$ will depend on the precise shape of the dispersion relation. However, in the vicinity of the Lifshitz transition, one must generally expect that $\mtr$ and $\mc$ will be substantially different. Interactions, of course, can make these two masses additionally different, even away from Lifshitz transitions.

Finally, in Fig.~\ref{fig:Hubbard_transport_mass} we present results for the transport mass $\mtot$, extracted from the Hubbard model results for $\langle K\rangle$, as defined by Eq.~\eqref{eq:transport_mass}.
We find that $\mtr(\delta)$ has a similar shape as $\mc(\delta)$ (shown in Fig.~\ref{fig:effective_mass}), but is much less renormalized. 
The cyclotron mass $\mc$ tends to be very large, reaching even $20m_e$ as we approach the LT. At the same time, the transport mass does not become larger than $3m_e$. 
This finding is in line with the observations from our simple $U=0$ example, but the difference between the two masses is greater. 
This is likely related to the effects of the interaction, which renormalizes both masses in the expected way: they both become larger with the increasing $U$.
At the doping $\delta=26\%$ relevant for the interpretation of the data in Ref.~\onlinecite{Legros2018}, we find $\mc/\mtot\approx 5$.
As already mentioned, this leads to the conclusion that the actual ratio of the scattering rate to the Planckian limit, $\alpha$, is around 5, rather than 1.

\begin{figure}
\centering    
    \includegraphics[width=1.0\columnwidth]{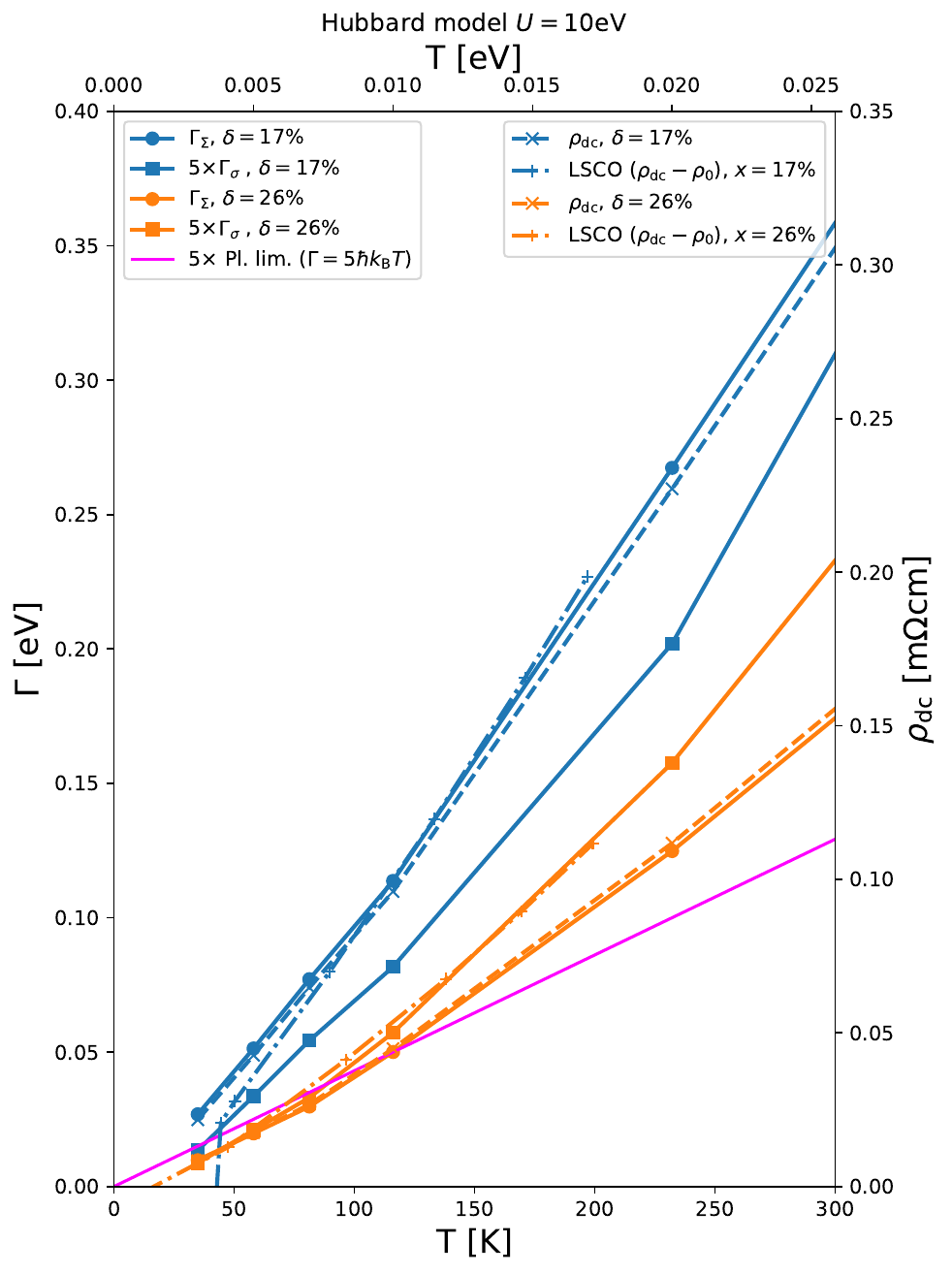}    
\caption{Two different estimates of scattering rate compared with the Planckian limit, as well as the theoretical and the experimental dc resistivity data.
The scattering rate extracted from the self-energy $\Gamma_\Sigma$ is almost perfectly proportional to the theoretical $\rho_\mathrm{dc}$, while the one extracted from the width of the Drude peak ($\Gamma_\sigma$) is not. Both $\rho_\mathrm{dc}$ and $\Gamma_\Sigma$ are roughly linear, but at $\delta=26\%$, one can observe some curvature, in both theory and experiment. $\Gamma_\sigma$ is much closer to Planckian limit than $\Gamma_\Sigma$ (note the different prefactors), but is much more curved.
}\label{fig:Gamma_vs_T}
\end{figure}

We cross-check this result by looking at another standard estimate of the effective scattering rate defined by\cite{Vucicevic2023,Hartnoll2022}
\begin{equation}
 \Gamma_\Sigma = -2\mathrm{Im}\Sigma(\omega=0)
\end{equation}
This estimate is derived by equating the Drude optical conductivity with the result of Kubo bubble calculation at weak-coupling\cite{Vucicevic2023}, assuming that the self-energy is purely local (which certainly holds at the level of our DMFT theory).
We plot the results as a function of temperature for two fixed dopings $\delta=17\%$ and $\delta=26\%$ in Fig.~\ref{fig:Gamma_vs_T}.
We find, indeed, that at $\delta=26\%$, $\Gamma_\Sigma$ is roughly linear and equal to about five times the Planckian limit.
At $\delta=17\%$, $\Gamma_\Sigma$ is significantly larger.
At both dopings, we compare $\rho_\mathrm{dc}(T)$ with $\Gamma(T)$ and find $\rho_\mathrm{dc}(T) \sim \Gamma(T)$ - the $\rho_{dc}$ closely tracks the effective scattering rate defined as $\Gamma_\Sigma$, which can then be used as a solid proxy for the dc resistivity\cite{Cha2020, ChaPNAS2020, Wu2022}. We also notice that neither $\rho_\mathrm{dc}(T)$ nor $\Gamma_\Sigma(T)$ are perfectly linear, especially at $\delta=26\%$. Non-linearity of $\rho_\mathrm{dc}(T)$ is also seen in the experimental data, as well as noted in the original analysis of this data in Ref.~\onlinecite{Cooper2009} - the resistivity was found to be the most linear in the range $17-23\%$ doping, and then progressively more quadratic with increasing doping.

\begin{figure*}
\centering
    
    \includegraphics[width=1.6\columnwidth]{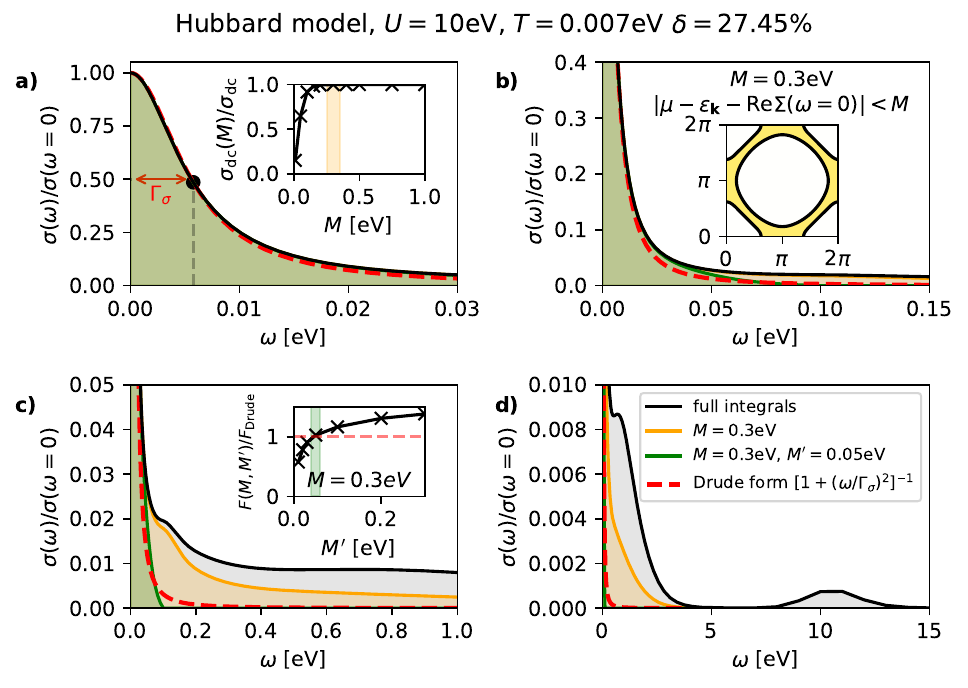}    
\caption{Partial optical conductivity compared to the full one and the Lorentzian fit to the Drude peak. Different panels focus on different energy ranges. See text for details. Inset to a): restricting the momenta with $M<\unit[0.3]{eV}$ leads to a loss of spectral weight at low frequency and an underestimation of $\sigma_\mathrm{dc}$. Inset to b): momenta included in the $M=\unit[0.3]{eV}$ calculation. Inset to c): f-sum changes as a function of the energy restriction $M'$, for a fixed $M$ ($F(M,M')=\int_0^\infty \mathrm{d}\omega \mathrm{Re}\sigma(\omega; M,M')$). At $M'=\unit[0.05]{eV}$, the obtained partial optical conductivity $\sigma(\omega; M,M')$ is in good agreement with the Lorentzian fitted to $\sigma(\omega)$ at lowest $\omega$'s, and has roughly the same f-sum. 
}\label{fig:Manalysis}
\end{figure*}

We emphasize that, at the level of DMFT, both $\rho_\mathrm{dc}(T)$ and $\Gamma_\Sigma(T)$ become quadratic as $T\rightarrow 0$\cite{Jarrell:1994ut,deng,VucicevicPRL2015,ding2017}. This is clearly consistent with our data - although $\rho_\mathrm{dc}(T)$ and $\Gamma_\Sigma(T)$ are roughly linear in the range of temperatures we investigate, the $T=0$ intercept of this linear law is negative; this means that at some lower temperature, the effective power-law exponent increases, as one must observe $\Gamma_\Sigma(T\rightarrow 0)=\rho_\mathrm{dc}(T\rightarrow 0)=0$.

\section{Effective Drude theories and the underlying assumptions of Planckian dissipation hypothesis}
\label{app:Drude_and_Planck}

The Planckian dissipation hypothesis (as formulated in \emph{Legros et al.}) relies on two major assumptions:
\begin{enumerate}
 \item One can describe \emph{all} conduction electrons using an effective Drude theory (which predicts a Lorentzian form of the optical conductivity, Eq.~\eqref{eq:Drude_optical}, and depends on only 3 effective parameters: effective mass, scattering rate and electron concentration).
 \item The scattering rate $\Gamma=\hbar/\tau$ is a linear function of temperature; as $\rho_\mathrm{dc}(T)$ curves are measured at a fixed doping, the electron concentration $n$ is a constant by definition and the effective mass $m^*$ is roughly constant; therefore, if $\Gamma\sim T$, it must be $\rho_\mathrm{dc}\sim T$.
 \end{enumerate}

The second assumption appears to be reasonably valid in our DMFT results (see Fig.~\ref{fig:Gamma_vs_T}).
The validity of the first assumption, however, needs to be checked.

The first question is whether the optical conductivity fits well the Drude form which is a Lorentzian peak (Eq.~\eqref{eq:Drude_optical}).
One can obtain the Kubo bubble expression for the optical conductivity by generalizing Eq.~\eqref{eq:Kubo_bubble} to finite frequencies. For the Hubbard model, and up to a constant prefactor, the generalized expression reads:
\begin{eqnarray}\label{eq:optical_cond_k}
\sigma(\nu) &\sim& \int_\mathrm{BZ} \mathrm{d}\mathbf{k} \;v^2_{x,\mathbf{k}}\int \mathrm{d}\omega \\ \nonumber
   && \times A_\mathbf{k}(\omega)A_\mathbf{k}(\omega+\nu)\frac{n_\mathrm{F}(\omega)-n_\mathrm{F}(\omega+\nu)}{\nu}
\end{eqnarray}
As already explained, at the level of DMFT, an additional simplification arises due to locality of self-energy \cite{khurana1990}, and one can write\cite{pruschke1993prb,jarrell1995,rozenberg1995,georges1996,Toschi2005,deng,VucicevicPRL2015}
\begin{eqnarray}\label{eq:optical_cond_epsilon}
\sigma(\nu) &\sim& \int_\mathrm{BZ} \mathrm{d}\varepsilon\; \Phi(\varepsilon)\int \mathrm{d}\omega \\ \nonumber
&& \times A(\varepsilon,\omega)A(\varepsilon,\omega+\nu)\frac{n_\mathrm{F}(\omega)-n_\mathrm{F}(\omega+\nu)}{\nu}
\end{eqnarray}
where $\Phi(\varepsilon)$ is the transport function (Eq.~\eqref{eq:transport_function}), and
$A(\varepsilon,\omega)=-\frac{1}{\pi}\mathrm{Im}\frac{1}{\omega+\mu-\varepsilon-\Sigma(\omega)}$
is the momentum-resolved spectral function, which now depends only on $\varepsilon_\mathbf{k}$ and not on $\mathbf{k}$ explicitly.

We compute the optical conductivity in the Hubbard model with relevant choice of parameters, and present it in Fig.~\ref{fig:Manalysis}.
We observe, as is well documented in previous studies\cite{jarrell1995,kokalj2017, Vucicevic2019,berthod}, that the optical conductivity has a very complicated frequency dependence, drastically different from a simple Lorentzian form. The intricate frequency-structure can only be fully appreciated by drawing $\sigma(\omega)$ at multiple zoom levels, each focusing on a different frequency range, as we do in Fig.~\ref{fig:Manalysis}. We identify four regions in the spectrum:
\begin{enumerate}
\item for $\omega<\unit[0.03]{eV}$, the Drude peak (roughly Lorentzian, comprising about 30$\%$ of the overall weight);
\item tiny bump at around $\omega \sim \unit[0.03]{eV}$, appearing at lower temperatures and higher dopings (similar features have been seen before\cite{Kim2026arxiv,Lin2009}, but we find in our solution that it is sensitive to numerical parameters of NRG; other features are much more robust in this sense; see App.~\ref{app:nrg_numerical} for the details of the NRG calculation used throughout this work);
\item the wide "Drude foot" in the range $\unit[0.2]{eV}\lesssim\omega\lesssim \unit[5]{eV}$, with the addition of the "mid-infrared peak" around $\sim \unit[1]{eV}$;
\item at $\omega \sim U = \unit[10]{eV}$ the "Hubbard peak" (comprising about 10$\%$ of the overall weight, but can be much more, depending on model parameters).
\end{enumerate}

The Lorentzian form Eq.~\eqref{eq:Drude_optical} cannot be reasonably fit to the entire optical conductivity. Therefore, the meaning of an effective Drude theory to describe the totality of the electrons in our model is questionable. Only the low frequency Drude peak can be well fit to a Lorentzian form (the red dashed curve on Fig.~\ref{fig:Manalysis}). However, if the effective Drude theory is expected to describe only a part of the optical conductivity spectrum, then it can no longer be thought of as describing \emph{all} the conduction electrons, and this would be inconsistent with the definition of the electron concentration $n$ employed in \emph{Legros et al.} Such an effective Drude theory can only be understood as describing the portion of electrons that contribute to the Drude peak (and to the Drude peak only). The number of such electrons is a priori unknown, yet the concentration $n$ is needed to extract the scattering rate from the $\rho_\mathrm{dc}$ alone (without the knowledge of the frequency dependence of $\sigma(\omega)$, as was done in \emph{Legros et al.}). Therefore, this theoretical approach might be impractical for analysis of experimental $\rho_\mathrm{dc}$ data. However, in other experiments, such as Brown et al.\cite{Brown2018}, the optical conductivity was (indirectly) accessible, and the scattering rate was, indeed, estimated as the half-width of the Drude peak. We emphasize that the underlying assumption of such an approach must be that it is possible to separate electrons into two groups: those that contribute to the Drude peak and those that do not, and that there is an effective Drude theory to describe the former. Otherwise, one could not identify the electrons in the Drude theory with any of the physical electrons, and one could not define the effective electron density $n$ and the effective mass $m^*$ separately, but only their ratio $m^*/n$.
Surely, one can forego the physical interpretation of the electrons in the effective Drude picture - this is, indeed, one possible approach\cite{Shekhter2026}.

Nevertheless, at the level of our DMFT theory we can try to identify the group of electrons that yield the Drude peak and thus give the physical meaning to the electrons in the effective Drude theory that describes it; this would allow us to extract $n$ and $m^*$ separately and have a complete Drude parametrization. One expects that it is the electrons with momenta near the Fermi surface and carrying low energies that primarily contribute to the Drude peak. To check whether this is true, we can restrict the integrations over $\mathbf{k}$ and $\omega$ in Eq.~\eqref{eq:optical_cond_k} when computing the optical conductivity. First, we require that the momenta are near the interacting Fermi surface, and thus satisfy
\begin{equation}
-M < \mu - \varepsilon_\mathbf{k}- \mathrm{Re}\Sigma(\omega=0) < M
\end{equation}
where $\varepsilon_\mathbf{k}$ is the dispersion in our Hubbard model, and $M$ is a parameter to be varied; $M\rightarrow0$ means we keep only the electrons precisely at the Fermi surface, and $M\rightarrow \infty$ means we keep all the electrons. This most easily translates to Eq.~\eqref{eq:optical_cond_epsilon} (that we actually implement) by replacing
\begin{equation}
\Phi(\varepsilon)\rightarrow\Phi(\varepsilon)\theta(M-|\mu-\varepsilon-\mathrm{Re}\Sigma(\omega=0)|)
\end{equation}
Next, we require that the electrons carry low energies, say below a threshold $M'$. This is implemented in Eq.~\eqref{eq:optical_cond_epsilon} by replacing
\begin{equation}
A(\varepsilon,\omega)\rightarrow A(\varepsilon,\omega)\theta(M'-|\omega)|)
\end{equation}
Further implementation details and the code used for this calculation are publicly available \cite{bubble}.
The resulting partial optical conductivity is labeled $\sigma(\omega; M,M')$; the full optical conductivity is recovered as $\sigma(\omega)=\sigma(\omega; M\rightarrow \infty,M'\rightarrow\infty)$. (If either of $M$, $M'$ is taken to infinity, we omit it from the argument list. Similar notation will be adopted for other quantities as well.)
The density of the portion of electrons we select this way can be computed by
\begin{eqnarray} \nonumber
n(M,M') &=& 2 \int \mathrm{d}\varepsilon \rho_0(\varepsilon) \theta(M-|\mu-\varepsilon-\mathrm{Re}\Sigma(\omega=0)|) \\
&& \times \int \mathrm{d}\omega A(\varepsilon,\omega) \theta(M'-|\omega|) n_\mathrm{F}(\omega).
\end{eqnarray}

We first vary $M$ while keeping $M'=\infty$. We find that taking $M<\unit[0.3]{eV}$ leads to a $\sigma(\omega; M)<\sigma(\omega)$ at low frequencies, including at $\omega=0$. This is shown in the inset in Fig.~\ref{fig:Manalysis}a. At $M=\unit[0.3]{eV}$, the entire Drude peak is perfectly well captured, but $\sigma(\omega; M)$ contains excess weight at relatively large frequencies that cannot be well fit to a Lorentzian form. 
%The portion of the BZ that we have selected this way is illustrated in inset.
We then vary $M'$ and find that reducing it "shaves off" the spectral weight at frequencies $\omega>M'$, without noticeably affecting the result at $\omega<M'$. We see that $\sigma(\omega; M=0.3\mathrm{eV}, M'=0.05\mathrm{eV})$ fits reasonably well to the Lorentzian form, and has overall weight roughly equal to that of the Lorentzian fitted to the lowest $\omega$'s (see inset in Fig.~\ref{fig:Manalysis}c). We conclude that it is indeed possible to select a portion of electrons that completely describe the Drude peak, and yield a partial optical conductivity that fits well to a Lorentzian form. We now compute $n(M,M')$ (with the above-mentioned values of $M$ and $M'$) and find that these electrons comprise only about 5\% of the total electron concentration in our model. We can now also extract the effective mass for those electrons (by looking at the f-sum of partial optical conductivity spectrum, $F(M,M')\approx F_\mathrm{Drude}$, 
%i.e. $m^*(M,M')=\pi n(M,M') e^2/(2 F(M,M'))$), and find that it is very small, about 0.28$m_e$. %WOULD NEED A FACTOR 2 IN FRONT, BUT THIS IS NOT WHAT WE COMPUTE ANYWAY
and using $m^*(M,M')/\mtot= (n(M,M')/n)/(F(M,M')/F)$, as well as knowledge of $\mtot$), and find that it is very small, about 0.28$m_e$. 
This effective Drude theory paints a picture drastically different from the one in \emph{Legros et al.}: 
the dc resistivity can be effectively represented by a very dilute gas of extremely light electrons, rather than a large number of very heavy electrons.

\begin{table*}
    \centering
    \begin{tabular}{cc|cc|cc|c}
         $\delta$ [\%] & $T$ [eV] & opt. $M$ [eV] & opt. $M'$ [eV] & $n(M,M')/n$ & $m^*(M,M')/m^*$ & $(m^*(M,M')/m^*)/(n(M,M')/n)$\\
         \hline \hline
         27.51 & 0.003 & 0.15 &  0.02 & 0.02   & 0.08  & 3.35\\
         26.77 &  0.02 & 0.75 &  0.14 & 0.1077 & 0.297 & 2.76\\
         \hline
         16.78 & 0.003 & 0.25 & 0.017 & 0.0192 & 0.09  & 4.99\\
         15.62 &  0.02 & 1    &  0.17 & 0.0802 & 0.282 & 3.51\\         
    \end{tabular}
    \caption{Data showing that the ratio $m^*/n$ cannot be considered independent of temperature if Drude theory is identified only with the low-frequency Drude peak in the optical conductivity. Results are obtained following the procedure explained in the text and illustrated in Fig.~\ref{fig:Manalysis}: $M$ is chosen such that Drude peak is fully preserved, and $M'$ is chosen such that high-frequency contributions are shaved off, and that partial optical conductivity has the same weight as the Lorentzian fitted to the lowest frequencies.
    At roughly a fixed doping, if temperature changes by roughly an order of magnitude (from $\approx \unit[35]{K}$ to $\approx \unit[232]{K}$), the effective mass increases by a factor of about 3.5, while the fraction of the electrons that contribute to the Drude peak by a factor of about 4-5, from about 2\% to about 10\%. The ratio $m^*/n$ that enters the Drude expression for dc resistivity drops by about 20-30\%, thus invalidating the idea that dc resistivity is simply proportional to the scattering rate $\rho_\mathrm{dc}\sim \Gamma$ (when measured as a function of temperature at a fixed doping).
    }
    \label{tab:n_vs_m_vs_T}
\end{table*}

We now compute the optical conductivity throughout the phase diagram and extract the scattering rate as the half-width of the Drude peak (to distinguish it from other methods of estimating $\Gamma$, we denote it $\Gamma_\sigma$). The results are shown in Fig.~\ref{fig:Gamma_vs_T}. We find that $\Gamma_\sigma$ is universally much smaller than $\Gamma_\Sigma$. However, it is not a particularly linear function of temperature, and it varies from about $0.5k_\mathrm{B}T$ to about $2k_\mathrm{B}T$, depending also on the doping. Comparing $\Gamma_\sigma(T)$ with $\rho_\mathrm{dc}(T)$ shows that they are not proportional. The reason is that the concentration $n$ and the effective mass $m^*$ appearing in the Drude formula can no longer be considered constants of temperature, thus $\rho_\mathrm{dc}$ is no longer simply proportional to $\Gamma$. 
We show this clearly in Table~\ref{tab:n_vs_m_vs_T}.
We conclude that defining a scattering rate via an effective Drude theory that describes only the lowest-energy part of $\sigma(\omega)$ (i.e. taking it to be $\Gamma_\sigma$) is inconsistent with the second assumption of the Planckian hypothesis, which states that $\Gamma \sim T$ leads to $\rho_\mathrm{dc}\sim T$. On the other hand, the approach based on $\Gamma_\Sigma$ has the desired property $\rho_\mathrm{dc} \sim \Gamma_\Sigma \sim T$, but we find that $\Gamma_\Sigma$ is much larger than the Planckian limit.

To summarize, the definition of an effective scattering rate ultimately depends on how an effective Drude theory is matched to the full theoretical (or experimental) results. There is an ambiguity in this matching, but once the assumptions about it are specified, the effective Drude parameters can be extracted through a well-defined procedure. However, under none of the assumptions do we find that the effective Drude mass can be extracted as the cyclotron mass. In addition, we find that identifying the effective Drude density with that of the totality of physical conduction electrons (one per unit cell minus the doping) is poorly justified. Therefore, the choice of \emph{Legros et al.} to use $n=(1-\delta)/{a^2 d}$ and $m^*=\mc$ in their analysis is both internally inconsistent and lacks a firm physical basis.

\section{Orbital- and band-resolved spectral functions in the Emery model and the effective single-band model}
\label{app:orbital_resolved}

\begin{figure*}
\centering
    \includegraphics[width=\columnwidth,trim=0 0 14cm 0cm, clip]{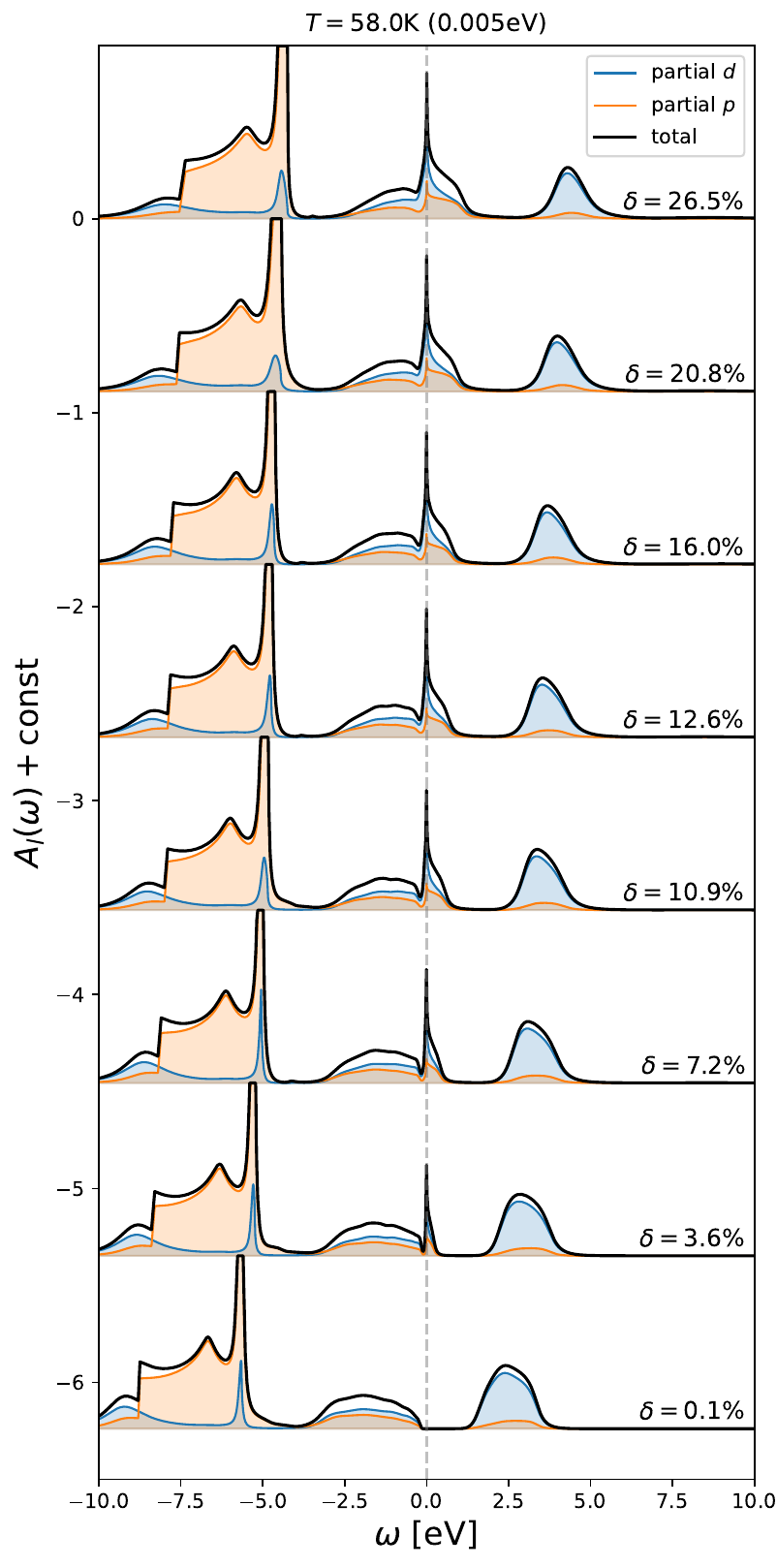}
    \includegraphics[width=\columnwidth,trim=0 0 14cm 0cm, clip]{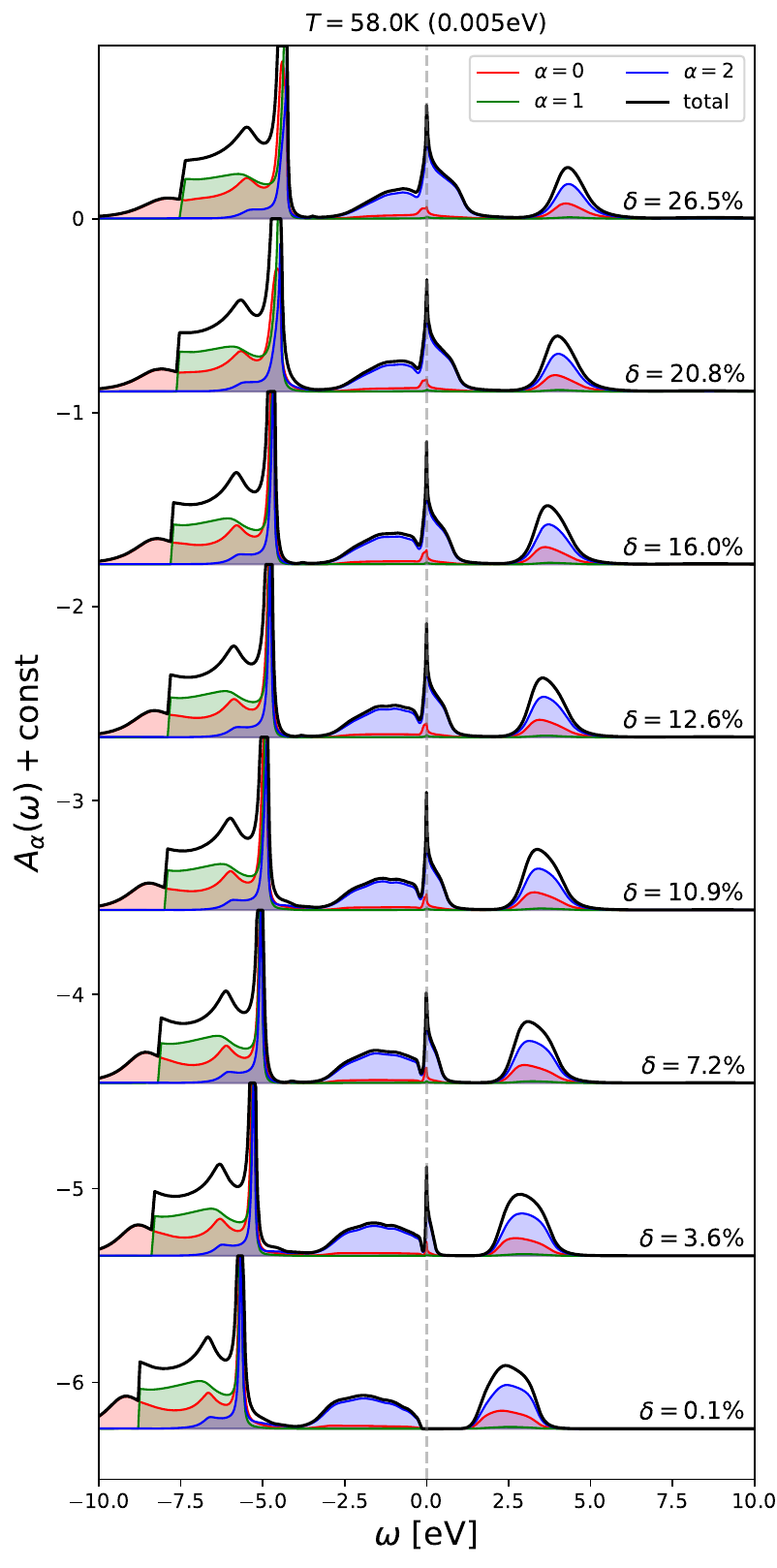}
\caption{
Left: Orbital-resolved spectral function in the Emery model for the T-structure, at various dopings for a fixed low temperature, as computed within DMFT. The $p_x$ and $p_y$ orbitals are degenerate and we only show their cumulative contribution. The ZRSB has roughly equal amount of $d$ and $p$ character. Right: non-interacting-eigenband-resolved spectral function in the same cases as on the left. The ZRSB has an almost pure $\alpha=2$ character, corresponding to the top band in the upper-right panel of Fig.~\ref{fig:dft}. This has implications for the formulation of an effective single-band model aimed at reproducing the physics of the Emery model at low energies.
}
\label{fig:Emery_orbital_resolved}
\end{figure*}

The orbital-resolved spectral function in the Emery model is defined as
\begin{equation}
A_l(\omega)= -\frac{1}{\pi}\mathrm{Im}G_{\mathrm{loc},ll}(\omega).
\end{equation}
The $p_x$ and $p_y$ orbitals are degenerate, and we define
\begin{equation}
A_p(\omega)\equiv \sum_{l={p_x,p_y}}A_l(\omega)=2 A_{p_x}(\omega)=2 A_{p_y}(\omega)
\end{equation}
We focus on the hole-doped T-structure, relevant for LSCO/LBCO.
In Fig.~\ref{fig:Emery_orbital_resolved} (left) we see, as was observed before\cite{Emery1987,Tseng2025}, that the band at the Fermi level has a mixed character---the $d$ and $p$ orbitals contribute to it roughly equally. This is a good illustration of the fact that the physical meaning of the single orbital in the Hubbard model is not that of pure copper $d$ orbital. As can be seen in Fig.~\ref{fig:dft} the orbital character of the band at the Fermi level in DFT (as well as in the downfolded Emery band structure) is a (momentum-dependent) mix of $d$ and $p$ orbitals. In the Emery model, it is the UHB and the LHB that have a predominant $d$-character. The sharp step between the LHB and the CTB appears to be a robust feature of the spectral function in the Emery model, observed before\cite{deMediciWangCaponeMillis2009,KowalskiPNAS2021,Tseng2025}.

The spectral function can also be decomposed into contributions associated with the eigenstates of the non-interacting Hamiltonian. We define
\begin{equation}
A_\alpha(\omega)=  -\frac{1}{\pi}\int_\mathrm{BZ} \frac{\mathrm{d}\mathbf{k}}{(2\pi)^2}  [\mathbf{P}^{-1}_\mathbf{k} \mathbf{G}_\mathbf{k}(\omega) \mathbf{P}_\mathbf{k}]_{\alpha\alpha}
\end{equation}
where $\mathbf{P}_\mathbf{k}$ is the basis-change matrix that diagonalizes the non-interacting Hamiltonian matrix at wave-vector$\mathbf{k}$, i.e.
\begin{equation}
 \mathbf{P}^{-1}_\mathbf{k} \mathbf{h}_\mathbf{k} \mathbf{P}_\mathbf{k} = \left( \begin{array}{ccc}
                                                                              \tilde{E}_{\alpha=0,\mathbf{k}} &0 &0 \\
                                                                              0 & \tilde{E}_{\alpha=1,\mathbf{k}}  &0 \\
                                                                              0 & 0 & \tilde{E}_{\alpha=2,\mathbf{k}}
                                                                            \end{array}\right)
\end{equation}
We plot the band-resolved spectral function $A_\alpha(\omega)$ in Fig.~\ref{fig:Emery_orbital_resolved} (right). We see that the band at the Fermi level has an almost pure $\alpha=2$ character, which corresponds to the top eigenband of the non-interacting Hamiltonian (see Fig.~\ref{fig:dft}). To describe the physics of the Emery model around the Fermi level, one would then expect that a single-band model with the dispersion $\tilde{E}_{\alpha=2,\mathbf{k}}$ would suffice. The partial spectral function $A_{\alpha=2}(\omega)$ captures also the upper Hubbard band (although not in its entirety), which suggests that the effective coupling in such a single-band model would be similar to what we have in our Hubbard model in Fig.~\ref{fig:Emery_vs_Hubbard_Aloc_Tstruct}, i.e. $U^\mathrm{Hubbard}=\unit[5]{eV}$.
However, we also see a sharp peak in $A_{\alpha=2}(\omega)$ at a large negative frequency; to capture this with a single-band model, a coupling to some "external" weakly dispersive states would likely be needed, i.e. one might need to include in the model an effective hybridization (retarded hopping).
Indeed, such terms naturally arise from downfolding, as was recently discussed in Ref.~\onlinecite{Profe2026}, but are rarely considered in practice.
Most importantly, the single-band dispersion $\tilde{E}_{\alpha=2,\mathbf{k}}$ is not the same as what one would normally have in the effective Hubbard model. The difference is significant both in shape and the bandwidth, as can be seen in Fig.~\ref{fig:dft}.

\section{Velocity matrix in the Emery model}
\label{app:velocity_matrix}

To derive the current operator, and the associated velocity matrix in our models, we make use of the Peierls substitution.
The derivation is trivial in the case of the Hubbard model, but is somewhat more involved in the case of the Emery model, which is why we present it here.

The Peierls phase between two points in space is
\begin{eqnarray}\label{eq:Peirels_phase}
f(\mathbf{r},\mathbf{r}') &=& \frac{e}{\hbar}\int_{\mathbf{r}}^{\mathbf{r}'} \mathbf{A}(\bar{\mathbf{r}}) \cdot \mathrm{d}\bar{\mathbf{r}}.
\end{eqnarray}
For uniform static electric field $ \mathbf{A}(\mathbf{r}) = -\mathbf{E} t $
and we write
\begin{eqnarray}
f(\mathbf{r},\mathbf{r}')= f(\mathbf{r}'-\mathbf{r}) = \frac{e}{\hbar} \mathbf{A} \cdot (\mathbf{r}'-\mathbf{r})
\end{eqnarray}
However, to be able to take derivatives with respect to local vector potential $\mathbf{A}(\mathbf{r})$ we need to retain the spatial dependence of $\mathbf{A}$. There is some ambiguity, as this can be done in multiple ways that are equally valid. We choose
\begin{eqnarray}
f(\mathbf{r},\mathbf{r}')&=& \frac{e}{\hbar} \mathbf{A}(\mathbf{r}_\mathrm{u.c.}(\mathbf{r})) \cdot (\mathbf{r}'-\mathbf{r})
\end{eqnarray}
where $\mathbf{r}_\mathrm{u.c.}(\mathbf{r})$ is the position vector of the unit cell to which the vector $\mathbf{r}$ belongs.
This choice affects the definition of the local current operator, but should not affect the uniform current-current correlation function which involves a sum over all $\mathbf{r}$.
The tight-binding matrix is then modified due to the presence of a uniform electric field in the following way:
\begin{equation}
 \mathbf{H}_0[\mathbf{A}] = \mathbf{H}_0[\mathbf{A}=0] \circ
 e^{i\mathbf{f}},
\end{equation}
where $\circ$ denotes element-wise multiplication and $\mathbf{f}$ is a matrix constructed out of Peierls phases between all pairs of orbitals.

We start from the Hamiltonian written in the basis of $d,p_x,p_y$-orbitals (denoted by $l=0,1,2$, respectively). For the sake of convenience, we define a row vector of creation operators
\begin{equation}
\mathbf{\Psi}^\dagger_{\sigma,\mathbf{r}} = ( c^\dagger_{l=0,\sigma,\mathbf{r}}, c^\dagger_{l=1,\sigma,\mathbf{r}}, c^\dagger_{l=2,\sigma,\mathbf{r}} )
\end{equation}
where $\mathbf{r}$ denotes the position of the unit cell and points to the $d$-orbital within it.
The positions of orbitals are:
\begin{equation}
 \begin{array}{cc}
   l=d, &\mathbf{r}, \\
   l=p_x, &\mathbf{r}-a\mathbf{e}_x/2, \\
   l=p_y, &\mathbf{r}-a\mathbf{e}_y/2
 \end{array}
\end{equation}
where $a$ is the lattice spacing, and unit vectors are $\mathbf{e}_x=(1,0)$, and $\mathbf{e}_y=(0,1)$. The conventions that we follow are illustrated in Fig.~\ref{fig:real_space_illustration}.

The non-interacting part of the Hamiltonian reads
\begin{eqnarray}
 \hat{H}_0 = \sum_{\sigma,\mathbf{r}\mathbf{r}'} \mathbf{\Psi}^\dagger_{\sigma,\mathbf{r}} \mathbf{h}_{\mathbf{r}\mathbf{r}'} \mathbf{\Psi}_{\sigma,\mathbf{r}'}
\end{eqnarray}
where $\mathbf{h}_{\mathbf{r}\mathbf{r}'}$ is a $3\times3$ matrix in the orbital space with the following elements:
\begin{widetext}
\small
\begin{eqnarray}\label{eq:Horig_real_space2}
 [\mathbf{h}_{\mathbf{r}\mathbf{r}'}]_{00} &=&
        \varepsilon_d \delta_{\mathbf{r}\mathbf{r}'} \\ \nonumber
 [\mathbf{h}_{\mathbf{r}\mathbf{r}'}]_{01} 
        &=&
        t_{pd}(\delta_{\mathbf{r}\mathbf{r}'}e^{-i\frac{ea}{2\hbar}A_x(\mathbf{r})}-\delta_{\mathbf{r}+a\mathbf{e}_x,\mathbf{r}'}e^{i\frac{ea}{2\hbar}A_x(\mathbf{r})}) \\ \nonumber
 [\mathbf{h}_{\mathbf{r}\mathbf{r}'}]_{02} 
        &=&
        t_{pd}(\delta_{\mathbf{r}\mathbf{r}'}e^{-i\frac{ea}{2\hbar}A_y(\mathbf{r})}-\delta_{\mathbf{r}+a\mathbf{e}_y,\mathbf{r}'}e^{i\frac{ea}{2\hbar}A_y(\mathbf{r})}) \\ \nonumber
 [\mathbf{h}_{\mathbf{r}\mathbf{r}'}]_{10} 
        &=&
        t_{pd}(\delta_{\mathbf{r}\mathbf{r}'}e^{i\frac{ea}{2\hbar}A_x(\mathbf{r})}-\delta_{\mathbf{r}-a\mathbf{e}_x,\mathbf{r}'}e^{-i\frac{ea}{2\hbar}A_x(\mathbf{r})}) \\ \nonumber
 [\mathbf{h}_{\mathbf{r}\mathbf{r}'}]_{11} 
        &=&
        \varepsilon_p \delta_{\mathbf{r}\mathbf{r}'} + t'_{pp}(
              \delta_{\mathbf{r}+a\mathbf{e}_x,\mathbf{r}'}e^{i\frac{ea}{\hbar}A_x(\mathbf{r})} 
          + \delta_{\mathbf{r}-a\mathbf{e}_x,\mathbf{r}'}e^{-i\frac{ea}{\hbar}A_x(\mathbf{r})}
            ) \\ \nonumber
 [\mathbf{h}_{\mathbf{r}\mathbf{r}'}]_{12} 
        &=&
        t_{pp}( \delta_{\mathbf{r}\mathbf{r}'}e^{i\frac{ea}{2\hbar}(A_x(\mathbf{r})-A_y(\mathbf{r}))}
               -\delta_{\mathbf{r}+a\mathbf{e}_y,\mathbf{r}'}e^{i\frac{ea}{2\hbar}(A_x(\mathbf{r})+A_y(\mathbf{r}))} 
                       -\delta_{\mathbf{r}-a\mathbf{e}_x,\mathbf{r}'}e^{i\frac{ea}{2\hbar}(-A_x(\mathbf{r})-A_y(\mathbf{r}))}
               +\delta_{\mathbf{r}-\mathbf{e}_x+\mathbf{e}_y,\mathbf{r}'}e^{i\frac{ea}{2\hbar}(-A_x(\mathbf{r})+A_y(\mathbf{r}))}
              )\\ \nonumber
 [\mathbf{h}_{\mathbf{r}\mathbf{r}'}]_{20} 
        &=&
        t_{pd}(\delta_{\mathbf{r}\mathbf{r}'}e^{i\frac{ea}{2\hbar}A_y(\mathbf{r})}-\delta_{\mathbf{r}-a\mathbf{e}_y,\mathbf{r}'}e^{-i\frac{ea}{2\hbar}A_y(\mathbf{r})}) \\ \nonumber
 [\mathbf{h}_{\mathbf{r}\mathbf{r}'}]_{21} 
        &=&
        t_{pp}( \delta_{\mathbf{r}\mathbf{r}'}e^{i\frac{ea}{2\hbar}(-A_x(\mathbf{r})+A_y(\mathbf{r}))}
                -\delta_{\mathbf{r}-a\mathbf{e}_y,\mathbf{r}'}e^{i\frac{ea}{2\hbar}(-A_x(\mathbf{r})-A_y(\mathbf{r}))} 
                -\delta_{\mathbf{r}+a\mathbf{e}_x,\mathbf{r}'}e^{i\frac{ea}{2\hbar}(A_x(\mathbf{r})+A_y(\mathbf{r}))}
                +\delta_{\mathbf{r}+a\mathbf{e}_x-a\mathbf{e}_y,\mathbf{r}'}e^{i\frac{ea}{2\hbar}(A_x(\mathbf{r})-A_y(\mathbf{r}))}
              ) \\ \nonumber
 [\mathbf{h}_{\mathbf{r}\mathbf{r}'}]_{22} 
            &=&
        \varepsilon_p \delta_{\mathbf{r}\mathbf{r}'} + t'_{pp}(
              \delta_{\mathbf{r}+a\mathbf{e}_y,\mathbf{r}'}e^{i\frac{ea}{\hbar}A_y(\mathbf{r})}
            + \delta_{\mathbf{r}-a\mathbf{e}_y,\mathbf{r}'}e^{-i\frac{ea}{\hbar}A_y(\mathbf{r})}
            ) \\ \nonumber
\end{eqnarray}
\normalsize
\end{widetext}

We now proceed with computing the current operator as the derivative with respect to vector potential.
First, we apply the Fourier transformation to diagonalize each element of $\mathbf{h}$ and obtain the following form:
\begin{equation}\label{eq:quadratic_part}
 \hat{H}_0 = \sum_{\sigma,\mathbf{k}}\mathbf{\Psi}^\dagger_{\sigma,\mathbf{k}} \mathbf{h}_\mathbf{k} \mathbf{\Psi}_{\sigma,\mathbf{k}}
\end{equation}
with $\mathbf{\Psi}^\dagger_\mathbf{k} = ( c^\dagger_{l=0,\sigma,\mathbf{k}}, c^\dagger_{l=1,\sigma,\mathbf{k}}, c^\dagger_{l=2,\sigma,\mathbf{k}} )$ and
\begin{widetext}
\begin{equation}
 \mathbf{h}_\mathbf{k} = \left(
 \begin{array}{ccc}
  \varepsilon_d & 2t_{pd}\sin\frac{ a(k_x-\frac{e}{\hbar}A_x)}{2} & -2t_{pd}\sin\frac{a(k_y-\frac{e}{\hbar}A_y)}{2} \\
  \mathrm{H.c.} & \varepsilon_p+2t'_{pp} \cos (a(k_x-\frac{e}{\hbar}A_x)) & -4t_{pp} \sin\frac{a(k_x-\frac{e}{\hbar}A_x)}{2}\sin\frac{a(k_y-\frac{e}{\hbar}A_y)}{2} \\
  \mathrm{H.c.} & \mathrm{H.c.} &\varepsilon_p+2t'_{pp} \cos (a(k_y-\frac{e}{\hbar}A_y))
 \end{array}
 \right).
\end{equation}

The current is computed as 
\begin{eqnarray}
 \mathbf{j}(\mathbf{q=0}) &=& -\left. V^{-1}\frac{\partial \mathbf{H_0}}{\partial \mathbf{A}(\mathbf{q=0})}\right|_{\mathbf{A}\rightarrow 0} \\ \nonumber
 j_x(\mathbf{q=0}) &=& -\left.2V^{-1}\sum_{\mathbf{k}}\mathbf{\Psi}^\dagger_{\mathbf{k}} (\partial_{A_x} \mathbf{h}_{\mathbf{k}}) \mathbf{\Psi}_{\mathbf{k}}\right|_{\mathbf{A}\rightarrow 0},
\end{eqnarray}
with
\begin{equation}
 \partial_{A_x} \mathbf{h}_\mathbf{k} = \left(
 \begin{array}{ccc}
  0 & -\frac{ea}{\hbar}t_{pd}\cos\frac{ a(k_x-\frac{e}{\hbar}A_x)}{2} & 0 \\
  \mathrm{H.c.} & \frac{2ea}{\hbar} t'_{pp} \sin (a(k_x-\frac{e}{\hbar}A_x)) & \frac{2ea}{\hbar}t_{pp} \cos\frac{a(k_x-\frac{e}{\hbar}A_x)}{2}\sin\frac{a(k_y-\frac{e}{\hbar}A_y)}{2} \\
  \mathrm{H.c.} & \mathrm{H.c.} & 0
 \end{array}
 \right).
\end{equation}
\end{widetext}
Setting the field $\mathbf{A}$ to zero, we obtain the final expression:
\begin{eqnarray}
 j_x(\mathbf{q=0}) &=& \frac{2}{a^2c}\frac{1}{N}\sum_{\mathbf{k}}\mathbf{\Psi}^\dagger_{\mathbf{k}} \mathbf{v}_{x,\mathbf{k}} \mathbf{\Psi}_{\mathbf{k}}
\end{eqnarray}
with
\begin{equation}\label{eq:vxk_from_momentum2}
 \mathbf{v}_{x,\mathbf{k}} = \frac{ea}{\hbar}\left(
 \begin{array}{ccc}
  0 & t_{pd}\cos\frac{ak_x}{2} & 0 \\
  \mathrm{H.c.} & - 2 t'_{pp} \sin (ak_x) & -2t_{pp} \cos\frac{ak_x}{2}\sin\frac{ak_y}{2} \\
  \mathrm{H.c.} & \mathrm{H.c.} & 0
 \end{array}
 \right).
\end{equation}
We define the reduced velocity as
\begin{equation}\label{eq:v_matrix_definition_app}
 \tilde{\mathbf{v}}_{x,\mathbf{k}} = 
 -\frac{\hbar}{ea}\left. (\partial_{A_x} \mathbf{h}_{\mathbf{k}}) \right|_{\mathbf{A}\rightarrow 0}.
\end{equation}
In the corresponding Eq.~\eqref{eq:v_matrix_definition} in the main text, there is no factor $1/a$ because we have switched to reduced momentum $a\mathbf{k}\rightarrow \mathbf{k}$.

\section{NRG numerical parameters and algorithmic details}
\label{app:nrg_numerical}

The numerical renormalization group calculations have been performed with the NRG Ljubljana code \cite{nrglj}. 

The discretization has been performed using the algorithm from Ref.~\cite{resolution} using parameter $\Lambda=2$ and $N_z=8$ interleaved discretization grids. All functions were sampled on an adaptive mesh, which was initialized as a uniform geometric mesh with parameter $r=1.1$, and which has been subsequently refined using the following algorithm: during a pass over all existing data points, one quantifies the accuracy of linear extrapolation from two preceding data points to the current data point. If the relative error is large (higher than $10^{-3}$), a new mesh point is added to the mesh (midway between the previous and the current point). We found that this procedure vastly improves the convergence in the Emery model which features a square-root van Hove singularity which requires a mesh refinement in its vicinity. The same refined mesh was used for all quantities (hybridization function, computed raw correlators, estimated self-energy, local and lattice spectral functions).

The NRG iteration made use of the full-density matrix algorithm \cite{Weichselbaum2007}. In truncation, we retained states up to 10$E_N$, where $E_N$ is the characteristic energy scale at $N$-th step of the iteration, but no more than 10000, and never less than 500. 

The broadening of raw spectra has been performed using the modified log-Gaussian kernel with parameter $\alpha=0.15$.  The self-energy was computed using the improved estimator introduced in Ref.~\cite{Kugler2022}. 

Linear mixing of hybridization functions with $\alpha_\mathrm{mix}=0.5$ has been used to improve DMFT self-consistency cycle convergence. 

The results do not depend significantly on any of the parameter choices, beyond the typical systematic error in the NRG procedure which is in the few percent range \cite{bulla2008,resolution}.

\begin{acknowledgments}
We acknowledge useful discussions with Malte Rösner, Jeremija Kovačević, Matija Čulo, Neven Barišić, Nigel Hussey and Jernej Mravlje.
We acknowledge discussions on numerical integration of singular functions with Samir El Shawish.
Malte Rösner, Yusuke Nomura and Chia-Nan Yeh have contributed to the work on wannierization.
Chia-Nan Yeh, Miguel Morales and Malte Rösner have contributed to the computation of the values of the bare Coulomb interaction.
We acknowledge contributions of Upendra Kumar in the early stages of the work.
Computations were performed on the PARADOX supercomputing facility
(Scientific Computing Laboratory, Center for the Study of Complex
Systems, Institute of Physics Belgrade) and IJS F1 department cluster. J.~V. acknowledges funding
provided by the Institute of Physics Belgrade, through the grant by
the Ministry of Education, Science, and Technological Development of
the Republic of Serbia, as well as funding by the European Research Council, grant ERC-2022-StG: 101076100.
R.\v Z. is supported by the Slovenian Research and Innovation Agency (ARIS)
under Program P1-0416.
\end{acknowledgments}

\bibliography{refs.bib}
\end{document}